%% file: main.tex
\documentclass[a4paper,twocolumn,11pt,unpublished]{quantumarticle}
\ifdefined\pdfoutput\pdfoutput=1\fi
\usepackage[utf8]{inputenc}
\usepackage[english]{babel}
\usepackage[T1]{fontenc}
\usepackage{array}[=2016-10-06]
\usepackage{graphicx,booktabs,amssymb,amsthm,mathtools}
\usepackage[numbers,sort&compress]{natbib}
\usepackage{microtype}
\usepackage{xcolor}
\usepackage[colorlinks=true,allcolors=blue,hypertexnames=false]{hyperref}
\newtheorem{theorem}{Theorem}
\newtheorem{proposition}[theorem]{Proposition}
\newtheorem{corollary}[theorem]{Corollary}
\newcommand{\Tr}{\operatorname{Tr}}
\newcommand{\norm}[1]{\lVert#1\rVert}
\newcommand{\unknown}{\textnormal{UNKNOWN}}

\newcommand{\diag}{\operatorname{diag}}

\begin{document}
\title{Readout Dependent Parameter Identifiability of Quantum Kernels}
\input{submission_metadata}
\begin{abstract}
A circuit parameter can change measured features without changing
their projected Gaussian kernel. We characterize this lost dependence
and distinguish recovery by metric design from nonlinear feature
expansion. With analytic feature coefficients, a fixed finite data
basis affinely independent on the input domain, a fixed positive
definite metric, and fixed bandwidth, the coefficient Gram matrix
determines all-input kernel equivalence, global coordinate inactivity,
and generic differential rank. An exact four-qubit family exhibits
a sharp boundary: diagonal metrics give rank two, dense metrics reach
but cannot exceed three, and no fixed affine expansion crosses this
ceiling. One appended nonlinear scalar coordinate suffices for rank
four and is minimal when the original features are retained. A single
pairwise product achieves this with an explicit nondiagonal metric;
a scalar sum of two products achieves it with a diagonal metric.
Both use existing ideal local means. Exact certificates establish
ranks in 54 selected circuit/readout settings. Matched finite-shot
comparisons show that minimal representation size need not give the
best sensitivity among the tested designs. Under a factorized
measurement distribution, conditional product estimation cannot
increase variance at a fixed design. Execution audits further
separate kernel identifiability from compiled cost.
\end{abstract}
\maketitle

\section{Introduction}
A quantum kernel specifies which distinctions in a quantum feature
map are available to a classical learner \cite{schuld,huang}.
A circuit parameter may change the prepared state and its measured
expectation values while leaving all pairwise kernel values unchanged.
Counting circuit parameters, or even identifying all directions that
change measured features, therefore need not determine the dimension
of the resulting kernel family. This distinction matters when choosing
a readout, designing a metric on its features, or removing redundant
parameters: each operation acts on a different object.

We study parameter identifiability for projected Gaussian kernels
built from ideal expectation values. Our central question is which
parameter distinctions survive the specified readout and metric,
and how dependence lost at the kernel level can be recovered.
The answer has three parts: an exact description of kernel equivalence,
limits on what a fixed metric can recover, and a separate assessment
of statistical resolution and executable cost. Figure~\ref{fig:overview}
summarizes these connections.

\input{overview_figure}

\paragraph{Characterizing the kernel family.}
For features of the form $f_\theta(x)=C(\theta)h(x)+b(\theta)$,
we use the coefficient Gram matrix $G(\theta)=C(\theta)^TMC(\theta)$
to characterize equality of kernels on all input pairs, global
coordinate inactivity, and generic differential rank
(Theorem~\ref{thm:gram}). The assumptions are a fixed metric
$M\succ0$, fixed bandwidth $\gamma>0$, analytic coefficients on an
open connected parameter box, and linear independence of
$\{1,h_1,\ldots,h_A\}$ on the actual input domain.
The resulting rank counts locally identifiable parameter combinations
at regular points. Sensitivity to each individual coordinate need not
imply full rank, and full generic rank need not imply global injectivity.
For the exact rational-frequency subclass specified in
Sec.~\ref{sec:gram}, this description also yields a finite decision
procedure.

\paragraph{Separating metric recovery from representation recovery.}
The coefficient description identifies directions invisible to every
fixed positive definite metric and bounds the rank attainable by any
single metric. An exact four-qubit Hubregtsen/local Z family makes
the distinction explicit: positive diagonal metrics give rank two,
whereas dense metrics reach but cannot exceed rank three
(Proposition~\ref{prop:recoveryboundary}). No fixed affine feature
expansion crosses the same ceiling
(Corollary~\ref{cor:affineceiling}). A nonlinear expansion does:
one appended scalar coordinate is sufficient and minimal when the
original four features are retained and the metric is fixed
independently of inputs and trainable parameters.

The number of appended coordinates must be distinguished from the
number of product terms. A single pairwise product $f_0f_2$ attains
rank four with an explicit nondiagonal metric
(Theorem~\ref{thm:oneproduct}); with a diagonal metric, any single
pairwise product gives rank at most three. Nevertheless, the single
scalar coordinate $f_0f_2+f_1f_3$ already attains rank four with a
positive diagonal metric (Appendix~\ref{app:oneproduct}). Appending
the two products separately is another sufficient construction.
Table~\ref{tab:recoverymechanism} distinguishes these designs.
They are nonlinear functions of existing exact local means in this
family, so the recovery does not require additional ideal measurement
information. The minimum concerns appended representation dimension,
not measurement count, global parameter recovery, or optimal precision.

\paragraph{Exact evidence and operational consequences.}
We provide exact rank certificates for 54 selected circuit/readout
settings, using distinct proof routes for the fixed-frequency and
product-angle cases. The latter are not covered by the fixed finite
basis theorem. Matched finite-shot comparisons on previously frozen
panels separate rank recovery from local resolution: among the tested
fixed designs, the smallest sufficient representation need not have
the best sensitivity. These comparisons use common measurement
outcomes and shot budgets; they do not optimize over all metrics
within each feature class. Under a factorized Z distribution,
conditional product estimation cannot increase variance or worsen
the corresponding noise-normalized local sensitivity at a fixed design
(Corollary~\ref{cor:productsensitivity}). Execution audits separately
compare certified redundancies with compiled work after global
normalization. Rank recovery alone establishes neither predictive
improvement nor execution savings.

\paragraph{Relation to prior work.}
Our equivalence relation is defined by a fixed readout and kernel
metric, rather than by symmetries of the input data
\cite{larocca,meyer}. Parametric equivalence checking preserves
circuit unitaries \cite{peham,rosswesley}; dead-gate elimination
analyzes contributory measurement outcomes and their use by a
classical host \cite{deadgate,deadgatejournal}. QuCLEAR transfers
Clifford processing into observables or classical postprocessing
\cite{quclear}, while Paulihedral exploits Pauli structure for
scheduling and synthesis \cite{paulihedral}. Kernel equivalence can
hold even when both the unitary and individual measured features change.
Table~\ref{tab:contracts} compares these preservation targets.

\input{contracts_table}

State-geometry methods characterize redundant parameters and the
dimension of accessible quantum states \cite{funcke,haug}.
Fourier analysis supplies finite feature representations
\cite{schuldFourier}, while constraints on coefficients and circuit
harmonic matrices describe structure beyond the frequency spectrum
\cite{mhiri2025constrained,CHM}. Building on these representations,
we study the coefficient Gram that determines pairwise feature
distances. Its equivalence classes can differ from those defined by
quantum states or individual feature coefficients.

Readout-dependent invariance also has familiar special cases.
A common output unitary cancels in fidelity kernels, and embedding
placement can erase gate dependence \cite{salmenpera}. For projected
kernels, common local coordinate rotations preserve isotropically
weighted local Pauli features; their universal connected gauges follow
from a locality-preserving unitary classification
\cite{cpr2019locality} (Sec.~\ref{sec:gaugeclass}). Changing the
observables or metric can break these invariances. Observable and
metric learning already have precedents
\cite{chenMeasure2025,changMetric2022}; our contribution is to
characterize recoverable dependence and its limits for a specified
kernel, with exact certificates and separate statistical and
execution checks.

\paragraph{Scope.}
The finite decision guarantee applies only to the exact subclass
specified above. Generic differential rank describes local smooth
dimension at regular points, not a global coordinate system.
Internal transport is sound but incomplete. The execution audit is
not a full Paulihedral or QuCLEAR implementation, and no matched
QFI-pruning benchmark is reported. The approximate output-symmetry
bound is given in Appendix~\ref{app:bound}.

\section{Kernel specification and equivalence}
Let $\rho_\theta(x)$ be an $n$-qubit density operator, and let
$O_1,\ldots,O_m$ be distinct nonidentity Pauli observables. Define
\begin{align}
 f_\theta(x)_j&=\Tr[\rho_\theta(x)O_j],\\
 M&=\diag(w_1,\ldots,w_m),\qquad w_j>0,\\
 \delta_\theta&=f_\theta(x)-f_\theta(x'),\nonumber\\
 k_\theta(x,x')&=\exp[-\gamma\delta_\theta^T M\delta_\theta],\label{eq:kernel}
\end{align}
with fixed $\gamma>0$. These are positive semidefinite kernels obtained
by applying an ordinary Gaussian kernel to $M^{1/2}f_\theta(x)$.
Input-dependent feature normalization, adaptive metrics, shot estimators,
and physical noise are not implicit in Eq.~\eqref{eq:kernel}.
The Pauli-partner criterion below uses the displayed diagonal metric.
The coefficient Gram theorem and the design study also allow a fixed
dense $M\succ0$; each chosen metric remains independent of $\theta$.

Two circuit representations are equivalent under this specification
if their kernels agree for all permitted inputs and all assignments
of the retained and shared parameters. A symbol is globally
kernel-inactive when the kernel is independent of it throughout that
domain. A numerical zero derivative on a finite sample is not a
certificate of this property. An uncertified symbol is \unknown,
not necessarily active.

If a common feature transformation satisfies
$f_t(x)=A_t f_0(x)+\tau_t$ and $A_t^T M A_t=M$, then it preserves
Eq.~\eqref{eq:kernel}. The common translation cancels in feature
differences. We use the linear Pauli-adjoint case below. The isometry
identity itself is elementary; the certification problem is to establish
that a particular circuit occurrence realizes such an action uniformly
in the other inputs and symbols.

\input{geometry_scope}

\input{fourier_gram}
\input{recovery_design}

\section{Circuit structure and exact certificates}
\label{sec:structure}
\begin{theorem}[Universal output Pauli criterion]
\label{thm:output}
For a common output rotation $V(t)=\exp(-itP/2)$, where $P$ is a
Hermitian Pauli word, the weighted feature distance in
Eq.~\eqref{eq:kernel} is invariant for all pairs of density operators
and all $t$ if and only if every measured $O_j$ anticommuting with $P$
has its Pauli partner $iPO_j$, up to sign, in the readout with the
same weight.
\end{theorem}
\begin{proof}
Conjugation fixes commuting observables and rotates each anticommuting
pair in a real two-dimensional plane. Closure and equal paired weights
therefore preserve the weighted distance. Conversely, extend the
diagonal metric to all nonidentity Pauli coordinates with zero weights
on unmeasured coordinates. An open neighborhood of the maximally mixed
state permits arbitrary sufficiently small traceless Hermitian
perturbations. Universal distance invariance implies $T(t)^TWT(t)=W$
on the full coordinate space. Differentiating gives $J^TW+WJ=0$;
each rotating two-coordinate block requires equal weights. Positive
measured weights force inclusion of the partner.
\end{proof}
Necessity uses \emph{all} density operators. It need not hold on a
restricted reachable-state family, and it is not a completeness theorem
for arbitrary internal circuit parameters.

\input{gauge_classification}

\begin{theorem}[Sound restricted transport and deletion]
\label{thm:transport}
An input-independent Pauli-rotation occurrence may be transported to
the output through commuting Pauli rotations and supported fixed
Clifford gates. If its transported generator satisfies
Theorem~\ref{thm:output}, deleting the occurrence preserves the kernel
pointwise in all inputs and parameter assignments. Any finite sequence
of such certified deletions is sound. A shared symbol is certified
inactive only when no occurrence of that symbol remains.
\end{theorem}
\begin{proof}
Commuting rotations can be exchanged regardless of the later angle's
data or parameter dependence. A Clifford $C$ obeys
$C e^{-itP/2}=e^{-itCPC^\dagger/2}C$. These identities move the
occurrence to the output without changing the unitary. Theorem
\ref{thm:output} then justifies its kernel-specific deletion.
Induction establishes the sequence claim, including tied assignments
of shared symbols. Absence of every occurrence removes the symbol's
dependency.
\end{proof}

The implemented rule supports fixed Hadamard and CNOT transport,
blocks unsupported noncommuting rotations, and checks equal metric
weights exactly rather than with a numerical tolerance. Its
operational steps and complexity, the source-to-IR precondition, and
the conditions under which deletion preserves optimizer trajectories
are given in Appendix~\ref{app:transport_impl}.

\subsection{A source-derived internal separation}
\label{sec:witness}
\begin{figure*}[t]
\centering
\includegraphics[width=.96\textwidth]{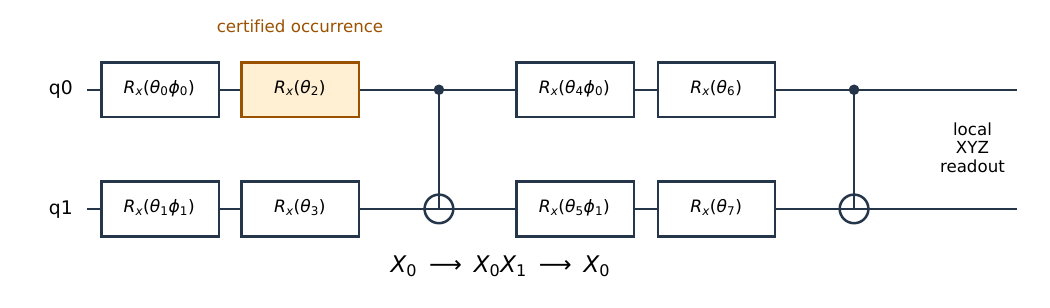}
\caption{An existing source circuit, not a newly selected favorable
grammar. Gates are chronological from left to right;
$\phi_q=\arccos(x_q)$. The marked early rotation is transported through
two CNOTs and later overlapping data rotations, then discarded only
under the specified readout and metric. The resulting common output
rotation changes states and features, not the ideal kernel. The
transport uses standard circuit identities; it does not establish that
all more powerful symbolic optimizers would miss this factorization.}
\label{fig:witness}
\end{figure*}

Figure~\ref{fig:witness} shows the two-qubit, two-layer ChebyshevRx
circuit in the frozen external-source inventory. Transporting the early
$R_{x,0}(\theta_2)$ produces
\begin{equation}
 \begin{aligned}
 X_0&\ \xrightarrow{\mathrm{CX}_{0,1}}\ X_0X_1\\
 &\ \xrightarrow{\text{later }X\text{ rotations}}\ X_0X_1
 \ \xrightarrow{\mathrm{CX}_{0,1}}\ X_0.
 \end{aligned}\label{eq:path}
\end{equation}
The intermediate generator commutes with the later data-dependent
$X$ rotations. Hence $U_\theta(x)=R_{x,0}(\theta_2)U_{\rm del}(x)$.
For generic $\theta_2$, this deletion is not unitary equivalence up
to phase and does not preserve individual $Y_0,Z_0$ expectations.
It nevertheless preserves full local $XYZ$ distances with equal
weights. It precedes a genuinely later overlapping data gate and so
lies outside a syntactic input-independent suffix.

In a seven-input witness at $\theta_2=0.73$, the maximum feature
change is $0.36885$ and the maximum state infidelity is $0.12741$,
whereas the maximum kernel discrepancy is $3.33\times10^{-16}$.
Equation~\eqref{eq:path}, not those finite samples, supplies the
certificate. The two- and four-qubit two-layer settings contain two
such globally inactive symbols each. Boundary effects and same-native
gate affine biases are counted separately below.

\input{global_normal_form}
\input{dimension_theory}
\input{extension_theory}
\input{family_extension}
\input{kyriienko_gram_results}

\section{Exact dimension census}
\label{sec:census}
\input{dimension_results}
\input{extension_results}

The 30-setting Fourier audit and twelve Kyriienko scaling
certificates were reproduced in a pinned Linux environment, with all
certificate comparisons agreeing. This is cross-environment
reproduction of the same implementations, not independent
implementation or formal verification. Table~\ref{tab:reproscope}
separates these full reruns from the dedicated local Gram calculation
and the earlier saved-evidence replay.
Appendix~\ref{sec:census-traceability} connects each census setting
to its upper bound, lower certificate, and reproduction command.

\input{shot_design}

\section{Execution cost audit}
\label{sec:cost}
A certificate concerns a kernel function; executable work also depends
on compilation, gate fusion, differentiation and measurement schedules.
In the two-layer ChebyshevRx example, the boundary and transport arms
retain $7$ and $5$ source symbols, yet both require three trainable
Pauli-angle tasks after global normalization. The exact kernel
dimension $r=4$ counts parameter dependence across all inputs, not the
number of rotation tasks at one input.

\input{global_results}

\subsection{Timing, survey, and learning controls}
The earlier adjacent-fusion study used 12 configurations, seven arms,
three NumPy derivative backends and three data problems, with five split
seeds and five repetitions: 18,900 cold fits and a separate 3,600-fit
reuse study. Serial cold-fit median reductions of 22.20\% and 12.38\%
are specific to that earlier baseline; they are not gains over global
normalization. Adjoint reverse work is unchanged, and the fastest
tested backend remains boundary-adjoint. The 102-slot circuit/readout
survey finds four extra-symbol slots but no additional final fused
operation or angle-task reduction. Full protocols, preserved timings,
the backend figure, and the 32-symbol mechanism breakdown are in
Appendix~\ref{app:resource_details}. Ideal derivative checks and analytic
fixed-shot counts are not hardware or fixed-accuracy shot experiments.
Appendix~\ref{app:legacy_controls} retains the moons/Wine, placement,
and fidelity-candidate controls. Appendix~\ref{app:visionad}
reports an image anomaly-detection study of the same PQK before and
after parameter optimization. This task provides a direct downstream
test because query-to-normal-support kernel similarities define the
anomaly score: changing the kernel can change the predicted ordering
of normal and anomalous images. We use it to examine whether parameter
optimization improves that ordering, not to test our metric or
feature-based rank recovery. Kernel dependence alone does not
establish predictive improvement.
These controls do not establish a general predictive or adaptive-search
advantage.

\section{Limits and implications}
The coefficient Gram representation requires a fixed finite data basis
independent of trainable parameters. General products $\theta\phi(x)$
fall outside that construction; the successful ChebyshevRx and ParamZ
certificates do not prove a decision procedure for this larger class.
Even within the exact fixed-frequency subclass, finite decidability
does not imply a practical expansion or minor budget.

Generic differential rank is a local regular-point statement. A
nonlinear factorization may require several charts, and a rank deficit
need not identify an individually deletable coordinate. The loss-gradient
space changes with the parameter point; optimizer state, regularizers
and finite-step reparameterizations need their own treatment.

Proposition~\ref{prop:gauge} describes only connected, universal-state
gauges for complete isotropic weight sectors, where it uses the
classification of Ref.~\cite{cpr2019locality}, and for local Z.
Disconnected symmetries beyond qubit permutations, anisotropic or
partial readouts, and reachable-state-specific gauge groups remain
outside that description. The grouped
single-layer result does not extend to arbitrary Hubregtsen depths.
For the Kyriienko-type family, the gauge upper bounds ($r\le3n-2$, and
$r\le2n$ for local Z) are proved for every $n$, but exact matching
minors exist only for even $n\le8$ and rely on computer arithmetic
that is not formally verified. Floating-point ranks agree for local
XYZ through $n=12$; no odd size is exactly certified, and an all-size
lower bound remains open.

The 54-setting census is selected source evidence, not a population
prevalence estimate. The strongest implemented resource baselines leave
zero additional angle-task savings in these tests. This neither proves
optimal execution cost nor establishes generic speedup, lower hardware
shot counts, quantum advantage, or better predictive performance.
The ideal finite-shot study of Sec.~\ref{sec:shotdesign} uses exact-model
design information and matched contrasts; it is neither a shot-only
learning procedure nor a noisy-device experiment. Its sampled ranks
do not extend the exact census, and its mixed sensitivity results do
not establish generic design superiority. In the Hubregtsen example,
the added products expand the kernel representation without adding
independent ideal measurement information. Their marginal-product
estimator relies on a product Z-outcome distribution, which is not
guaranteed for other circuits or correlated device noise. Conventional
compilation, measurement optimization and kernel-alignment subsampling
\cite{sahin2024subsampling} address different resource questions.
Appendix~\ref{app:additional_scope} retains the explicit transport,
search-descendant and predictor-storage limitations.

\section{Conclusion}
Parameters that a circuit, its state, or its measured features
distinguish may be indistinguishable to the specified kernel. For
fixed finite data bases, the coefficient Gram matrix makes this
precise: it characterizes all-input kernel equivalence, global
inactivity, and the identifiable dimension. This dimension can be
certified exactly, as in all 54 census settings, the grouped Z-readout
family, and the Kyriienko-type family at even $n\le8$. It depends on
the readout, is not monotone in it, and can fall below the number of
individually active coordinates. This diagnosis also guides intervention:
metric changes recover some lost dependence, whereas an exact
fixed-quadratic-metric ceiling can be exceeded by product features.
In our exact example those features are computable from existing means;
two products within the input groups suffice with a diagonal metric,
and a single product suffices with an explicit nondiagonal metric
(Theorem~\ref{thm:oneproduct}), whereas every fixed affine
expansion retains the rank-three ceiling
(Proposition~\ref{prop:recoveryboundary} and Corollary~\ref{cor:affineceiling}).
The held-out shot study separates restored rank from statistical
resolution. A matched comparison of zero, one, two, and six products
shows why a minimal representation need not maximize finite sample
sensitivity. At a fixed design with factorized Z outcomes, conditional
product estimation cannot worsen the local sensitivity
(Corollary~\ref{cor:productsensitivity}); strict improvement and better
prediction do not follow. It is also not execution cost: in the
tested scope, the strengthened baselines leave no additional
angle-task savings. Three problems remain open. Circuits with
products $\theta\phi(x)$ of parameters and data lie outside the fixed-basis
construction. For the Kyriienko-type family, no odd size is exactly
certified, and equality $r=3n-2$ for all $n$ still lacks a matching
lower bound. Finally, extending the ideal-model finite-shot design
study to estimated design information, calibrated statistical decisions,
and physical noise remains open.

\section*{Data and code availability}
The reproducibility package is publicly available at
\url{https://github.com/parkfinity/readout_kernel_identifiability}.
The integrated package accompanying this manuscript is pinned to
\href{https://github.com/parkfinity/readout_kernel_identifiability/tree/98c922ec225a8a6ac00c4533de3660a03b73a038}
{commit \texttt{98c922e}}.

The package contains code, frozen circuit inputs, exact certificates,
reference results, environment specifications, and instructions for
the dimension and resource audits, readout and metric recovery, and
finite-shot and product-feature controls.
The exact single product checks, matched product experiment and its
frozen inputs, and certificate index for the 54 settings
(Appendices~\ref{app:oneproduct} and~\ref{sec:census-traceability})
are included under \texttt{experiments/minimal\_product/}.
The repository documents which commands perform fresh calculations
and which reanalyse archived results.

The package does not provide an end-to-end rerun of every historical
appendix experiment. Raw industrial images, original QPU jobs, and
the large historical timing archive are not included.
Public benchmark data originate from the sources identified in
Appendix~\ref{app:evidence}; third-party code and data retain their
original terms.
Table~\ref{tab:reproscope} records the earlier cross-environment
certificate checks, not a new validation of every repository workflow.

\section*{Author contributions and AI disclosure}
Hyong Keun Park designed the research, developed the proofs, and
performed the experiments. Jahoon Jeong and Joonkyo Shim reviewed
the work. Hyunsoo Yoon supervised the research.

We acknowledge the use of large language models (LLMs) to assist in
the preparation of this manuscript. Specifically, LLMs assisted with
research planning, drafting and revising text, literature discovery,
experimental code development and debugging, and organizing and
analyzing results, including calculation checks. The authors carefully
reviewed and edited all AI-assisted outputs and take full
responsibility for the final content and scientific integrity of
this work.

\section*{Acknowledgments}
No project-specific research grant was received for this work.
The authors thank Yonsei University for facilitating access to IBM
quantum computing resources. Hyong Keun Park acknowledges
scholarship support through the ICONS interdisciplinary research
scholarship program of the Institute of Convergence Science,
Yonsei University, and its support for interdisciplinary research.

\section*{Competing interests}
Hyong Keun Park is employed by POSCO. POSCO had no institutional
involvement in this research. The other authors declare no
competing interests.

\begingroup
\raggedright
\bibliographystyle{quantum}
\bibliography{references}
\endgroup

\onecolumn\newpage
\appendix
\input{appendices}
\input{dimension_appendix}
\input{extension_appendix}
\input{legacy_controls}
\input{gram_appendix}
\input{followup_appendix}
\input{resource_details}
\input{additional_scope}
\input{spectral_context}
\input{transport_implementation}
\input{kyriienko_scaling_appendix}
\input{recovery_appendix}
\input{product_control}
\input{one_product_appendix}
\input{census_traceability}
\end{document}

%% file: submission_metadata.tex
\author{Hyong Keun Park}
\affiliation{Department of Industrial Engineering, Yonsei University, Republic of Korea}
\affiliation{POSCO, Republic of Korea}
\author{Jahoon Jeong}
\affiliation{Department of Industrial Engineering, Yonsei University, Republic of Korea}
\author{Joonkyo Shim}
\affiliation{Department of Industrial Engineering, Yonsei University, Republic of Korea}
\author{Hyunsoo Yoon}
\email[Corresponding author: ]{hs.yoon@yonsei.ac.kr}
\affiliation{Department of Industrial Engineering, Yonsei University, Republic of Korea}
\date{October 19, 2026}

%% file: overview_figure.tex
\begin{figure*}[!t]
\centering
\includegraphics[width=\textwidth]{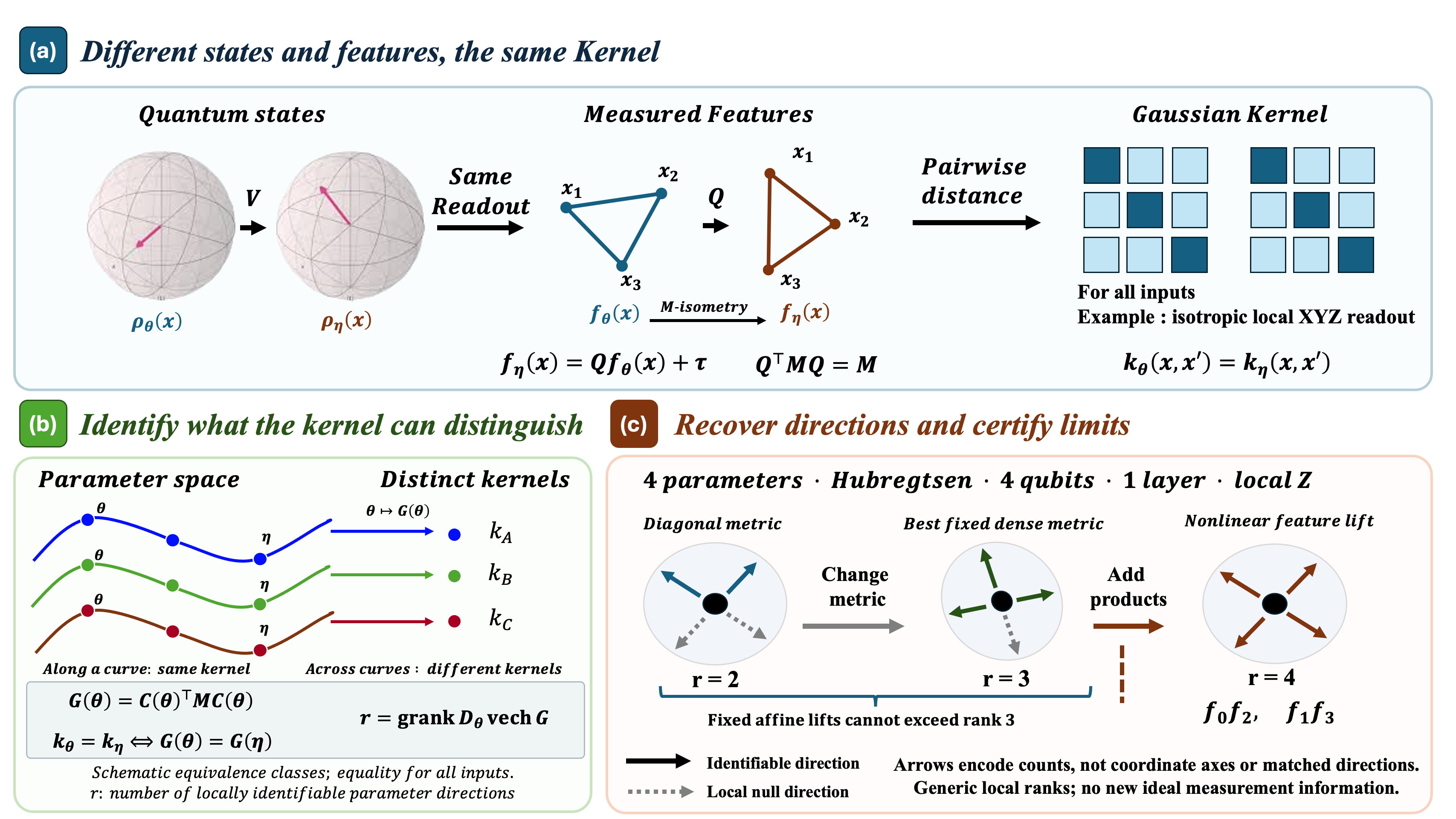}
\caption{Readout-dependent kernel equivalence, identifiable
 dimension, and recovery limits.
(a) Different states and measured features can define the same
Gaussian kernel when $f_\eta(x)=Qf_\theta(x)+\tau$ with
$Q^\top M Q=M$. The triangles illustrate an isotropic metric;
an arbitrary output unitary need not preserve the kernel.
(b) Schematic equivalence classes under equality on all input pairs.
Under the assumptions of Theorem~\ref{thm:gram}, equality of
$G=C^TMC$ characterizes kernel equivalence, and its generic
differential rank gives the local identifiable dimension.
The curves do not specify the global topology of a particular family.
(c) For the four-qubit Hubregtsen/local Z family, diagonal metrics
give rank two, dense metrics attain at most three, and fixed affine
lifts cannot cross this ceiling. The displayed two-product lift
attains rank four with a positive diagonal metric; it is a sufficient
construction, not a minimum number of appended scalar coordinates.
One scalar coordinate also suffices: either $f_0f_2$ with the
nondiagonal metric of Theorem~\ref{thm:oneproduct}, or
$f_0f_2+f_1f_3$ with a positive diagonal metric
(Appendix~\ref{app:oneproduct}). These lifts use existing exact
local means. All ranks are generic; arrows count identifiable and
null directions, not matched coordinate axes. Finite-shot resolution,
predictive gains, and execution savings require separate assessment.}
\label{fig:overview}
\end{figure*}

%% file: contracts_table.tex
\begin{table*}[t]
\caption{Objects and certification targets of related approaches.
Entries describe the cited works' objectives, not impossibility
results for other tasks. In the last row, $r$ is generic kernel
differential rank; the fixed-basis assumptions and exact decision
subclass are stated in Sec.~\ref{sec:gram}.}
\label{tab:contracts}
\centering
\begin{tabular}{@{}>{\raggedright\arraybackslash}p{.20\textwidth}>{\raggedright\arraybackslash}p{.22\textwidth}>{\raggedright\arraybackslash}p{.31\textwidth}>{\raggedright\arraybackslash}p{.19\textwidth}@{}}
\toprule
Approach & Object preserved or characterized & Principal output & Measurement specification\\
\midrule
Parametric equivalence\newline Peham; Ross--Wesley \cite{peham,rosswesley}
& Unitary equivalence, including global-phase variants
& Symbolic checks; finite cutoffs for supported all-parameter identities
& No particular readout required\\[4pt]
Dead-gate elimination \cite{deadgate}
& Contributory measurement-outcome distribution
& Removal of gates irrelevant to the specified outcomes
& Specified measurement outcomes\\[4pt]
Clifford extraction and absorption\newline QuCLEAR \cite{quclear}
& Observable expectations or classically corrected outcomes
& Circuit reduction with Clifford processing transferred to measurement or classical computation
& Observables or postprocessing may change\\[4pt]
State geometry\newline Funcke; Haug \cite{funcke,haug}
& State-manifold dimension and effective quantum dimension
& Redundant-parameter identification and dimension-preserving pruning
& State geometry before a chosen kernel readout\\[4pt]
Fourier/coefficient analysis\newline Schuld; Campbell \cite{schuldFourier,CHM}
& Feature spectra and coefficient-space geometry
& Encoding-dependent modes; harmonic matrices, coefficient Jacobians and training kernels
& Observable enters the coefficient representation\\[4pt]
This work: coefficient Gram identifiability
& Projected Gaussian kernel at fixed readout and metric
& $G=C^TMC$: all-input equivalence, global coordinate inactivity, and exact certificates for $r$
& Fixed readout and metric define the equivalence\\
\bottomrule
\end{tabular}
\end{table*}

%% file: geometry_scope.tex
\subsection{State geometry is not kernel sensitivity}
\label{sec:qfi_scope}
The relation to state geometry can be stated without identifying two
different notions of inactivity. For a differentiable normalized pure
state $|\psi_\theta(x)\rangle$, use the SLD-normalized QFI
\begin{equation}
 F_{ij}(x)=4\operatorname{Re}\langle\partial_i\psi_x|
 (I-|\psi_x\rangle\langle\psi_x|)|\partial_j\psi_x\rangle.
\end{equation}
The real quantum geometric tensor convention in Ref.~\cite{haug}
differs by the factor four, which does not affect rank. At a fixed
parameter assignment, let $F_{\mathcal X}=\sum_{x\in\mathcal X}F(x)$
for a finite input set, and let $J_{\mathcal X}$ contain the gradients
of all pairwise kernel entries on that set. If the differentiable
kernel depends on $\theta$ only through these states and uses a fixed
readout and metric, then
\begin{equation}
 \ker F_{\mathcal X}\subseteq\ker J_{\mathcal X},\qquad
 \operatorname{rank}J_{\mathcal X}\leq\operatorname{rank}F_{\mathcal X}.
 \label{eq:qfi_inclusion}
\end{equation}
Indeed, a real direction $v$ in the left null space has zero horizontal
state tangent at every input, since $v^TF_{\mathcal X}v$ is a sum of
nonnegative squared norms. Each density operator therefore has zero
directional derivative, and the chain rule gives the inclusion.
This is a local, sample-specific implication, not a new general
completeness result or a global inactivity certificate.

The inclusion can be strict even for one qubit. Consider
\begin{align}
 |\psi_\theta(x)\rangle&=R_y(\theta)R_y(x)|0\rangle,\\
 F_{\theta\theta}(x)&=1,\qquad
 k^{\rm fid}_\theta(x,x')=\cos^2[(x-x')/2].
\end{align}
The state QFI has rank one, although the fidelity kernel is globally
independent of $\theta$. The full XYZ feature vector is
$(\sin(\theta+x),0,\cos(\theta+x))$, whose squared Euclidean
distance is $2-2\cos(x-x')$, also independent of $\theta$.
With only Z readout, the squared distance instead is
\begin{equation}
 d_Z^2=[\cos(\theta+x)-\cos(\theta+x')]^2.
\end{equation}
Its $\theta$ derivative equals 2 at $(\theta,x,x')=(0,0,\pi/2)$.
Thus changing the readout can restore kernel sensitivity without
changing the prepared-state QFI at all. An input-independent output
unitary cancels in the fidelity overlap, not necessarily in the state
geometry. The compute--uncompute \emph{output state} can itself be
parameter independent, but it is a different state family from
$|\psi_\theta(x)\rangle$. The exact identities above are checked in the
reproduction script; they clarify the preservation contract, rather
than constituting a matched implementation of state-based pruning.

%% file: fourier_gram.tex
\section{Finite feature Gram identifiability}
\label{sec:gram}
The relevant finite representation is a Gram matrix of \emph{feature
coefficients}, not a sampled data Gram matrix. Let $\Theta\subset\mathbb R^K$
be an open connected parameter box and suppose
\begin{equation}
 f_\theta(x)=C(\theta)h(x)+u(\theta),\quad
 G(\theta)=C(\theta)^T M C(\theta),\label{eq:coefficientgram}
\end{equation}
where $h:D\to\mathbb R^A$ is fixed, $C,u$ are real analytic, and
$M\succ0$ and $\gamma>0$ are fixed. Define $d_\theta^2$ as the
squared $M$-distance of features and $k_\theta=\exp(-\gamma d_\theta^2)$.
We require $\{1,h_1,\ldots,h_A\}$ to be linearly independent as
functions on the \emph{actual} input domain $D$. Equivalently, the
affine hull of $h(D)$ is $\mathbb R^A$.

\begin{theorem}[Coefficient Gram identifiability]\label{thm:gram}
Under these assumptions, for any $\theta,\eta\in\Theta$,
\begin{equation}
 \begin{gathered}
 k_\theta(x,x')=k_\eta(x,x')\quad\forall x,x'\in D\\
 \Longleftrightarrow\quad G(\theta)=G(\eta).
 \end{gathered}
\end{equation}
For every parameter direction $v$, the kernel directional derivative
vanishes for all input pairs if and only if $D G(\theta)[v]=0$.
Consequently, coordinate $\theta_j$ is globally inactive for the kernel if
and only if $\partial_jG\equiv0$, and the generic kernel differential
rank is
\begin{equation}
 r=\operatorname{grank}D_\theta\operatorname{vech}G(\theta)
 \le \min\{K,A(A+1)/2\}.\label{eq:gramrank}
\end{equation}
Here $\operatorname{vech}$ lists the upper triangular entries once.
\end{theorem}
\begin{proof}
The offset $u$ cancels, so $d_\theta^2=\Delta h^T G(\theta)\Delta h$.
Fix $x_0$ and put $y(x)=h(x)-h(x_0)$. These vectors span $\mathbb R^A$.
If a symmetric $H$ obeys $\Delta h^TH\Delta h=0$ for all pairs, then
$y(x)^THy(x)=0$ and polarization gives $y(x)^THy(x')=0$ for all
$x,x'$. Spanning implies $H=0$. This proves injectivity of the linear
map from $G$ to squared distances. The exponential is injective and
has nonzero derivative, establishing both equivalence and the
directional statement. Integration along coordinate intervals proves
the global inactivity assertion. The finite dimensional injective
distance map preserves differential rank, giving Eq.~\eqref{eq:gramrank}.
Analyticity makes the maximum rank generic on the connected box.
\end{proof}

Linear independence of $h$ alone is insufficient: $h(x)=(1,x)^T$
has independent components, but $G=\diag(1,1)$ and $G'=\diag(2,1)$
give the same distances. Constant modes must therefore be separated
into $u$. Likewise, formal independent variables cannot replace
dependent coordinates of a restricted data manifold.

For Pauli circuits with fixed frequencies, $h$ consists of real sine and
cosine modes with distinct nonzero frequency vectors (one representative
per sign pair). These functions and $1$ are independent if the
transformed data domain contains an open set. The finite basis
representation links identifiability over all input pairs to
coefficient algebra. Its relation to existing spectral methods is
discussed in Appendix~\ref{app:spectral_context}.

\input{gradient_corollary}

\begin{corollary}[Finite decision procedure on an exact subclass]
\label{cor:gramdecision}
In addition to Theorem~\ref{thm:gram}, suppose an exact finite
description of $C$ is available whose entries are finite Laurent
polynomials in $e^{i\theta_j/q}$, for a known integer $q>0$, over
an effective algebraic number field containing the entries of $M$.
Then global coordinate inactivity and generic differential rank are
decidable by finitely many exact coefficient tests.

A sufficient circuit grammar is a finite composition of fixed
algebraic unitaries and Pauli rotations with angles
$a\cdot\theta+b\cdot z(x)+c$, where $a,b$ are rational,
$e^{ic/2}$ is algebraic, the initial density operator and observables
have algebraic entries, and $z(D)$ contains an open set. After exact
collection of the finite data frequencies, the real basis and its
coefficient conversion must be constructed over the same effective
field. Products of parameters with data variables are excluded.
\end{corollary}
\begin{proof}
For the stated grammar, multiplying the finitely many gate matrices
first produces finite joint Laurent expansions in the parameter and
data phases. Clearing rational frequency denominators and collecting distinct
data frequencies gives the required exact representation of $C$.
Entries of $G$, their derivatives,
and all Jacobian minors are finite Laurent polynomials in
$e^{i\theta_j/q}$ over an effective algebraic field (adjoining $i$).
Distinct parameter frequencies are independent on an open box.
Thus a derivative or minor vanishes identically exactly when every
coefficient vanishes. Test derivatives for inactivity and minors in
decreasing order for rank. There are finitely many of each.
\end{proof}
This is a termination result, not a polynomial time guarantee or a
claim that the present pilot implements exhaustive minor enumeration.
Analyticity alone, as in Theorem~\ref{thm:gram}, is not an effective
description and does not imply this decision guarantee.
Its expansion and minor counts can be prohibitive. Arbitrary analytic
angles, unspecified real constants, and products $\theta_j\phi(x)$
are outside this decision claim. In particular, the ChebyshevRx and
ParamZ examples still require the separate methods below.

\begin{corollary}[Dimension bound imposed by the readout]\label{cor:readoutbound}
Let $\rho=\min(m,A)$, where $m$ is the number of measured features.
Under Theorem~\ref{thm:gram},
\begin{equation}
 r\le\min\{K,A\rho-\rho(\rho-1)/2\}.
\end{equation}
\end{corollary}
\begin{proof}
The Gram matrix is positive semidefinite of rank at most $\rho$.
The positive semidefinite stratum of rank $q$ has dimension
$Aq-q(q-1)/2$: factor it by an $A\times q$ matrix of full column rank,
then quotient by $O(q)$. On the generic stratum of constant rank of
$C(\theta)$, the differential image lies in this stratum's tangent
space. Maximizing over $q\le\rho$ gives the bound.
\end{proof}
This standard bound from rank strata sharpens Eq.~\eqref{eq:gramrank}
when $m<A$. Its null directions may depend on $\theta$; they need
not be deletable coordinates or executable savings.

\subsection{Isometries and readout dependence}
If $C(\eta)=Q C(\theta)$ with $Q^TMQ=M$, then the kernels agree.
Conversely, equality of their coefficient Grams admits such a $Q$:
whiten by $M^{1/2}$, map corresponding column spans isometrically,
and extend orthogonally to the ambient feature space. Arbitrary
offsets independent of the data remain unidentifiable. This describes
equivalence of reachable features; the Pauli output criterion for all
states has a stronger universal quantifier.
For a fixed proposed feature transformation $Q$, the exact
condition on the reachable family is
$C(\theta)^T(Q^TMQ-M)C(\theta)=0$ for every $\theta$.
It does not generally imply isometry on the linear span of all
these column spaces: bilinear terms across parameter values are untested.
Appendix~\ref{app:followup} supplies a counterexample.
The Hubregtsen example below is directly represented by a diagonal
$G$ with entries $\sin^2\theta_0+\sin^2\theta_2$ and
$\sin^2\theta_1+\sin^2\theta_3$. A rank deficit need not make any
individual coordinate inactive, nor produce a global minimal chart.
The same example separates $D_\theta\operatorname{vech}G$ from the
coefficient Jacobian $D_\theta\operatorname{vec}C$, whose Gram matrix
of coefficient derivatives is the object of QNTK type built from circuit
harmonic matrices \cite{CHM}. With $h=(\cos x_0,\cos x_1)$, the four
nonzero entries $C_{q,q\bmod2}=-\sin\theta_q$ give
$\operatorname{rank}D_\theta\operatorname{vec}C=4$, one per retained
angle, whereas $\operatorname{rank}D_\theta\operatorname{vech}G=2$,
the number of input groups, at generic $\theta$.

\begin{proposition}[Full Pauli readout and nonmonotonicity]
\label{prop:IC}
For all nonidentity $n$-qubit Pauli words, with their usual
unnormalized convention and unit weights,
\begin{equation}
 d^2=2^n\Tr[(\rho-\rho')^2].
\end{equation}
For pure states this is $2^{n+1}(1-F)$, where $F=\Tr(\rho\rho')$.
For general subsets of Pauli readouts, kernel differential rank is
not monotone under adding measured observables.
\end{proposition}
\begin{proof}
Pauli orthogonality gives the first identity; the identity component
contributes nothing because the two states have equal trace. Purity gives the second. For
nonmonotonicity, the certified two-qubit examples in
Table~\ref{tab:dimension} give both directions: at two layers,
ChebyshevRx has $r_Z=6>r_{XYZ}=4$, whereas one-layer ParamZ has
$r_Z=0<r_{XYZ}=2$.
\end{proof}
Adding rows to $C$ adds positive semidefinite terms to $G$, but this
ordering does not order Jacobian ranks. Full readout is therefore not
generally a rank minimizer. The two-qubit readout of weight at most two
is information complete; for mixed states its distance also depends
on the two purities and is not just a function of an overlap.

%% file: gradient_corollary.tex
\begin{corollary}[Instantaneous kernel-loss gradient space]
\label{cor:gradientspace}
Under Theorem~\ref{thm:gram}, set $g(\theta)=\operatorname{vech}G(\theta)$
and $J_g=D_\theta g$. For a differentiable loss depending on $\theta$
only through finitely many fixed-input kernel evaluations,
\begin{equation}
 \nabla_\theta L\in\operatorname{im}J_g(\theta)^T
       =(\ker J_g(\theta))^\perp.
\end{equation}
This space has dimension $\operatorname{rank}J_g(\theta)\le r$,
with equality on the regular parameter set.
\end{corollary}
\begin{proof}
Each evaluated kernel is a smooth function of $g$, so the loss factors
as $L=\ell\circ g$. The chain rule gives
$\nabla_\theta L=J_g^T\nabla_g\ell$; the orthogonal-complement
identity follows from finite-dimensional linear algebra.
\end{proof}
This is an instantaneous constraint, not confinement to one fixed
$r$-dimensional linear subspace along training. The subspace can rotate
with $\theta$; finite steps, momentum and preconditioning need not stay
in it. Direct parameter regularization can add components outside it.
Nor does this imply $r$ circuit evaluations suffice for a gradient.

%% file: recovery_design.tex
\section{Recovering identifiability by readout and metric design}
\label{sec:recovery}
Changing the metric or the readout changes the kernel, unlike the
equivalence-preserving transformations above. The design objective is
not to make every source symbol active, but to recover dependence
already visible in the centered features and to identify its limits.

\subsection{What can a metric change recover?}
At a fixed parameter point, write $H_v=D C(\theta)[v]$. By
Theorem~\ref{thm:gram}, the all-input kernel derivative vanishes exactly
when
\begin{equation}
 C^T M H_v+H_v^T M C=0.\label{eq:recoverability}
\end{equation}
This test separates a symmetry of a particular metric from a loss of
dependence in the features themselves.

\begin{proposition}[Irrecoverability over fixed metrics]
\label{prop:metricrecover}
Assume the finite-basis hypotheses of Theorem~\ref{thm:gram}. At a
point where $f_\theta$ is nonconstant, a direction $v$ is invisible to
every fixed positive definite metric if and only if $H_v=0$,
equivalently $D_v f_\theta(x)$ is input-independent.
\end{proposition}
\begin{proof}
For an input pair set $a=f_\theta(x)-f_\theta(x')$ and
$b=D_v f_\theta(x)-D_v f_\theta(x')$. If $a^TMb=0$ on the open
positive definite cone, linearity in $M$ implies
$ab^T+ba^T=0$. For $a\ne0$ this gives $b=0$. When $a=0$, compare
both inputs to a third input with a different feature value, which
exists by nonconstancy. Thus $b=0$ for every pair. Affine independence
of the data basis implies $H_v=0$. The converse is immediate.
\end{proof}
The nonconstant assumption is essential: at $C=0$, every
squared-distance first derivative vanishes regardless of $H_v$.
Also, the intersection of nullspaces over all metrics need not equal
the nullspace of any \emph{single} metric. The bound
$r\le A(A+1)/2$ continues to constrain each fixed metric.

At a fixed parameter point, the minors of Eq.~\eqref{eq:recoverability},
viewed as a map of $v$, are polynomials in the metric entries.
Consequently, if a rank is attainable in an open positive definite
metric family, almost every metric in that family attains its maximum
rank. This applies separately within the diagonal family and to
finitely many parameter points simultaneously
(Appendix~\ref{app:recovery}). Random anisotropy is therefore a
necessary baseline; generic rank alone does not imply conditioning
or good finite-shot sensitivity.

\subsection{An exact boundary between metric and feature recovery}
The question here is whether a fixed metric on the existing features
can recover all four feature-visible directions, or whether the
representation itself must change.
For the four-qubit, one-layer Hubregtsen source circuit, the local Z
features are $f_q(x)=-\sin\theta_q\cos x_{q\bmod2}$ for $q=0,\ldots,3$.
Its diagonal controlled-rotation tail changes none of these features
(the circuit derivation is given at Eq.~\eqref{eq:hubfeatures}).
The four retained angles vary independently; the remaining tail
angles do not enter these features. Assume that
$(\cos x_0,\cos x_1)$ ranges over a set with nonempty interior.
The following statement concerns generic differential rank, not
rank at every parameter point or globally unique parameter recovery.

\begin{proposition}[Fixed-quadratic ceiling and product recovery]
\label{prop:recoveryboundary}
For this four-qubit Hubregtsen/local Z family with fixed
$\gamma>0$: (i) every fixed positive diagonal metric on the four
local means has generic kernel rank two; (ii) the maximum generic
rank over fixed dense positive definite metrics on these means is
three, attained by Eq.~\eqref{eq:recoverymetric}; (iii) appending
$f_0f_2$ and $f_1f_3$ gives generic rank four for every fixed positive
diagonal metric on the resulting six features. These products equal
the corresponding ZZ means in this family and need no additional
ideal measurement information.
\end{proposition}
\begin{proof}
Write $a=\sin\theta_0$, $b=\sin\theta_1$,
$c=\sin\theta_2$, and $d=\sin\theta_3$. For diagonal weights, the coefficient
Gram on $h=(\cos x_0,\cos x_1)^T$ is
\[
 G=\diag(w_a a^2+w_c c^2,\ w_b b^2+w_d d^2).
\]
Its two entries have generically independent derivatives, proving (i).
For every dense metric the same Gram is symmetric $2\times2$, giving
rank at most three. With $u=(1,2,3,4)^T$,
\begin{equation}
 M=(32I+uu^T)/158\label{eq:recoverymetric}
\end{equation}
is positive definite with unit trace, and an exact nonzero rational
minor certifies rank three. Its value is
$-\frac{835727837184}{978271035584375}$ at the rational-phase point in
Appendix~\ref{app:recovery}, proving (ii).

The state before the diagonal tail is a product and the tail commutes
with every Z string. Consequently $f_{qr}(x)=f_q(x)f_r(x)$ exactly.
These ZZ features are a nonlinear lift of the original four means,
not new ideal information that was absent from those means.
For (iii), put $(u,v)=(\cos x_0,\cos x_1)$.
The data functions $1,u,v,u^2,v^2$ are independent. Four entries of
their coefficient Gram are
\begin{equation}
 \begin{aligned}
 \beta_u&=w_a a^2+w_c c^2,&\beta_{u^2}&=w_{ac}a^2c^2,\\
 \beta_v&=w_b b^2+w_d d^2,&\beta_{v^2}&=w_{bd}b^2d^2.
 \end{aligned}\label{eq:twoproductrecovery}
\end{equation}
The $(a,c)$ Jacobian block has determinant
$4acw_{ac}(w_a a^2-w_c c^2)$, and the $(b,d)$ block has the
analogous generically nonzero determinant. Multiplying by the four
cosine derivatives preserves generic nonvanishing, proving rank at
least four. Only the four pre-tail angles occur, giving the matching
upper bound.
\end{proof}
The metric selection study retained all six ZZ products and also
has exact minor certificates. Two products are sufficient; we do not
claim their minimality over all allowed feature/metric designs.

The following result sharpens this sufficiency statement when a
fixed metric that is not diagonal is allowed, while retaining the
same four original means.
\begin{theorem}[Recovery with one product and restricted minimality]
\label{thm:oneproduct}
Retain the four local means of Proposition~\ref{prop:recoveryboundary}
and allow appended scalar features and a positive definite metric
fixed independently of inputs and trainable parameters.
The minimum number of appended coordinates needed to attain generic
kernel differential rank four is one. Specifically, append only
$f_0f_2$ and use the metric $M_1/7$, where
\begin{equation}
 M_1=\begin{pmatrix}
 2&1&0&0&0\\
 1&2&0&0&0\\
 0&0&1&0&0\\
 0&0&0&1&0\\
 0&0&0&0&1
 \end{pmatrix}.
 \label{eq:oneproductmetric}
\end{equation}
This metric is positive definite, has unit trace after division by
seven, and gives generic rank four for any fixed $\gamma>0$.
\end{theorem}
\begin{proof}
Put $u=\cos x_0$, $v=\cos x_1$, and use
$a,b,c,d$ as in the preceding proof. The lifted feature vector is
$(-au,-bv,-cu,-dv,ac u^2)^T$. The functions $1,u,v,u^2$ are
linearly independent on the actual domain: a polynomial
$A+Bu+Cv+Du^2$ vanishing on an open rectangle first has $C=0$
by varying $v$, and then $A=B=D=0$ by varying $u$.
The eigenvalues of $M_1$ are $3,1,1,1,1$. In the fixed basis
$h=(u,v,u^2)^T$, its unnormalized coefficient Gram is
\begin{equation}
 G_1=\begin{pmatrix}
 2a^2+c^2&ab&0\\
 ab&2b^2+d^2&0\\
 0&0&a^2c^2
 \end{pmatrix}.
 \label{eq:oneproductgram}
\end{equation}
For the four entries in the order
$s=(G_{1,11},G_{1,12},G_{1,22},G_{1,33})$, exact differentiation gives
\begin{equation}
 \det D_{(a,b,c,d)}s=-8a^2cd(2a^2-c^2).
 \label{eq:oneproductminor}
\end{equation}
At $\theta_0=\cdots=\theta_3=\pi/6$, multiplication by
$\prod_{j=0}^3\cos\theta_j$ gives the angle Jacobian minor
$-9/128$. For the metric with unit trace it is
$-9/(128\cdot7^4)=-9/307328$, still nonzero.
Theorem~\ref{thm:gram} therefore gives rank at least four on a
generic set; the four retained angles give the matching upper bound.
With no appended coordinate, the symmetric $2\times2$ coefficient
Gram of the original features bounds every fixed metric's rank by
three. Zero additions are therefore insufficient and one suffices.
\end{proof}

The minor uses the data basis $(u,v,u^2)$, not
$(\cos x_0,\cos x_1,\cos 2x_0)$; changing that basis rescales its
value, not its nonvanishing. The theorem is a local differential
statement, not global injectivity. For example, the distinct positive
amplitude assignments
\[
 \begin{gathered}
 (a,b,c,d)=(\tfrac12,\tfrac14,\tfrac12,\tfrac12),\\
 (\tfrac1{2\sqrt2},\tfrac{\sqrt2}4,\tfrac1{\sqrt2},\tfrac{\sqrt2}4)
 \end{gathered}
\]
give the same $G_1$, and both are regular points.
The minimality claim retains the original features and varies only
appended coordinates and fixed metrics; it does not optimize over
replacement representations, metrics that depend on parameters, or
kernel families. Nor does it prove that two scalar additions are
necessary for a diagonal metric. A \emph{single pairwise product}
with a diagonal metric gives at most three coefficient combinations,
but one scalar sum $f_0f_2+f_1f_3$ can already have rank four
(Appendix~\ref{app:oneproduct}).

\begin{table}[t]
\caption{Recovery limits for the four-qubit Hubregtsen family,
retaining all four local Z means. Metrics are fixed and positive
definite; ranks are generic. Proposition~\ref{prop:recoveryboundary}
and Theorem~\ref{thm:oneproduct} establish the metric ceiling and
product recovery. Appendix~\ref{app:oneproduct} gives the diagonal
one-product bound and scalar-sum construction. The sum is one
appended coordinate containing two product terms; it is not an
experimental arm in Table~\ref{tab:minimal_product_matched}.
The six-product row is the earlier experimental design. All rows
use the same ZZZZ basis, and the lifts are available from exact
local means.}
\label{tab:recoverymechanism}
\centering
\begin{tabular}{@{}>{\raggedright\arraybackslash}p{.76\columnwidth}r@{}}
\toprule
Added coordinates and metric & Rank\\
\midrule
None; any positive diagonal metric & 2\\[3pt]
None; best dense metric & 3\\[3pt]
One product $f_i f_j$; positive diagonal metric & $\leq3$\\[3pt]
One product $f_0f_2$; explicit nondiagonal metric & 4\\[3pt]
One scalar $f_0f_2+f_1f_3$; positive diagonal metric & 4\\[3pt]
Two coordinates $(f_0f_2,f_1f_3)$; positive diagonal metric & 4\\[3pt]
Six pairwise products; experimental diagonal metric & 4\\
\bottomrule
\end{tabular}
\end{table}
\begin{corollary}[Fixed affine expansion cannot cross the ceiling]
\label{cor:affineceiling}
For the four local features of Proposition~\ref{prop:recoveryboundary},
let $\widetilde f=Lf+b$ have any finite output dimension, with $L,b$
fixed independently of inputs and trainable parameters. With any
fixed $\widetilde M\succ0$ and fixed bandwidth, its kernel has
generic differential rank at most three. Thus no such affine
expansion achieves rank four, whereas the nonlinear constructions in
Proposition~\ref{prop:recoveryboundary} and Theorem~\ref{thm:oneproduct}
suffice.
\end{corollary}
\begin{proof}
The offset cancels and
\begin{equation}
 \Delta\widetilde f^T\widetilde M\Delta\widetilde f
 =\Delta f^T L^T\widetilde M L\Delta f.
\end{equation}
The effective metric may be singular, but the data basis is still
$(\cos x_0,\cos x_1)$. Its symmetric $2\times2$ coefficient Gram
has only three entries, bounding the differential rank by three.
\end{proof}
This is a restriction on fixed affine lifts, not on arbitrary
input-dependent processing or jointly trainable postprocessing.

At finite shots, direct ZZ parity averages and products of local Z
sample means are different estimators of the same ideal features in
this family. We compare them on identical outcomes in
Sec.~\ref{sec:productcontrol}. General correlated states do not obey
this product identity. The result establishes a limit of fixed
quadratic metrics on the original features, not a necessity for
additional quantum measurement information. Equal basis count alone
does not imply equal estimation precision.

On ten previously studied source circuits from the YZ-CX,
Hubregtsen, and Kyriienko constructors, fixed diagonal local XYZ
weights evenly spaced from $1$ to $4$ attain
generic rank $K$, certified by exact Fourier Gram minors. In
particular, the two- and four-qubit Kyriienko ranks change from
$4$ to $6$ and from $10$ to $12$. These are redesigned kernels on
existing circuits, not additional entries in the original 54-setting
census. The broader 30-setting metric survey is numerical; ten
single-basis designs also have exact full-rank certificates. Simple
random metrics and uniform-basis controls already attain many of the
same ranks, so these results do not establish a sophisticated selector's
superiority.

The next two sections give the circuit-specific certificates and
dimension census supporting this identifiability picture
(Secs.~\ref{sec:structure} and \ref{sec:census}). We then return to
our third question in Sec.~\ref{sec:shotdesign}: how well can the
retained or restored directions be resolved with finite shots?
Exact rank is an algebraic prerequisite, not a guarantee of statistical
resolution or predictive improvement.

%% file: gauge_classification.tex
\subsection{Connected universal readout gauges}
\label{sec:gaugeclass}
Here a gauge is a common output unitary preserving the readout distance
for every pair of density operators, not just the reachable states.
Proposition~\ref{prop:gauge} concerns the identity component of the
gauge group, modulo global phase; disconnected symmetries are discussed
after the proof.
\begin{proposition}[Isotropic Pauli readout gauges]\label{prop:gauge}
Consider unit-weight readout of a set of nonidentity Pauli words.
(i) A unitary is a universal gauge if and only if its adjoint action
preserves the span of the measured Pauli words.
(ii) For the complete set of Pauli words of weights $1,\ldots,k$ with
$1\le k<n$, including local XYZ ($k=1$), the connected gauge consists
of products of single-qubit unitaries~\cite[Sec.~4.2]{cpr2019locality}.
(iii) For all nonidentity Paulis ($k=n$), every unitary is a gauge.
(iv) For local Z readout, the connected gauge consists of
computational-basis diagonal unitaries.
\end{proposition}
Part~(ii) is not new. By part~(i), its gauges are the unitaries whose
conjugation preserves the space of $k$-local operators, and part~(ii)
is the qubit, identity-component case of the classification of
locality-preserving unitaries in Ref.~\cite[Sec.~4.2]{cpr2019locality},
which also includes site permutations. Our contribution here is the
reduction in part~(i) from universal kernel invariance to preservation
of the measured Pauli span; parts~(iii) and (iv) are elementary.
\begin{proof}
(i) Universal invariance is equivalent to preserving the measured Pauli
subspace: the adjoint action is orthogonal in Hilbert--Schmidt
coordinates, and the metric is its orthogonal projector.

(ii) Reference~\cite{cpr2019locality} shows that unitaries preserving
the $k$-local operators are generated by single-site unitaries and site
permutations; the identity component of that group consists of the
products of single-site unitaries. For completeness we sketch the
infinitesimal step, which is the commutator and weight-growth argument
of Ref.~\cite{cpr2019locality}. Expand a Hermitian generator $H$ in
Pauli words. For a fixed measured $O$, different anticommuting words
$P$ produce different words $iPO$, so leakage outside the measured
subspace cannot cancel between coefficients of $H$. Let $P$ have weight
$w\ge2$. If $w>k$, a single-site $O$ on its support anticommuting with
$P$ gives $iPO$ of weight $w>k$. If $2\le w\le k<n$, let $O$
anticommute with $P$ on one support site and carry $k+1-w$ further
nonidentity factors outside that support; then $O$ has weight
$k+2-w\le k$ but $iPO$ has weight $k+1$. Thus no weight-$w\ge2$ term
is allowed in $H$, while single-site generators preserve every complete
weight sector.

(iii) For $k=n$ the measured span is the full traceless subspace.

(iv) For local Z, any $X$ or $Y$ factor in $P$ leaks under commutation
with that site's Z; exactly the Z-only strings remain. Their span
integrates to all diagonal unitaries.
\end{proof}
Qubit permutations preserve every Pauli weight and hence each readout
above. For $k<n$ and for local Z they are gauges outside the identity
component, consistent with Ref.~\cite{cpr2019locality}. Anisotropic
weights, incomplete weight sectors, other discrete symmetries, and
gauges specific to restricted reachable states require separate
analysis. In particular, entangling diagonal gates preserve local Z
features, whereas generic entangling generators are excluded for $k<n$.

%% file: global_normal_form.tex
\subsection{Global normal form and parameter combinations}
\label{sec:normalform}
Write $\phi_j(x)=\arccos x_j$ for the two inputs in Fig.~\ref{fig:witness}.
Signed Clifford conjugation and commuting same-Pauli fusion give
\begin{align}
 U_\theta(x)&=R_{X_0}(a)R_{X_1}(b)R_{X_0X_1}(c),\\
 a&=\theta_0\phi_0+\theta_2,\nonumber\\
 b&=(\theta_1+\theta_5)\phi_1+\theta_3+\theta_7,\nonumber\\
 c&=\theta_4\phi_0+\theta_6.\label{eq:normal}
\end{align}
All three generators commute. Starting from $|00\rangle$, the two local
Bloch vectors are $(0,-\sin a\cos c,\cos a\cos c)$ and
$(0,-\sin b\cos c,\cos b\cos c)$. Thus unit-weight local XYZ distance is
\begin{align}
 d^2(x,x')={}&2[\cos^2c(x)+\cos^2c(x')]\nonumber\\
 &-2\cos c(x)\cos c(x')\nonumber\\
 &\quad\times\big\{\cos[a(x)-a(x')]\nonumber\\
 &\hspace{16mm}+\cos[b(x)-b(x')]\big\}.\label{eq:closed}
\end{align}
The Gaussian kernel therefore factors through
\begin{equation}
 \beta=A\theta=(\theta_0,\theta_1+\theta_5,\theta_4,\theta_6)^T.
 \label{eq:beta}
\end{equation}
The factorization gives an upper bound of four on the number of necessary
combinations. The matching lower bound is proved below, establishing
minimality among smooth factorizations for this fixed initial state and
readout. Five retained symbols therefore overcount the effective
representation. The common biases in $a,b$
cancel from pairwise differences, while the bias in $c$ generally does
not. This separation uses the specified initial state and readout;
it is not the universal-state quantifier of Theorem~\ref{thm:output}.

An independently evaluated source circuit and dense matrix product
agree in 1,000 random two-input conditions: maximum unitary error up to
global phase is $7.36\times10^{-16}$, distance error is
$7.11\times10^{-15}$, and kernel error is $9.44\times10^{-16}$.
The analytic identities, rather than these finite probes, explain the
factorization. This representative circuit is consequently an
explanatory example, not evidence that global Pauli optimization fails.

Reparameterization does not automatically preserve ordinary gradient
updates. If $L(\theta)=\ell(A\theta)$, then
$\nabla_\theta L=A^T\nabla_\beta\ell$ and an ascent step induces
\begin{equation}
 \begin{gathered}
 \beta_{t+1}=\beta_t+\eta AA^T\nabla_\beta\ell(\beta_t),\\
 AA^T=\diag(1,2,1,1).
 \end{gathered}
 \label{eq:pushforward}
\end{equation}
Twenty steps with this metric preserve the source-coordinate kernel
trace to $6.67\times10^{-16}$ in a fixed diagnostic; using the identity
metric instead changes the final Gram matrix by $5.52\times10^{-3}$.
This diagnostic is not an accuracy comparison. Other optimizers and
coordinate regularizers need their own transformation rules.

%% file: dimension_theory.tex
\subsection{A matching lower bound on kernel dimension}
\label{sec:dimensiontheory}
At a fixed parameter assignment, let the kernel-relevant differential
rank be the dimension of the span of the functions
$\partial_{\theta_j}d^2(\cdot,\cdot;\theta)$ over input pairs.
Multiplication by the common nonvanishing function
$-\gamma e^{-\gamma d^2}$ gives the same rank for the Gaussian kernel.
We use its generic value $r$ on the connected analytic parameter domain;
special parameter assignments may have smaller rank.

\begin{proposition}[Exact representative dimension]
For Eq.~\eqref{eq:normal}, initialization $|00\rangle$, and unit-weight
local XYZ readout on $x\in(-1,1)^2$, the generic kernel-relevant rank is
exactly $r=4$. Four is also the minimum dimension of a global smooth
factorization of this kernel family through its source parameters.
\end{proposition}
\begin{proof}
Equation~\eqref{eq:beta} gives a four-dimensional factorization. Its
four constant null directions in source coordinates are
$e_2,e_3,e_7,e_1-e_5$. For a matching lower bound set
$\beta^*=(1,1,1,0)$ and use four input pairs with
$(p,p',q,q')=(\phi_0(x),\phi_0(x'),\phi_1(x),\phi_1(x'))$ given by
\begin{align}
 (p,p',q,q')/\pi\in\{& (1/6,1/2,1/6,2/3),\nonumber\\
 &(1/6,1/3,1/3,1/2),\nonumber\\
 &(1/6,2/3,1/6,5/6),\nonumber\\
 &(2/3,1/6,1/6,1/2)\}.
 \label{eq:rankpoints}
\end{align}
All are feasible because $0<\phi_j<\pi$ and $x_j=\cos\phi_j$.
Let $J$ have rows $\nabla_\beta d^2$ at these pairs. Differentiating
Eq.~\eqref{eq:closed} exactly yields
\begin{equation}
 \det J=\frac{\pi^3(2\sqrt3-39)}{1536}\ne0.
 \label{eq:rankminor}
\end{equation}
The explicit matrix is given in Appendix~\ref{app:dimension}.
This analytic minor is nonzero at one point, hence gives generic rank
at least four. Any smooth factorization through $m$ real variables
would, by the chain rule, bound the rank of these four kernel evaluations
by $m$, requiring $m\ge4$. The global linear map in
Eq.~\eqref{eq:beta} attains the bound.
\end{proof}

This is not a unique global chart: discrete symmetries and singular
points can remain. It is also not a lower bound on quantum gates,
measurement settings, or shots. The theorem concerns the fixed
initialization and readout, not the universal-state quantifier of the
common-output theorem.

For other circuits we use two separate certificate types. Exact
identities for independent constant null directions give an upper bound;
a nonzero minor of the source-coordinate derivative matrix at feasible
inputs and one shared parameter assignment gives a lower bound.
Numerical singular values may select a minor but do not certify its
nonzero value. Equally, activity of each individual coordinate does not
exclude a null combination of coordinates.

%% file: extension_theory.tex
\subsection{Nonlinear redundancy with no inactive retained coordinate}
\label{sec:nonlinearhub}
The distinction is not limited to redundant linear combinations.
Consider the frozen four-qubit, one-layer Hubregtsen source circuit
\cite{hubregtsen2022,squlearn}
with two input features and local Z readout. It prepares $|+\rangle$
on each qubit, applies $R_z(x_{q\bmod2})$ and $R_y(\theta_q)$,
and ends with controlled-Z rotations carrying four further symbols.
Those controlled rotations commute with every measured Z and are
already removed by the strengthened observable-commutant baseline.
The four retained features are exactly
\begin{equation}
 f_q(x)=-\sin\theta_q\cos x_{q\bmod2},\qquad q=0,\ldots,3.
 \label{eq:hubfeatures}
\end{equation}
Exact Heisenberg expansions of the frozen compiled IR reproduce
each identity, including its sign and source-parameter index.

\begin{proposition}[A nonlinear two-coordinate factorization]
For this source configuration and unit-weight local Z readout,
all four retained source coordinates are individually kernel-active.
The minimum dimension of a smooth kernel factorization and the
generic differential rank are both two, whereas any factorization
through a constant linear map of the retained coordinates requires
rank four.
\end{proposition}
\begin{proof}
Let $c_j=\cos x_j-\cos x'_j$. Equation~\eqref{eq:hubfeatures} gives
\begin{align}
 d^2_\theta(x,x')&=\beta_0 c_0^2+\beta_1 c_1^2,\label{eq:hubdistance}\\
 \beta_0&=\sin^2\theta_0+\sin^2\theta_2,\nonumber\\
 \beta_1&=\sin^2\theta_1+\sin^2\theta_3.\label{eq:hubbeta}
\end{align}
This is a global smooth factorization through two real variables.
Set the four retained angles to $\pi/4$ and choose input pairs
$((0,0),(\pi/4,0))$ and $((0,0),(0,\pi/4))$ in $(-1,1)^2$.
The distance Jacobian minor in source columns $\theta_0,\theta_1$
is diagonal with entries $(1-1/\sqrt2)^2$, hence
\begin{equation}
 \det J=(1-1/\sqrt2)^4=\frac{17}{4}-3\sqrt2>0.
 \label{eq:hubminor}
\end{equation}
Analyticity and the chain rule therefore give generic rank and
minimum smooth factor dimension two. Each source derivative
$\sin(2\theta_q)c_{q\bmod2}^2$ is not identically zero, so every
retained coordinate is active. A constant null vector $v$ would
require $v_0\sin2\theta_0+v_2\sin2\theta_2=0$ and
$v_1\sin2\theta_1+v_3\sin2\theta_3=0$ for all independently
variable angles. Each coefficient must vanish. A rank-deficient
constant linear factor would supply a nonzero such vector and is
therefore impossible.
\end{proof}

Here an exact constant-null-space calculation returns the valid
upper bound four, not the minimal nonlinear dimension two. This
source-derived example delimits the completeness of that upper-bound
method; it does not establish a general nonlinear reduction algorithm.
The two input features are reused across four qubits, and the final
controlled rotations are invisible to the selected readout. The
factorization is thus a readout-specific reduction to a weighted
classical cosine-feature kernel, not evidence of quantum advantage
or a newly efficient hardware compiler. Nor does ordinary gradient
descent in $\beta$ reproduce the original finite-step optimizer in
$\theta$ without a separately specified update map.

%% file: family_extension.tex
\subsection{A dimension formula for a restricted Z-readout family}
\label{sec:hubfamily}
The nonlinear example extends to a structural family, rather than to
every depth or readout of the named source constructor.
\begin{proposition}[Grouped single-layer Z features]\label{prop:hubfamily}
Let $n\ge d\ge1$, and let $\sigma:\{1,\ldots,n\}\to\{1,\ldots,d\}$
be surjective. Prepare $|+\rangle^{\otimes n}$, apply
$R_y(\theta_q)R_z(x_{\sigma(q)})$ on each qubit, and then any diagonal
unitary. Use local Z readout with fixed positive weights $w_q$, fixed
$\gamma>0$, and $x\in(-1,1)^d$. The generic kernel rank and minimum
global smooth factor dimension are $d$. Every retained $\theta_q$
is individually active, and any constant-linear factorization of
these $n$ coordinates has rank at least $n$.
\end{proposition}
\begin{proof}
The diagonal tail commutes with Z. The measured features are
$f_q=-\sin\theta_q\cos x_{\sigma(q)}$, whence
\begin{align}
 d^2_\theta(x,x')&=\sum_{j=1}^d\beta_j(\theta)
                    (\cos x_j-\cos x'_j)^2,\\
 \beta_j(\theta)&=\sum_{q:\sigma(q)=j}w_q\sin^2\theta_q.
\end{align}
This is a global smooth $d$-coordinate factorization. Set every
$\theta_q=\pi/4$, choose one representative $q_j$ per group and input
pairs $(0,(\pi/4)e_j)$. The distance Jacobian minor in columns $q_j$
is diagonal with entries $w_{q_j}(1-1/\sqrt2)^2>0$.
Thus the generic rank and minimum factor dimension equal $d$.
Each coordinate derivative is a nonzero function. If a constant
vector $v$ annihilates every distance derivative, independent variation
of angles in each group gives $v_qw_q=0$ for every $q$. A
rank-deficient constant-linear factor would provide such a nonzero
vector, a contradiction.
\end{proof}
The four-qubit, two-input, one-layer Hubregtsen instance is a member
with $w_q=1$. Data re-uploading, nondiagonal final gates, tied angles,
and other readouts are outside this formula. It proves a kernel
dimension law, not deletion of every retained angle or shot savings.

\subsection{Kyriienko-type gauge bounds and scaling certificates}
\label{sec:kyrifamily}
For $n\ge2$, consider Eq.~\eqref{eq:kyriform} with $n$ Euler blocks
and $K=3n$ independent angles, a fixed data-only encoding
$D(x)$, a fixed entangler $E$, and unit readout weights. The source
constructor uses the forward nearest-neighbor CNOT entangler
$E=E_{\rm o}E_{\rm e}$: $E_{\rm e}=\prod_j\mathrm{CX}_{2j,2j+1}$
acts first and $E_{\rm o}=\prod_j\mathrm{CX}_{2j+1,2j+2}$ second.
For $n=2,4$ this is the entangler of Sec.~\ref{sec:gramkyri}. Let
$\mathfrak{lu}=\bigoplus_q\mathfrak{su}(2)_q$ be the Lie algebra of
traceless single-qubit generators, that is, of products of
single-qubit unitaries with the global phase ignored, and let
$\mathfrak s_E=\mathfrak{lu}\cap E^\dagger\mathfrak{lu}E$
contain the local output generators that $E$ carries back into the
Euler blocks.
\begin{proposition}[Kyriienko-type upper bounds]\label{prop:kyribounds}
(i) If $E$ commutes with $Z_0$ and $X_{n-1}$, then isotropic local XYZ
and complete weight-at-most-$k$ readout with $k<n$ satisfy
\begin{equation}
 r\le3n-2.\label{eq:kyrifamilyupper}
\end{equation}
For the forward nearest-neighbor CNOT entangler,
$\mathfrak s_E=\mathrm{span}\{iZ_0,iX_{n-1}\}$, so
$\dim\mathfrak s_E=2$ for every $n\ge2$.
(ii) If $E$ is any CNOT network, then under local Z readout with any
positive weights the trailing angles $c_0,\ldots,c_{n-1}$ are globally
kernel-inactive, and $r\le2n$.
(iii) For complete Pauli readout ($k=n$), $r=0$ for every fixed $E$.
\end{proposition}
\begin{proof}
(i) By Proposition~\ref{prop:gauge}(i), output rotations about $Z_0$
and $X_{n-1}$, which preserve every Pauli weight, are readout gauges.
Commuting through $E$, they become left
rotations of the first and last Euler blocks. On a regular chart
($\sin b_q\ne0$) the Euler map is locally invertible, so they give two
independent parameter tangents annihilating every $\partial d^2$;
analyticity extends the vanishing higher minors.

For the dimension, conjugation by $\mathrm{CX}_{c,t}$ permutes signed
Pauli words: it fixes $Z_c$ and $X_t$ and maps $X_c\mapsto X_cX_t$,
$Y_c\mapsto Y_cX_t$, $Z_t\mapsto Z_cZ_t$, and $Y_t\mapsto Z_cY_t$.
Distinct words cannot cancel, so $Y\in\mathfrak{lu}$ satisfies
$EYE^\dagger\in\mathfrak{lu}$ if and only if each single-qubit word
in $Y$ has a weight-one image. Each layer acts on disjoint pairs and
maps nonidentity restrictions to nonidentity restrictions. A two-site
image under $E_{\rm e}$ occupies sites $2j,2j+1$, which lie in
different pairs of $E_{\rm o}$, so its weight stays at least two. A
weight-one image thus requires a word fixed by $E_{\rm e}$: $Z_q$ for
even $q<n-1$, $X_q$ for odd $q$, or any word on the unpaired site
$n-1$ for odd $n$. Then $E_{\rm o}$ maps $Z_q\mapsto Z_{q-1}Z_q$ for
even $q\ge2$, $X_q\mapsto X_qX_{q+1}$ for odd $q<n-1$, and
$Z_{n-1},Y_{n-1}$ to two-site words for odd $n$, while fixing $Z_0$
and $X_{n-1}$.

(ii) A CNOT network permutes computational basis states, so
$\Delta(c)=E[\bigotimes_qR_z(c_q)]E^\dagger$ is diagonal and
$U_\theta(x)=\Delta(c)E[\bigotimes_qR_x(b_q)R_z(a_q)]D(x)$. Diagonal
unitaries commute with every $Z_j$, so no local Z feature depends on
$c$, and the $n$ columns $\partial_{c_q}d^2$ vanish identically.
Equivalently, Theorem~\ref{thm:transport} carries each generator $Z_q$
to a Z string commuting with the readout.

(iii) With unit weights, $d^2=2^n\Tr[(\rho_x-\rho_{x'})^2]$, and the
data-independent unitary after $D(x)$ cancels from this
Hilbert--Schmidt distance; readout assumptions are essential.
\end{proof}
Since connected universal gauges are products of single-qubit
unitaries (Proposition~\ref{prop:gauge}(ii), from
Ref.~\cite{cpr2019locality}) and $\dim\mathfrak s_E=2$ exactly, they
alone cannot certify $r<3n-2$ for the source entangler. Matching lower
bounds are certified only at fixed sizes. A separate scaling
computation extends the frozen source-to-IR route to $n=6,8$; these
six settings are not added to the 54-setting census. At one rational
parameter point it finds exact nonzero minors of the
$\partial_\theta d^2$ Jacobian over 45 input pairs, of orders equal to
the entries of Table~\ref{tab:kyriscaling}, and it reproduces the
census ranks at $n=2,4$. With Proposition~\ref{prop:kyribounds}, these
minors certify $r=3n-2$ for local XYZ ($n=2,4,6,8$) and
weight-at-most-two readout ($n=4,6,8$), and $r=2n$ for local Z
($n=2,4,6,8$).

\begin{table}[t]
\caption{Certified generic ranks for the Kyriienko-type family: exact
nonzero minors match Proposition~\ref{prop:kyribounds}. Rows $n=2,4$
repeat census values (Table~\ref{tab:extension}); rows $n=6,8$ form a
separate scaling extension outside the 54-setting census. The dagger
marks $k=n$, where part (iii) gives $r=0$.}
\label{tab:kyriscaling}
\begin{tabular}{rrrrr}
\toprule
$n$ & $K$ & $r_{XYZ}$ & $r_{\le2}$ & $r_Z$\\
2&6&4&$0^\dagger$&4\\
4&12&10&10&8\\
6&18&16&16&12\\
8&24&22&22&16\\
\bottomrule
\end{tabular}
\end{table}

These certificates cover even $n$ only; for odd $n$ every qubit
encodes both inputs, and no odd size has an exact certificate.
Floating-point Jacobian ranks equal $3n-2$ for local XYZ at every
$n=2,\ldots,12$, and for complete weight-at-most-$k$ readout at all
six pairs $(n,k)$ with $2\le k<n\le5$. At $(5,3)$, five seeded
trials give numerical rank $13$, stable across the tested rank
tolerances (Appendix~\ref{app:kyriscaling}). This is numerical
evidence only, not an exact certificate for odd $n$.
Equality for all $n$ remains open. A block-cover lower-bound route
works when a finite direct-summand condition on the data modes holds,
but that condition fails at $n=5,7$, where the cover modes are
already dependent, and at $n=12$ through a Chebyshev frequency
resonance, although the numerical rank there is still $3n-2$
(Appendix~\ref{app:kyriscaling}).

%% file: kyriienko_gram_results.tex
\subsection{Exact ranks for the Kyriienko settings}
\label{sec:gramkyri}
The frozen one-layer Kyriienko constructors have
\begin{equation}
 U_\theta(x)=E\Bigl(\bigotimes_{q=0}^{n-1}
 R_z(c_q)R_x(b_q)R_z(a_q)\Bigr)D(x),\label{eq:kyriform}
\end{equation}
with $K=3n$. Here $D$ is the fixed-frequency encoding,
$E=\mathrm{CX}_{0,1}$ for $n=2$, and $E$ is the chronological
chain $(\mathrm{CX}_{0,1},\mathrm{CX}_{2,3},\mathrm{CX}_{1,2})$
for $n=4$. Disjoint gates commute to give this form exactly.
Both $Z_0$ and $X_{n-1}$ remain unchanged under conjugation by $E$.
Local XYZ and complete weight-at-most-two readouts with unit weights
are invariant under arbitrary local output rotations. Thus left
rotations about these two axes give independent one-dimensional
kernel gauge orbits. On a regular Euler chart ($\sin b_q\ne0$),
the Euler map is locally invertible, so the two source-parameter
tangent directions are independent and $r\le3n-2$.
Analyticity extends the vanishing higher minors from this open set.

For a matching lower bound we compute real data-Fourier coefficients
and their parameter derivatives using exact Gaussian rational
arithmetic at
\begin{equation}
 \theta_j=2\arctan t_j,\qquad t_j=(j+2)/(j+5).
\end{equation}
All sines and cosines are rational. Nonzero exact Gram Jacobian minors
of orders $4,10,10$ prove $r\ge4,10,10$ for the two-qubit XYZ,
four-qubit XYZ, and four-qubit weight-at-most-two settings. Together
with the gauge upper bounds, these minors fix all three ranks exactly.
These are three settings of the existing census, not three new
independent families. Full rational matrices and determinants are
retained in the accompanying audit; no floating rank threshold is used.

The resulting ranks are exactly $4,10,10$, closing all 54 settings
across the two dimension studies. The second gauge is a
parameter-dependent Euler direction, not another globally dead
source coordinate. These closures consequently do not change the
reported inactive-coordinate counts or establish gate or shot savings.
Implementation and independent source checks are given in
Appendix~\ref{app:gram}.

%% file: dimension_results.tex
\subsection{A restricted certificate and dimension pilot}
\label{sec:dimensionpilot}
A separate follow-up fixes two source modules, ChebyshevRx and
ParamZFeatureMap, at two and four qubits, one and two layers, and the
same three readouts: eight circuit configurations and 24 circuit/readout
settings. These settings use integer angle coefficients without
ambiguous fixed-angle restoration. They form a selected subset of the
102-setting inventory, not a new population sample or a full rerun.
The selection criterion, pair budget and precision were fixed before
calculating this pilot's results.

We expand the fixed-readout squared distance by Pauli Heisenberg
propagation, retaining exact Gaussian-rational coefficients and shared
data expressions. Exact cancellation of all coefficients of a parameter
direction certifies inactivity; a nonzero formal expression is not used
to infer activity. A separate interval statevector and tangent engine
evaluates derivatives at actual input pairs. It certifies activity only
when the resulting interval excludes zero. The same calculation provides
nonzero Jacobian minors at a single shared parameter assignment. We do
not use random all-zero tests to accept identities.

Of 162 source-symbol entries, 30 have exact inactivity identities and
132 have feasible nonzero derivative witnesses; none remains undecided
in this restricted pilot. All 18 previously certified entries in this
subset are recovered. Exact constant-null-space upper bounds match
the feasible-minor lower bounds in every setting
(Table~\ref{tab:dimension}). Each entry uses at most 24 input pairs.
The activity witnesses and the 22 nontrivial minors replay successfully
at increased interval precision; the other two kernels are exactly
constant and have rank zero. Appendix~\ref{app:dimension} gives the
arithmetic and source-check scope.

\begin{table}[htbp]
\caption{Matching lower and upper bounds on generic kernel differential
rank in the restricted pilot. $K$ counts source parameters; $r_{XYZ}$,
$r_Z$ and $r_{\le2}$ refer to isotropic local XYZ, local Z, and all Pauli
words of weight at most two. These are kernel ranks, not executable
angle-task counts or independent population observations.}
\label{tab:dimension}
\begin{tabular}{lrrrrrr}
\toprule
Family & $n$ & $L$ & $K$ & $r_{XYZ}$ & $r_Z$ & $r_{\le2}$\\
ChebyshevRx &2&1&4&3&4&2\\
 &2&2&8&4&6&3\\
 &4&1&8&7&8&7\\
 &4&2&16&12&14&12\\
ParamZ &2&1&2&2&0&2\\
 &2&2&4&4&2&4\\
 &4&1&4&4&0&4\\
 &4&2&8&8&4&8\\
\bottomrule
\end{tabular}
\end{table}

Two distinctions are operationally important. First, for two-qubit,
two-layer ChebyshevRx with local Z readout, all eight individual source
symbols are active, but $r=6$ because two shared combinations remain
redundant. Local XYZ and full two-qubit Pauli readout instead give
$r=4$ and $r=3$. Thus coordinate activity, parameter-combination rank,
and the measurement geometry cannot be substituted for one another.

Second, all 12 newly identified inactive entries occur in ParamZ/local Z.
For one layer the local Z features vanish, giving six inactive entries
across the two sizes and constant kernels. For two layers the first
encoded rotations affect Z expectations but the last six data-dependent
Z rotations commute with the readout. These are elementary
state/readout-specific identities outside the earlier rule's
input-independent-parameter candidate list, not evidence of a novel
advantage over an appropriately extended commutant baseline.
No new training-time or gate/shot reduction is measured in this pilot.

The method is sufficient, not a general complete decision procedure.
Exact coefficient cancellation can prove an identity without assuming
independence of every lifted angle. Conversely, activity is checked only
on the actual input manifold. This distinction prevents a nonzero value
in an unrealizable independent-angle extension from becoming a false
activity verdict. The larger survey's unresolved constant provenance
and full-scope implementation are not silently resolved by this pilot.

%% file: extension_results.tex
\subsection{Three additional source families and structural closure}
\label{sec:extension}
We next fixed 30 disjoint combinations of circuits and readouts from YZ-CX,
Hubregtsen, and Kyriienko source modules
\cite{squlearn,hubregtsen2022,kyriienko2021}: ten source circuit
configurations under the same three readouts. The first two modules
use two/four qubits and one/two layers; Kyriienko retains the inventory's
default one-encoding-layer, one-variational-layer construction.
Exact half-integer coefficients from controlled-rotation decompositions
are admitted, without restoring approximate constants to multiples of
$\pi$. Scope, witnesses and resource definitions were fixed before the
run; the nonlinear analysis in Sec.~\ref{sec:nonlinearhub} followed
the observed rank gap.

The bounded Fourier pass completed in 13 settings. Across 252 source
symbol entries it certified 25 exact inactive identities and 195
feasible activity witnesses, leaving 32 entries undecided because
symbolic expansion exceeded its fixed budget. We retain those outcomes
rather than interpreting a budget failure as activity. A separate
structural closure then recovered all 32 remaining inactive entries
using a stronger baseline: global signed Pauli normalization,
common-output isometry deletion, and observable-commutant deletion
that also permits data-dependent rotation angles. All 57 previously
inactive entries in this subset are recovered, and no new inactive
coordinate lies beyond this baseline.

Combining its exact angle-coefficient upper bound with feasible
Jacobian minors closes the rank in 26 settings. The explicit nonlinear
factorization in Sec.~\ref{sec:nonlinearhub} closes one further setting.
At that stage three Kyriienko bounds remained open; activity of every
retained coordinate does not settle these combination ranks. The
post hoc coefficient Gram minors and output-gauge proof in
Sec.~\ref{sec:gramkyri} close those three intervals, completing all
30 ranks in this extension and all 54 across the two studies.
For the 26 matching structural bounds, this baseline already attains
the minimum smooth parameter-factor dimension of the tested kernel
family. This does not certify optimal gate, measurement, or shot counts.

\begin{table}[htbp]
\caption{Generic kernel differential rank in the 30-setting extension.
$K$ is the original symbol count. Exact values combine feasible
lower bounds with structural or symbolic upper bounds. Stars mark
the three new coefficient Gram/gauge closures. The dagger
uses the nonlinear factorization of Eq.~\eqref{eq:hubbeta}.
Rank zero means independence from trainable parameters, not necessarily
a constant function of the inputs.}
\label{tab:extension}
\begin{tabular}{lrrrrrr}
\toprule
Family & $n$ & $L$ & $K$ & $r_{XYZ}$ & $r_Z$ & $r_{\le2}$\\
YZ-CX &2&1&4&3&2&2\\
 &2&2&8&7&6&6\\
 &4&1&8&6&4&6\\
 &4&2&16&14&12&14\\
Hubregtsen &2&1&2&0&2&0\\
 &2&2&4&2&4&2\\
 &4&1&8&8&$2^\dagger$&8\\
 &4&2&16&16&12&16\\
Kyriienko &2&1&6&$4^*$&4&0\\
 &4&1&12&$10^*$&8&$10^*$\\
\bottomrule
\end{tabular}
\end{table}

In all 30 settings, zeroing the additionally Fourier-certified
inactive coordinates after this strengthened baseline produces zero
incremental reduction in Pauli rotation tasks, trainable angle tasks,
or the corresponding two-shift feature-evaluation schedule. These are
ideal operation counts, not executed hardware shots or a fixed-error
shot guarantee. In particular, the nonlinear two-coordinate example
does not constitute another successful internal gate-deletion family.
It instead shows why individual activity and constant-linear null-space
analysis can both miss a nonlinear kernel factorization.

%% file: shot_design.tex
\section{Finite shot sensitivity and estimator controls}
\label{sec:shotdesign}
The question here is whether restored ideal dependence is well
resolved at a fixed shot budget, not whether rank alone improves
prediction. Grouped Pauli measurement and shot allocation affect
finite sampling error, so the number of observables alone is not a
measure of measurement cost \cite{crawford2021efficient}. We fix the
measurement schedules below rather than optimize over all grouping
or shot allocation strategies.
The study fixes three seed panels, each with three design points
and five held out parameter/input probes, for YZ-CX and Kyriienko
local XYZ on two qubits and Hubregtsen local Z and Z+ZZ on four qubits.
Inputs are synthetic continuous values; no labels or predictive
accuracy are used. Each design selects one metric from the same
bank of 34 candidates at 1,024 shots and keeps it fixed at all evaluation
points and budgets. We compare isotropy, fixed diagonal anisotropy,
untuned random diagonal/dense metrics, selection based on rank and a
variant that accounts for noise. Both selectors prioritize rank; ties
are resolved using retained singular values of the distance Jacobian,
respectively without and with noise whitening. All use ideal model information.

For 16 disjoint input pairs, let $J$ be the distance Jacobian and
$Q$ an orthonormal basis of parameter directions visible in the
centered reference features. Let $\nu_p$ be the exact variance of an
unbiased split shot distance estimate at the unperturbed point.
The local diagnostic for a total displacement of $\epsilon=0.1$ is
\begin{equation}
 W=\epsilon\,\diag\bigl((2\nu_p)^{-1/2}\bigr)JQ.
 \label{eq:shotwhiten}
\end{equation}
The factor two accounts for independent estimates at the two
parameter states in the local limit. The smallest retained singular
value is reported alongside rank; the weakest sensitivity over the
\emph{whole} target space is zero if rank is deficient. Reference
spaces, directions, inputs and selected metrics are frozen before
server evaluation. This prevents each method from choosing favorable
test directions and does not mistake rotation of a nullspace for
an increase in total rank.

The norm of a displacement is the Euclidean norm in the original
gate angle coordinates (radians), with $Q^TQ=I$ in those coordinates.
Thus singular values of $W$ depend on the coordinates and their scale;
they are comparable here because the parameterization, $Q$, and
$\epsilon$ are held fixed. They are not the classical or quantum
Fisher information of the full measurement distribution, and no
invariance under reparameterization is asserted. A zero kernel
derivative concerns the mean distance statistic only: raw outcome
probabilities or its sampling variance can still depend on that
direction.

\begin{table*}[tp]
\caption{Held out sensitivity normalized by noise at 4,096 requested
shots per input and parameter state. Each triple gives seed panel
medians of five paired ratios, not a ratio of medians.
The statistic is the smallest retained singular value of
Eq.~\eqref{eq:shotwhiten}; ratios above one favor the selector that accounts for noise.
The final column instead gives the median weakest sensitivity over
the whole reference space. Hubregtsen local Z remains rank deficient;
its retained sensitivity ratio to a diagonal metric of rank two would
compare different ranks and is omitted. These are ideal model
diagnostics, not measured classification or hardware gains.}
\label{tab:shotdesign}
\small
\begin{tabular}{@{}lcccc@{}}
\toprule
Circuit/readout & Target rank & Versus fixed diagonal & Versus rank based & Whole space weakest\\
\midrule
YZ-CX 2q/local XYZ & $4/4$ & $1.77,1.65,2.33$ & $1.00,1.30,1.61$ & $0.331,0.425,0.415$\\
Kyriienko 2q/local XYZ & $6/6$ & $1.75,2.03,3.22$ & $1.00,1.24,1.10$ & $0.135,0.254,0.275$\\
Hubregtsen 4q/local Z & $3/4$ & not comparable & $1.00,1.00,1.00$ & $0,0,0$\\
Hubregtsen 4q/Z+ZZ & $4/4$ & $1.22,1.32,0.91$ & $1.00,1.00,1.00$ & $0.0082,0.0188,0.0052$\\
\bottomrule
\end{tabular}
\end{table*}

Table~\ref{tab:shotdesign} shows improvements over fixed anisotropy
for YZ-CX and Kyriienko at the level of panel medians, but not uniform
improvement at every point. The selectors based on noise and rank
choose the same metric in eight of twelve panel/readout designs.
Even in the four differing designs, some held out points deteriorate.
For Hubregtsen Z+ZZ the fourth direction becomes identifiable but
the weakest sensitivity over the whole space is only $0.0052$ to $0.0188$ at
4,096 requested shots: a fourth direction exists mathematically but
is weakly resolved by this local diagnostic normalized by noise.
This statistic is neither a detection probability nor classification
accuracy. One panel's
median ratio to fixed anisotropy is below one. The result therefore
supports diagnosis that accounts for recovery, not generic superiority of a new
metric learning algorithm.

All four budgets, $256$, $1{,}024$, $4{,}096$ and $16{,}384$, are
retained. Joint multinomial sampling uses 256 replicates and preserves
correlations within each basis; local XYZ splits the total across three
bases, while Z and Z+ZZ share ZZZZ outcomes. Empirical/predicted
variance ratios have pooled median $0.997$ and 5th to 95th percentiles
$[0.864,1.143]$. These are descriptive calibration statistics, not
independent hypothesis tests. Ideal matched contrasts specific to each method
are permitted symmetrically in the finite displacement check, so it
remains an oracle diagnostic rather than practical detection power.
Appendix~\ref{app:shotprotocol} gives the estimator, full budget table,
measurement accounting, and replay across platforms.

\subsection{A classical product feature control on the same outcomes}
\label{sec:productcontrol}
This control asks a separate question: can the same ideal lifted
representation be estimated more precisely from the same outcomes?
For the Hubregtsen family in Sec.~\ref{sec:recovery}, we additionally
form every ZZ feature by multiplying the two local Z sample means
\emph{within each independent shot half}. This changes only the
estimator, not the ideal lifted representation or its rank. Joint
Z outcome probabilities factorize, so these products are unbiased;
the distance estimator using independent halves is then unbiased as well.
Under this factorization, the product estimator is the conditional
expectation of the direct parity estimator given the counts for each qubit.
Conditional variance therefore cannot increase for a fixed
metric and a fixed linear distance contrast. The proof and exact
covariances appear in Appendix~\ref{app:productcontrol}. This is a
classical estimation control exploiting a known model restriction,
not a new quantum measurement advantage.

\begin{corollary}[Local sensitivity under product estimation]
\label{cor:productsensitivity}
Under the product Z distribution assumptions of
Appendix~\ref{app:productcontrol}, fix the metric, shot allocation,
parameter point, disjoint input pairs, target basis $Q$, and
displacement $\epsilon$. Use independent shot halves and independent
estimates across pairs. Suppose both estimators have strictly positive
distance variance for each pair. For $W$ in Eq.~\eqref{eq:shotwhiten},
\begin{equation}
 W_{\rm product}^T W_{\rm product}
 \succeq W_{\rm direct}^T W_{\rm direct}.
 \label{eq:productinformation}
\end{equation}
Every ordered singular value is therefore nondecreasing, while the
rank and nullspace are unchanged. In particular, the smallest retained
sensitivity cannot decrease; a missing target direction remains missing.
\end{corollary}
\begin{proof}
Conditional variance gives
$0<\nu_{{\rm product},p}\le\nu_{{\rm direct},p}$ for every pair.
The ideal representation, and hence $JQ$, is identical. Writing its
$p$th row as $a_p^T$, the difference in Eq.~\eqref{eq:productinformation}
is
\begin{equation}
 \frac{\epsilon^2}{2}\sum_p
 \left(\frac{1}{\nu_{{\rm product},p}}-
       \frac{1}{\nu_{{\rm direct},p}}\right)a_pa_p^T\succeq0.
\end{equation}
The eigenvalue min--max principle gives the ordering of singular values.
Positive row weights preserve the rank and nullspace of $JQ$.
\end{proof}
This is a weak comparison in the local limit at the same fixed design.
It neither guarantees strict improvement nor compares selectors that
choose different metrics. Table~\ref{tab:shotdesign} retains the
original direct parity sensitivity results; the paired variances below
concern finite displacements and fixed contrasts, a distinct statistic.

A supplementary local check reuses all three frozen panels, five
held out points, four target directions, six metric arms and four
shot budgets. Both estimators use identical outcomes and the
\emph{original direct parity contrast}; no metrics or contrasts are
retuned. At 4,096 shots, the three panel medians of the exact
product/direct contrast variance ratio are $0.895,0.829,0.869$ for
the frozen arm that accounts for noise. Lower is better; these correspond to
about $10.5\%,17.1\%,13.1\%$ variance reductions for that comparison,
not reductions in hardware shots. The original local Z control with a dense
metric has rank three and remains distinct from both lifted
representations of rank four. The direct parity draws reproduce the saved server
samples within $5.6\times10^{-14}$; the new product arm has so far
been evaluated locally, not replayed on the server. These results
reinforce the distinction between nonlinear representation and its
estimation, rather than the superiority of direct ZZ measurement.

\input{minimal_product_results}

%% file: minimal_product_results.tex
\subsection{Matched finite shot check of minimal product recovery}
\label{sec:minimalproductshots}
We compare four untuned designs with unit trace on the same
Hubregtsen source on four qubits: no product with
$M=(32I+ww^T)/158$, $w=(1,2,3,4)^T$; one product
$f_0f_2$ with $M=(\left(\begin{smallmatrix}2&1\\1&2\end{smallmatrix}\right)
\oplus I_3)/7$ in order $(f_0,f_1,f_2,f_3,f_0f_2)$; two products
$(f_0f_2,f_1f_3)$ and all six products, with diagonal weights evenly spaced
from 1 to 4 and normalized to unit trace. The latter two use the earlier
fixed diagonal convention. Thus the comparison is between complete fixed
designs, not an isolated effect of feature count.

This matched study reuses all three previously frozen panels and their
five held out points, with 16 disjoint input pairs per point. It is not new
held out data or independent preregistration: the designs were specified
after the analytic result was available. No metric selection is performed.
The target is the Euclidean space of the original four angles, $Q=I_4$, with
$\epsilon=0.1$ in Eq.~\eqref{eq:shotwhiten}. All designs use the same
ZZZZ outcomes from one basis, split into two independent halves, at each of
the four budgets $256,1024,4096,16384$ per input and parameter state. Both
direct parity estimates and products of marginal means use the full exact
covariance; the latter are unbiased only under the verified factorized
Z distribution. No extra shots are allocated to extra products.

\begin{table*}[tp]
\centering
\caption{Weakest sensitivity over the whole target space, normalized by noise,
at 4,096 shots. Each triple contains the three panel medians over five matched
points. The control without products has a missing fourth direction, so its
value over the whole space is zero rather than its positive smallest retained
value. These are local exact model diagnostics, not detection probabilities
or QPU results.}
\label{tab:minimal_product_matched}
\small
\begin{tabular}{@{}rccc@{}}
\toprule
Products & Rank/target & Direct parity & Product marginals\\
\midrule
0 & 3/4 & $0,0,0$ & $0,0,0$ \\
1 & 4/4 & $0.003732,0.007171,0.002165$ & $0.003838,0.007445,0.002237$ \\
2 & 4/4 & $0.004962,0.007957,0.007381$ & $0.005498,0.009550,0.008281$ \\
6 & 4/4 & $0.004940,0.011580,0.005900$ & $0.005334,0.012630,0.006342$ \\
\bottomrule
\end{tabular}
\end{table*}

All 15 points give numerical ranks three without products and four for the
three lifted designs at the three recorded tolerances. This checks the
separate analytic result but is not itself an exact rank certificate.
Minimality with one product does not imply superior conditioning with finite
shots. At 4,096 shots its median weakest sensitivity in each panel is below
both larger lifts for both estimators, although pointwise comparisons include
both improvements and deteriorations. For a fixed design, product estimation
never increases the exact distance variance and never decreases a retained
ordered singular value of the noise whitened Jacobian, within numerical
tolerance; the missing direction in the control of rank three remains missing.

Independent dense propagation of the frozen source gate program agrees with
all ten analytic features within $2.2\times10^{-15}$ and with factorized
ZZZZ probabilities within $9.5\times10^{-16}$ across the 15 points.
Eight tests include exhaustive feature covariance and conditional expectation
checks with two shots. In paired sanity checks with 256 replicates, the pooled
empirical/exact distance variance ratio has median $1.001$ and 5th to 95th
percentiles $[0.840,1.188]$; these describe correlated simulation cells,
not confidence intervals. The protocol, frozen original inputs, all 480
point/design/estimator/budget records, exact covariances, Jacobians and raw
paired samples are retained. All budgets are reported in that local
experiment bundle; no threshold based claim about required shots or savings
on hardware is made.

%% file: global_results.tex
\subsection{Stronger global normalization changes the interpretation}
\label{sec:globalresults}
A bounded follow-up applies signed Clifford extraction and fusion across
commuting intervening rotations to all 12 existing settings, without
changing constructors or readouts. Its input is the compiled IR; its
remaining Clifford frame is retained unless exactly identity up to
phase. In the two-layer ChebyshevRx examples, unitary-preserving
normalization alone reduces trainable Pauli-angle tasks from 4 to 3
and from 8 to 7. These reductions occur in \emph{both} the boundary and
transport arms (Table~\ref{tab:global}). They must not be credited to
the kernel certificate. A multi-qubit Pauli rotation is one angle task,
not necessarily one native physical gate.

\begin{table}[htbp]
\caption{Two-layer ChebyshevRx after global normalization.
Coefficient rank is the exact linear rank of the retained angle
coefficients in the displayed source-feature basis, not a proof of
minimal kernel dimension. B and T denote the earlier boundary and
transport arms.}
\label{tab:global}
\begin{tabular}{lrrrr}
& \multicolumn{2}{c}{2 qubits} & \multicolumn{2}{c}{4 qubits}\\
& B & T & B & T\\
\toprule
Symbols & 7 & 5 & 15 & 13\\
Angle tasks (old) & 4 & 4 & 8 & 8\\
Angle tasks (global) & 3 & 3 & 7 & 7\\
Coefficient rank & 6 & 4 & 14 & 12\\
\bottomrule
\end{tabular}
\end{table}

Across these 12 settings, no additional angle-task reduction appears
when comparing the globally normalized boundary and transport arms.
All source-to-normal-form feature, unitary, kernel and derivative
probes pass; Appendix~\ref{app:global} specifies the independent
rewrite checker and numerical scope. This is a new local structural
experiment, not a rerun of the Linux timing study. In particular,
the historical timing percentages below are \emph{not} speedups over
this stronger global-normalization baseline. The new representation
can absorb the representative internal explanation into standard
normalization followed by the readout-isometry test; exclusivity over
such a composition is not established.

The zero angle-task increment has a simple sufficient explanation.
For a fixed fused representation with angles
$\alpha_r(x,\theta)=u_r(x,\theta)+b_r(\theta)$, deleting certified
input-independent terms $b_r$ cannot remove a trainable-angle task
if $u_r$ still has a nonzero retained-parameter dependence. If this
holds for every trainable rotation, and gate grouping and measurement
bases are unchanged, the per-angle two-shift schedule keeps exactly
the same task count. This is an accounting statement for that schedule,
not a lower bound on all possible kernel evaluators. It applies to the
surviving product-angle terms in the two representative circuits.

%% file: appendices.tex
\section{Metric defect and finite perturbations}
\label{app:bound}
\begin{proposition}[Finite-angle Gram and ridge stability]
\label{prop:stability}
For a single common output Pauli rotation with closed readout, fix the
prefix, positive metric and RBF bandwidth. If $\partial_t f=Gf$, set
\begin{align}
 B&=G^TM+MG,\qquad
 b=e^{-1}\norm{M^{-1/2}BM^{-1/2}}_2,\\
 \eta&=\min\{1,b|t-s|\}.
\end{align}
Then each kernel entry changes by at most $\eta$, and on $N$ fixed
training inputs $\norm{K_t-K_s}_2\leq(N-1)\eta$.
For normalized ridge regression $\alpha_u=(K_u+N\lambda I)^{-1}y$,
$h_u(x)=k_u(x)^T\alpha_u$, and fixed $\lambda>0$,
\begin{equation}
 |h_t(x)-h_s(x)|\leq\frac{\norm{y}_2}{\sqrt N}\eta
 \left[\frac1\lambda+\frac{N-1}{N\lambda^2}\right].\label{eq:krr}
\end{equation}
\end{proposition}
The proof below integrates a uniform derivative
bound and applies the resolvent identity. It uses standard estimates.
The result compares fixed kernel fits, not independently evolving
optimization paths, and does not apply when the same symbol also
changes the prefix. Small regularization can make the bound loose.
The earlier 756 derivative-bound conditions and 48 finite-angle/Gram/KRR
checks exhibit no violations beyond their numerical tolerances; they
do not prove tightness or independent mathematical correctness.

The proposition concerns exact output transport only. It neither bounds
an approximate internal transport residual nor selects a pruning
threshold for the exact algorithm. Tightness at the runner's actual
regularization has not been established; no tolerance-based deletion
is authorized by this ancillary bound.

For a closed readout under one common output rotation,
$\partial_t f=Gf$. Set $d^2=\delta^TM\delta$ and
$B=G^TM+MG$. Then
\begin{align}
 \partial_t d^2&=\delta^TB\delta,\\
 |\partial_t d^2|&\leq\norm{M^{-1/2}BM^{-1/2}}_2d^2.
\end{align}
Differentiating $k=e^{-\gamma d^2}$ and using $ze^{-z}\leq1/e$
gives $|\partial_t k|\leq b$. For diagonal paired weights, the
whitened symmetric block has off-diagonal magnitude
$|w_i-w_j|/\sqrt{w_iw_j}$, so its operator norm is the largest
such quantity over rotating pairs. Integration on $[s,t]$ and
$0\leq k\leq1$ give the entry bound $\eta$ in
Proposition~\ref{prop:stability}.

The Gram difference is symmetric with zero diagonal and row sums in
absolute value at most $(N-1)\eta$. Its spectral norm obeys the
same bound. Positive semidefiniteness gives
\begin{equation}
 \norm{(K_u+N\lambda I)^{-1}}_2\leq(N\lambda)^{-1}.
\end{equation}
The resolvent identity bounds the inverse difference by
$(N-1)\eta/(N\lambda)^2$. Splitting the prediction difference
between the changed test vector and the changed inverse, and using
$\norm{k_t(x)-k_s(x)}_2\leq\sqrt N\eta$ and
$\norm{k_s(x)}_2\leq\sqrt N$, proves Eq.~\eqref{eq:krr}.

For several coordinates, one may add segment bounds along a specified
coordinate path only if the required uniform output-family assumptions
hold on each segment. This does not bound the divergence of two
independently trained parameter paths. The normalized ridge in
Proposition~\ref{prop:stability} corresponds to empirical squared loss
divided by $N$ plus $\lambda$ times squared RKHS norm. The numerical
training runner's $K+0.1I$ uses the unnormalized convention, equivalent
to $\lambda=0.1/N$; those conventions must not be interchanged.

\section{Evidence provenance and negative results}
\label{app:evidence}
The certificate validation and source-library audit are distinct from
the later cost studies. The certificate archive reports 360 designed
instances, 756 metric-defect conditions, and 20 paired synthetic
training runs. The library-coverage v1.2 archive contains the 34
executable settings and their 4,590 stored arrays. Reproduction checks
used original-source features, independent dense simulator tests and
finite differences where specified; they do not constitute a machine
checked proof or independent investigator replication.

The structural analysis adds the elementary local-tail
baseline and tracks gate provenance to distinguish same-native affine
terms from genuinely later data gates. The four crossing symbols are
$\theta_2,\theta_3$ in the two-qubit, two-layer ChebyshevRx case and
$\theta_6,\theta_7$ in its four-qubit counterpart. The remaining
certified boundary symbol in each case is not counted as a genuinely
internal data crossing. The representative numerical witness uses
seven inputs and deletes only the selected early occurrence.

\subsection{Earlier runtime evidence is not the new endpoint}
An earlier purpose-selected ChebyshevPQC transfer study used WDBC,
Ionosphere and Auto MPG with five seeds and three timing repetitions.
It observed a median paired time reduction of 23.00\% in its serial
per-parameter implementation, with gradient calls decreasing from
640 to 560. Later two-dataset, three-seed, three-repetition workloads
observed 26.90\% and 15.49\% reductions for raw-plus-metric versus
raw-plus-commutant batched and adjoint implementations, respectively.
Batched calls remained 60 while derivative entries decreased from
1,020 to 840; adjoint reverse passes remained 60 while reverse gate
updates decreased from 84,480 to 63,360.

These studies already included preparation and certification inside
the total timer, but their compiled-metric route certified \emph{before}
compilation. Their globally inactive ChebyshevPQC symbols are found
by elementary local tails as well. Thus none of these percentages is
the incremental benefit of the internal rule over the new fused
boundary baseline. Deleted shared occurrences can still differ, but
their incremental effect requires a separate measurement. The earlier
source-derived YZ-CX structure grid, with 72 runs, 6,624 kernel requests
and 3,240 cross-validation fits, showed zero additional execution
saving. It is retained rather than reinterpreted as a speedup.

\subsection{Image anomaly detection: task, protocol, and negative results}
\label{app:visionad}
\paragraph{Task and purpose.}
Image anomaly detection assigns a higher score to a defective or
otherwise anomalous image than to a normal image. Since the anomaly
score uses query-to-normal-support kernel similarities directly,
this task lets us examine whether changing a kernel by parameter
optimization improves a downstream ranking.
It evaluates image-level detection, not pixel-level localization.
It does not apply our rank recovery interventions, change the readout
or metric during training, or test exact parameter removal.
Thus it is a performance control, not evidence for or against the
predictive benefit of restoring an identified null direction.

\paragraph{Data and preprocessing.}
We use 46 categories from five publicly available industrial visual
anomaly-detection datasets: MVTec AD (15 categories)
\cite{mvtecAD2019}, MVTec AD 2 (8) \cite{mvtecAD22026},
MVTec LOCO AD (5) \cite{mvtecLOCO2022}, MPDD (6)
\cite{mpdd2021}, and VisA (12) \cite{visa2022}.
MVTec AD 2 uses its publicly labeled test partition, not the private
evaluation server. For each category and each of five seeds
($42,123,777,999,2024$), we draw four normal support images.
For the MVTec datasets and MPDD, support comes from the normal training
pool and query images from the test pool. For VisA, support is drawn
from the normal-image pool and queries comprise the remaining normal
images and all anomalous images; this is our episodic split, not a
claim to use an official VisA benchmark split. A frozen ImageNet-1K
pretrained ViT-B/16 produces 768-dimensional, per-image normalized
features from images resized to $224\times224$. A seeded Gaussian
projection reduces these to eight coordinates. Min--max scaling is
fitted to support images only, then applied with clipping to $[0,\pi]$
to define the input rotation angles. No backbone fine-tuning is used.

\paragraph{Evaluation protocol and comparison.}
Within each of the 230 category--seed episodes, 15 query images are
selected without consulting their labels for an unlabeled calibration
set $\mathcal C$; all remaining queries form the held-out evaluation
set. Calibration can therefore contain anomalous images even though
their labels are not used. This is \emph{normal support with unlabeled
target-pool calibration}, not purely normal-only training.
Sample paths and byte hashes are checked for overlap between
support, calibration, and evaluation. The primary arm is an
eight-qubit, two-layer entangling PQK with 48 trainable angles and
24 local XYZ expectation values. Its Gaussian kernel is
$k_\theta(x,x')=\exp[-\|f_\theta(x)-f_\theta(x')\|^2/24]$.
We compare this \emph{same model at initialization and after 30 Adam
updates} (learning rate $0.1$), with all other episode settings fixed.
The primary loss, with normal support $S=\{s_1,\ldots,s_4\}$, is
\begin{equation}
 \mathcal L(\theta)
 =-\frac{1}{6}\sum_{i<j}k_\theta(s_i,s_j)
 -\frac{1}{2}\operatorname{Var}_{c\in\mathcal C}
   \left[\frac{1}{4}\sum_{i=1}^{4}k_\theta(c,s_i)\right],
 \label{eq:visionloss}
\end{equation}
where the variance is the population variance over calibration images.
This exploratory label-free objective favors support similarity and
variation in calibration-to-support similarity. Increasing that
variance does not guarantee separation of anomalies from normal
images: it can also amplify normal appearance variation or move
arbitrary calibration samples away from support.
The label ``PQK/RBF'' denotes a
Gaussian function of quantum features in this arm, not a comparison
with a separate classical RBF model. The full 18-arm study crosses
three models (entangling PQK, no-entanglement control, and a
24-feature/48-parameter classical control), two training similarities
(dot product or Gaussian), and three variance weights
($0,0.5,1$). All arms use Gaussian similarity at evaluation.

The anomaly score is $a_\theta(q)=1-\frac14\sum_i k_\theta(q,s_i)$,
with normal label $0$ and anomaly label $1$. We measure the area
under the receiver operating characteristic curve (AUROC), using the
fixed final iterate without evaluation-label-based tuning of the
checkpoint or primary arm. We average seeds within each category
before taking the category macro-average. Five seeds are repeated
episodes, not five independent datasets. Intervals use 10,000 category
bootstrap resamples and are exploratory because categories within a
dataset may remain dependent.

\paragraph{Results and matched-cohort split comparison.}
Table~\ref{tab:visionprotocol} summarizes the primary entangling PQK
before and after optimization. Across all 46 categories, mean AUROC
decreases by $4.90$ percentage points. Restricting the same archived
image-level-split results to the eight MVTec AD 2 categories instead
gives a change of $+0.36$ points. Thus the overall negative change
must not be treated as the image-level baseline for this eight-category
cohort. These are optimization outcomes within the specified
pipeline, not effects of certified parameter removal.

\begin{table}[tbp]
\centering
\caption{Image anomaly-detection results for the primary entangling
PQK with Gaussian similarity and variance weight $0.5$. AUROC is in
percent; changes and their exploratory 95\% intervals are in
percentage points. Each row averages five seeds within category and
then categories equally. Intervals resample category-level paired
training changes 10,000 times. The second row is a subset of the first,
not an additional experiment. Only the last two rows have the same
category--seed cohort; their calibration and evaluation images differ.}
\label{tab:visionprotocol}
\begin{tabular}{@{}llrrrr@{}}
\toprule
Cohort (categories) & Query split & Episodes & Initial & Trained & Change [95\% interval]\\
\midrule
All five datasets (46) & Image-level & 230 & 51.22 & 46.32 & $-4.90\;[-7.62,-2.51]$\\
MVTec AD 2 only (8) & Image-level & 40 & 49.01 & 49.37 & $+0.36\;[-5.29,5.52]$\\
MVTec AD 2 only (8) & Filename-group & 40 & 48.10 & 49.38 & $+1.28\;[-5.57,7.88]$\\
\bottomrule
\end{tabular}
\end{table}

A follow-up in the eight MVTec AD 2 categories examines the possibility
that related views or lighting variants of the same putative object
are split between calibration and evaluation. The image-level split
can place files sharing a numeric filename stem but differing in
view/lighting suffix on both sides. We therefore group such files
within their parent good/bad directory, assign entire groups to one
side, and retain exactly 15 calibration images; unused images from
selected calibration groups are excluded from evaluation.
The backbone features, support, projection, scaling, model arms,
optimizer, and five seeds are unchanged, but optimization is repeated
using the new calibration set. This follow-up was designed after
observing the original results. On the matched eight-category cohort,
the grouped split gives a training change of $+1.28$ points versus
$+0.36$ points for the image-level split. Their descriptive difference
is $+0.92$ points, with exploratory category-paired bootstrap interval
$[-1.21,3.33]$. This comparison does not establish a systematic
benefit from grouping. Matching the cohort removes the category-set
mismatch, but calibration and evaluation membership still change
together; it does not isolate a causal effect of leakage correction.
Filename groups also remain proxies, not verified physical-object
identities.

\paragraph{Interpretation and evidence scope.}
Neither analysis establishes predictive improvement from parameter
optimization. The main entangling architecture has no newly certified
global inactivity in the audited setting, so the negative result
neither falsifies exact kernel preservation nor attributes failed
optimization to inactive parameters. The reduction from 768 to eight
input coordinates, scaling estimated from four support images,
clipping, the circuit, and the training objective can each influence
the outcome. This study does not isolate these factors, and its
results concern this fixed pipeline, not a general limitation of
quantum kernels or of identifiability recovery. The matched-cohort
table is a reaggregation of archived experiments: AUROC is recomputed
from all 270 saved primary-arm score arrays (230 original and 40
grouped episodes), with category--seed identities matched for the
eight-category comparison. The wider evidence audit checks stored
inputs and model evaluations. Neither operation is an independent
rerun of original-image feature extraction or optimization trajectories.

\subsection{Frozen incremental experiment}
The final local pilot uses the design named
\texttt{20261001\_boundary\_fused\_v3}. The seven arms are original
split IR, compiled split IR, compiled fused execution, local-tail
reduction, boundary reduction, boundary plus extra transport, and
observable-commutant reduction. Adjacent same-axis fusion is common
to all strong arms. Initial parameters and train/test data match
within every comparison. The fixed kernel bandwidth and optimization
schedule are not selected with test labels.

The completed full protocol has 2,700 cold paired blocks across seven
arms, or 18,900 fits. Here a cold block fixes circuit setting, dataset,
split seed, timing repetition and backend and executes all seven arms.
Reuse instead has \emph{only two arms}, boundary and transport.
Its 90 blocks are two ChebyshevRx settings times three datasets times
one seed (42) times five repetitions times three backends. Each block
executes 20 actual fits in each of its two arms, so
$90\times2\times20=3{,}600$. The horizons 1, 2, 5, 10 and 20 are
cumulative checkpoints within those 20 fits, not separate restarts. Cold timing
includes compilation, representation conversion, simplification,
certification, input encoding, initialization, learning and prediction.
Frozen-input loading, untimed numerical validation, warmup, disk I/O
and audits are excluded. No numerical kernel/gradient cache is used.

Every completed block checks stored kernels, derivatives, objectives,
parameters and predictions against its full reference. The additional
scalar-source replay covers repetition-zero serial cold references
for every task and setting, and repetition-zero serial reuse fits.
It checks saved update increments; it is not an autonomous retraining
experiment. Timing arithmetic is audited, not independently repeated.
The reproducibility README identifies the full archive and its hash.
Its 5,306 manifest-listed files and 22,500 array references were checked;
lossless identical matrices can share a stored file. Sixty-one compared
source and frozen-input files match the preserved release byte for byte.
The archive's selective replay covers 180 cold references and 120 reuse
fits. Its maximum replay kernel and gradient errors are
$2.44\times10^{-15}$ and $6.96\times10^{-13}$; these are not the
maximum differences between separately evolved stored training paths.
The latter maxima are $4.65\times10^{-11}$ for kernels,
$2.75\times10^{-10}$ for gradients, $2.69\times10^{-11}$ for
parameters, $8.31\times10^{-12}$ for objectives and
$2.22\times10^{-11}$ for predictions. The respective frozen limits
are $10^{-10},10^{-9},10^{-9},10^{-10},10^{-7}$.

The server uses Python 3.8.10, NumPy 1.24.4, Qiskit 1.2.4 and Linux
x86\_64, with BLAS/OMP thread counts set to one. The archive does not
identify the exact CPU model. Each fit uses 64 training and 64 test
inputs, ten alignment-ascent steps of size 0.1, RBF bandwidth 0.5,
and ridge $K+0.1I$. This is a fixed workload, not time to convergence.
The separate earlier pilot used four updates; its single split and
repetition do not support an independent runtime estimate.

\subsection{Repeated full timing and reuse}
\label{app:full_time}
For each matched pair the reduction is $100(1-T_{\rm full}/T_{\rm boundary})$.
Over ten updates, serial evaluator calls fall from 70 to 50 at two
qubits and 150 to 130 at four. Batched calls remain ten, while tangent
gate updates fall from 26,880 to 19,200 and from 134,400 to 116,480.
Adjoint reverse passes remain ten and reverse gate updates remain
3,840 and 8,960. Derivative-entry counts are not reverse-work counts.
Median boundary-adjoint times are 19.54 and 32.50 ms, compared with
50.80 and 187.11 ms for transport-serial. Backend selection therefore
matters more than the serial savings in these examples. The cold-fit
repetitions concern three datasets; reuse below uses only split seed 42.

Table~\ref{tab:full_time} gives both paired medians and reductions in
the sum of elapsed times. No observations are removed. The two summaries
need not agree: two-qubit batched Auto MPG has a positive median but
a $-3.33\%$ summed-time reduction. Its most negative pair is
$-146.71\%$. Adjoint extrema range from $-4.97\%$ to $64.04\%$
at two qubits, and from $-135.05\%$ to $8.23\%$ at four qubits.
The logs do not establish the cause of these timing excursions.

\begin{table}[t]
\caption{Full cold-fit time reductions in percent. IQR is the
25th--75th percentile range, not a confidence interval. Each row
summarizes 75 matched pairs for the two-layer ChebyshevRx settings.
All pairs, including negative reductions, are retained; Positive
denotes the number of pairs with a positive time reduction.}
\label{tab:full_time}
\begin{tabular}{llrrrr}
\toprule
$n$ & Backend & Median & IQR & Positive & Sum\\
2 & Serial & 22.20 & [20.93,23.12] & 75 & 22.11\\
2 & Batched & 2.70 & [1.73,4.82] & 65 & 1.15\\
2 & Adjoint & 0.61 & [$-0.36$,2.81] & 50 & 3.87\\
4 & Serial & 12.38 & [11.44,13.30] & 75 & 13.16\\
4 & Batched & 4.40 & [1.71,7.01] & 66 & 4.53\\
4 & Adjoint & $-0.18$ & [$-2.18$,1.26] & 34 & $-1.90$\\
\bottomrule
\end{tabular}
\end{table}

The serial dataset-specific medians (WDBC, Ionosphere, Auto MPG) are
22.34\%, 21.85\%, 22.20\% at two qubits and 12.36\%, 12.38\%,
12.46\% at four qubits. Both relative execution-order strata have
positive serial medians. Batched medians differ by order, from
2.42\% to 3.65\% at two qubits and from 2.88\% to 5.22\% at four
qubits. Future estimates of small effects should use explicitly
balanced pair ordering and monitor time-dependent load.

An exploratory bootstrap resamples five seed blocks within each of
the three fixed datasets, retaining all five timing repetitions in a
selected block. Across 4,000 resamples, the 2.5th--97.5th percentile
ranges of the pooled serial median are [21.65,22.34]\% and
[12.09,12.57]\%. Repeated splits overlap. These conditional sensitivity
ranges are not confidence intervals for new datasets or circuit families.

Median extra transport-certificate times are approximately 0.15 and
0.40 ms and are already inside the total timers. At reuse horizon 20,
the serial median reductions are 26.25\% and 13.32\% (15/15 positive
blocks at both sizes), and batched gives 5.70\% and 6.90\% (14/15
and 15/15). Both serial sizes already have 15/15 positive blocks at
horizon 1. Two-qubit batched retains a negative block even at horizon
20. Adjoint reuse medians are 2.57\% and 1.24\%, without reduced
reverse-gate work. These seed-42 reuse checkpoints share their fits
and are not independent observations or guaranteed break-even bounds.

\subsection{Single-pair pilot timing observations}
\label{app:pilot_time}
The corresponding single-pair cold times are 15.25/12.78 ms and
44.81/39.55 ms for serial boundary/transport, 7.28/7.02 ms and
14.04/13.65 ms for batched, and 7.23/7.21 ms and 11.34/11.28 ms for
adjoint. The extra certificate stage takes approximately 0.073--0.075 ms
and 0.214--0.218 ms, respectively; it is already included in the totals.
One pair cannot distinguish stable runtime benefit from timing noise.
No confidence interval is constructed by treating the three backends
or the two circuit sizes as independent repeated measurements.

These pilot observations are descriptive and are superseded for runtime
estimation by the separate full study above. The adjoint differences are not evidence of
reduced reverse propagation. Repetitions on the same family would not
by themselves establish cross-family scope or quantum execution savings.

\section{Fixed-scope resource survey}
\label{app:survey}
The source inventory was fixed before this follow-up execution, after
inspection of the earlier pilot. It is exploratory, not preregistered.
The ten modules use their default constructors, including seed zero
for random circuits, with two input features and explicit qubit/layer
arguments where supported. Constructor signatures, supplied arguments
and resulting parameters are recorded. ChebyshevTower, Qiskit circuit
wrapper and QCNN builder remain excluded for the original unselected
mandatory-configuration reasons. No favorable replacement circuit is
constructed for them. Eighteen layer-argument duplicates remain in
the 120-slot record but are not rerun.

The baseline suffix rule deletes a parameter-only readout isometry if
the whole remaining suffix consists of input-independent readout
isometries. A second sufficient case allows disjoint suffix gates when
every overlapping suffix gate is an input-independent isometry confined
to the candidate support. The confinement condition is essential:
an entangler could otherwise spread a generator into a subsequent,
initially disjoint data gate. Rotations use the exact Pauli criterion;
fixed Hadamard/CNOT actions use exact signed-permutation closure and
weight matching. Ordinary restricted boundary transport and fusion
follow this elementary rule. No completeness claim is made for it.

The shared-angle simulator evaluates shifted expectation features,
forms their angle derivatives, and accumulates all classical chain
factors for every shared symbol occurrence. Validation uses two fixed
probe seeds, five inputs per seed and nonzero parameter assignments.
Qiskit source features are checked at these base points; central finite
differences are computed from the compiled interpreter, not by rerunning
every perturbed source circuit. There are 1,626 parameter-coordinate
finite-difference probes across combinations of circuit, readout, and seed.
Maximum feature-adapter and feature Jacobian differences are
$3.89\times10^{-15}$ and $1.78\times10^{-15}$.
Kernel Jacobian and objective-gradient finite-difference differences
are at most $7.01\times10^{-10}$ and $9.38\times10^{-9}$,
below the fixed $2\times10^{-7}$ finite-difference tolerance.
The two-update trajectory and ridge-prediction comparisons give maxima
$4.45\times10^{-16}$ and $1.67\times10^{-15}$ where applicable.
These are finite numerical checks, not a new theorem or hardware test.

An initial run raised degenerate-centered-alignment exceptions for
ParamZFeatureMap with local Z readout at both qubit sizes. The failed
run is retained. A disclosed handling amendment adds feature and kernel
Jacobian checks and classifies those two slots separately, without
running undefined objective-gradient training. No circuit, readout,
seed, baseline or favorable-outcome threshold was changed in response.
The completed run contains 82 trainable alignment-probe slots, 18
parameter-free slots and two kernel-only slots, with no failure of
an applicable numerical check. Neither repeated probes nor additional
readouts are treated as independent positive circuit families.

Four slots remove two extra symbols each: two- and four-qubit,
two-layer ChebyshevRx with local XYZ or weight-at-most-two Pauli
readout. The same source symbols appear under the two readouts, so
eight slot-counted symbols are not eight distinct source discoveries.
All 102 distinct slots have zero incremental fused-operation and
trainable-angle reduction. Consequently this survey does not trigger
a new large timing campaign. Its local macOS run is separate from the
Linux full-study timings; server replay is provided as executable code,
and a subsequently returned Linux survey has also been audited.
The server archive contains 308 manifest-listed files and 3,252 array
references. Replays cover 204 source-feature probe batches, 612 arm
evaluations and 1,626 finite-difference coordinates. All applicable
checks pass. Six HighDim two-layer slots have different fixed-angle
coefficients in their saved compiled IR from the local survey, although
their operation structure matches and numerical replays pass. They
must not be described as byte-identical compiled circuits.

For the two primary local XYZ ChebyshevRx settings, direct comparison
of the saved full-timing and survey archives finds identical final
programs, retained symbol lists, input-expression lists, observables
and weights for both boundary and transport. The definitions of the
baseline procedures remain different; their final executable objects
coincide in these particular cases. This check resolves the structural
consistency question without repeating the 18,900 fits. It does not
resolve timing variability or validate any timing claim for the new
global-normalization arm.

\section{Source identity and preprocessing}
\label{app:source}
{\raggedright
The external circuit definitions are frozen sQUlearn source bytes
\cite{squlearn}, pinned at commit
\texttt{0f6cb0982fdbe80dcd0ea6cb0acd68c59e362de4}.
The manifest includes per-file hashes and source URLs. The lightweight
host does not install or exercise the complete sQUlearn runtime;
it uses explicit dependency stubs and deferred annotations for two
pinned modules while retaining their original bytes. Execution and
symbolic compilation use Qiskit 1.2.4. The ten executable definitions are
ChebyshevPQC, ChebyshevRx, HighDimEncodingCircuit, HubregtsenEncodingCircuit,
KyriienkoEncodingCircuit, MultiControlEncodingCircuit, ParamZFeatureMap,
RandomEncodingCircuit, RandomLayeredEncodingCircuit, and YZ\_CX\_EncodingCircuit.
Exact module paths and constructor arguments are in the packaged inventory.
These are ten modules, not ten empirically independent positive families.\par}

The timing study reuses frozen inputs from the earlier audited transfer
run. Each split uses 64 training and 64 test rows; classification splits
are stratified. Median imputation, standardization, and two-component
full-SVD PCA are fitted on the training rows only. PCA scores $z$ are
mapped to $x=\tanh(z/2)$ before any $\arccos(x)$ encoding. Test rows
use the saved training transforms. Regression targets are standardized
using the training mean and population standard deviation; classification
targets use their signed labels. Alignment uses centered $K$ and
$(y-\bar y\mathbf1)(y-\bar y\mathbf1)^T$, normalized by their Frobenius
norms. The label mean in training objectives uses training rows only.
The simulated arithmetic is float64/complex128. Frozen-input metadata
stores split indices, transform parameters and source hashes. Missing
entity-group identifiers are not silently replaced by row independence.

\section{Global normalization audit and checker}
\label{app:global}
The follow-up fixes the earlier 12 source configurations and local XYZ
readout. It is a fixed-inventory diagnostic, not a new population survey.
For each configuration, 20 random five-input probes compare source
features, unitary matrices up to global phase, kernels and gradients.
One probe per configuration additionally checks every source-coordinate
kernel derivative by central finite differences with step $10^{-6}$.
Across the configurations, maximum unitary, feature, kernel and
objective-gradient errors are $6.48\times10^{-16}$,
$1.67\times10^{-15}$, $1.23\times10^{-15}$ and
$2.96\times10^{-14}$, respectively. The finite-difference error is
at most $3.84\times10^{-10}$, below the $10^{-6}$ audit limit;
all other errors are required to be below $10^{-9}$.
The separate 1,000-condition closed-form check and 20-step optimizer
diagnostic are reported in Sec.~\ref{sec:normalform}.

The normalizer records each signed Clifford conjugation, its current
frame, and any commuting-fusion target. A separate standard-library-only
checker uses explicit Pauli multiplication and H/CNOT rules rather than
the normalizer's Qiskit tableau operations. It validates each recorded
merge, the identity status of the final frame, the readout partner/weight
conditions for recorded common-term deletions, and the retained-symbol
dependencies. It checks all 12 configurations; unit tests reject a
corrupted sign and altered output program. Ninety random mixed Pauli/H/CNOT
circuits additionally test signed normalization against dense matrices.
This is a checker of the supplied IR rewrites, not a formal verification
of the source adapter, the mathematical theorem, or optimizer software.
No tolerance is used to accept a symbolic rewrite or metric equality.

The global scan permits fusion only across commuting rotations, does
not absorb nonidentity Clifford frames into the readout, and does not
resynthesize Pauli rotations to a hardware basis. Hence its task counts
are not native-gate counts. The coefficient-rank diagnostic combines
identical source feature expressions exactly; it is not a complete
test of general functional independence or reachable-state kernel rank.
This normalization alone establishes neither completeness beyond its
rules nor global minimality. The separate representative rank theorem
in Sec.~\ref{sec:dimensiontheory} supplies a matching minimal smooth
dimension for that specific kernel. Large-qubit certificate time/memory scaling and
a controlled multi-session timing comparison of these new arms remain
separate experiments, not consequences of the small diagnostic.

\section{Completeness and scope checklist}
The universal output criterion quantifies over all density operators;
the internal algorithm checks only its supported gate grammar and
transport rules. It is neither necessary for inactivity on a particular
input manifold nor complete for circuits with cancellations outside
that grammar. A successful certificate is sufficient; a failure to
certify conveys no activity conclusion.

Independent review must examine exact metric equalities, signed
Clifford conjugation conventions, shared symbols, source-to-IR
semantics, and the deterministic optimizer contract. Sign-free
transport is used only to decide deletion and never to evaluate a
retained transformed rotation. Physical noise, finite-shot bias and
variance, data-dependent metrics and continuation-safe search remain
outside the present theorem and implementation.

%% file: dimension_appendix.tex
\section{Dimension certificates and feasible witness checks}
\label{app:dimension}
For the representative minor in Eq.~\eqref{eq:rankminor}, define
$B=J\diag(\pi^{-1},\pi^{-1},\pi^{-1},1)$. Its entries lie in
$\mathbb Q(\sqrt3)$:
\begin{equation}
 B=\begin{pmatrix}
 0&0&\sqrt3/12&-\sqrt3/2\\
 \sqrt3/24&\sqrt3/24&\sqrt3/12&0\\
 -\sqrt3/4&-1/2&-11/24+\sqrt3/2&-1/2\\
 -\sqrt3/4&-1/4&11/24+\sqrt3/2&1/2
 \end{pmatrix}.
\end{equation}
Exact expansion gives $\det B=(2\sqrt3-39)/1536$. Since $\sqrt3<2$,
the determinant is strictly negative without a numerical rank threshold.
This explicit certificate supplements rather than relies on a
finite-difference SVD or symbolic zero heuristics for transcendental
expressions.

The 24-setting pilot consumes frozen compiler IRs from the earlier
survey, validates integer coefficients and supported data expressions,
and does not snap approximate constants to multiples of $\pi$.
It supports repeated rational multiples of $x_j$ or $\arccos x_j$,
with the same source variable used at every occurrence. Fixed
initialization and metric are part of each input hash.
Heisenberg propagation produces finite sums of
Gaussian-rational coefficients times exponentials of polynomial phases.
The distance includes both self-square terms and the cross-input term.
Grouping identical phases and differentiating with all shared
occurrences preserves the source semantics. Exactly vanishing formal
coefficients suffice for an identity; no converse or general
polynomial-phase independence theorem is used.

Real and imaginary rational coefficient rows form a matrix whose exact
null space supplies constant invariant directions. Independent rows
give a certified upper bound $K-\dim\mathcal N$ on differential rank.
For the lower bound, one parameter vector with rational entries of
denominator 17 and 24 pairs of input vectors with denominator 19 are
fixed per setting using seed 20261002. All input coordinates are in
$(-1,1)$. A separate statevector engine propagates source-parameter
tangents, including every shared occurrence. It evaluates sine,
cosine, square roots and arccos via interval operations; arccos is
implemented as $\operatorname{atan2}(\sqrt{1-x^2},x)$ on this interior.
The implementation uses mpmath interval arithmetic at 50 decimal
digits. A replay at 65 digits confirms all 132 activity witnesses and
22 nonzero minors. Interval Gaussian elimination divides only by
pivots whose intervals exclude zero. Floating-point elimination is
used solely to choose candidate rows and columns.

Each certificate stores the source hash, the common parameter point,
actual input pairs, derivative endpoint rationals, selected minor, and
exact null vectors. The symbolic and interval engines use separate
propagation implementations. Sixteen unit tests include invalid
source hashes, infeasible independent-angle witnesses, unsupported
near-Clifford constants, data sharing and domain violations.
The implementation handles an mpmath 1.3.0 interval-complex conjugation
issue by explicitly constructing conjugates from real and imaginary
intervals. The arithmetic library and these implementations are not
proof-assistant verified; higher precision replay does not replace
independent software or mathematical review.

A further numerical diagnostic reconstructs the pinned external
source circuits rather than just the IR. Across 48 input-pair probes,
maximum feature and distance discrepancies are
$1.56\times10^{-15}$ and $3.20\times10^{-14}$; 324 source-coordinate
finite differences agree to $2.38\times10^{-9}$, within the declared
$10^{-6}$ diagnostic limit. These are consistency checks, not the
acceptance rule for an exact identity or a nonzero interval witness.
The local run took 82.65 seconds in its recorded environment; this is
certificate-generation cost, not a training-speedup benchmark.

The pilot's 12 new inactive entries have elementary explanations.
Write $z_q=x_{q\bmod2}$ for the source's cyclic assignment of two input
features. For one-layer ParamZ, each qubit has $R_z(\theta_qz_q)H|0\rangle$,
so its Z expectation is zero. For two layers it has
$R_z(\theta_{q+n}z_q)H R_z(\theta_qz_q)H|0\rangle$,
whose Z expectation is $\cos(\theta_qz_q)$; only the final rotation is
irrelevant to Z readout. Thus the two-layer kernel is not generally
constant. These facts also show why the new entries should not be
credited as advantages over a baseline that already propagates
readout commutants through data-dependent rotations.

No randomized polynomial-identity tests, unrestricted finite-grid
fallback or 102-setting completeness claim is used here. More
general source constants, dependent data expressions outside the
implemented grammar, and HighDim's differing fixed-angle compiler
representations remain outside this pilot's exact semantic contract.

%% file: extension_appendix.tex
\section{Extension protocol and nonlinear certificate}
\label{app:extension}
The 30-setting extension uses exact integer or half-integer IR
coefficients and the frozen source data expressions. Its Fourier
calculation is limited to 20 seconds per setting, two million
pre-convolution term pairs per observable, and 50,000 accumulated
distance terms. Limits on time are machine-dependent; saved evidence
replay is not the same test as repeating a budget-limited discovery run.
The first execution saved 21 settings before a timer/exception-handling
race stopped it. Disarming the timer before handling the budget
exception allowed resumption without changing those 21 records, the
scope, or witness points. Both code snapshots and retained-file hashes
are archived.

Using seed 20261003, source parameters are integers from 2 through 24
divided by 17; input coordinates are integers from $-15$ through 15
divided by 19. Each setting evaluates at most 24 input pairs at one
shared parameter point with 50-digit interval arithmetic. Evaluation
can stop after at least two pairs once the original upper/lower bounds
match and all coordinates have a verdict. Structural closure is a
separate post-run analysis, not a replacement of timed-out Fourier
records. The inherited experimental-arithmetic qualifications of
Appendix~\ref{app:dimension} apply unchanged.

All 195 activity witnesses and 27 nonzero minors replay at 65-digit
precision; the three zero-rank cases use exact structural identities.
Numerical source/IR checks cover 60 input pairs and 504 finite-difference
coordinates. Across 600 random resource-arm probes, maximum kernel
and squared-distance-gradient discrepancies are respectively
$1.23\times10^{-15}$ and $1.29\times10^{-14}$. The source finite-
difference check differs from the interval derivative midpoints by
at most $5.32\times10^{-9}$. These checks do not constitute a formal
proof of the entire source compiler or a physical-device experiment.

For the nonlinear Hubregtsen example, exact Fourier coefficients for
each of the four compiled features match Eq.~\eqref{eq:hubfeatures}.
The controlled rotations commute with all measured Z operators, so
the proof is independent of their four shared circuit assignments.
The factorization follows by grouping the repeated input coordinate
in the squared distance, not by promoting a numerical rank estimate
to an identity. In 1,000 separately sampled original-source input
pairs, the largest kernel discrepancy against the closed form is
$6.67\times10^{-16}$; at 100 parameter points the largest derivative
discrepancy is $7.50\times10^{-16}$. These tests check implementation
consistency, whereas Eqs.~\eqref{eq:hubfeatures}--\eqref{eq:hubminor}
supply the mathematical upper and lower bounds.

\subsection{Returned server audit}
The separately returned server archive contains five audit outputs:
evidence replay, original-source comparison, structural closure,
the nonlinear certificate, and its source-level numerical check.
All report success. Parsed replay, closure and exact-certificate
records agree identically with the local outputs; only numerical
error diagnostics differ in the two source checks. The run-manifest
hash matches all 74 retained files, the four reported script hashes
match the distributed scripts, and the nonlinear source check links
to the exact returned proof file by hash.

The server replay validates 195 activity witnesses and 27 nonzero
minors across 30 settings at 65-digit precision. Source comparisons
again cover 504 finite-difference coordinates and 600 resource-arm
probes. Maximum kernel and distance-gradient discrepancies are
$1.34\times10^{-15}$ and $1.26\times10^{-14}$; the finite-difference
comparison is below $4.13\times10^{-9}$. In the nonlinear check,
1,000 input pairs and 100 gradient points give maximum kernel and
gradient discrepancies below $5.56\times10^{-16}$ and
$7.50\times10^{-16}$. All satisfy the unchanged validation thresholds.

These are reexecutions of saved-evidence checks, not independent
new datasets, a second timed discovery run, or new quantum-device
measurements. The archive does not record the server's package and
operating-system versions, so full environment provenance remains
incomplete. Structural closure still gives 195 active and 57 inactive
coordinates; adding the nonlinear result closes 27 of 30 rank
intervals, leaving the same three Kyriienko cases unresolved.
No incremental angle-task or gate-task saving is introduced by this
reproduction.

%% file: legacy_controls.tex
\section{Archived learning and candidate-reuse controls}
\label{app:legacy_controls}
These controls come from archived earlier experiments.
We reaggregate the saved per-trial records and inspect their source
protocols; no new training or QPU execution is claimed here. Three
CSVs and their hashes accompany the reproduction script. These results
are separate from the compiler benchmarks.

\subsection{Matched classical feature learning}
The supervised control uses five qubits, ten trainable PQK parameters
and five local Z expectations. The classical map has five features
$z_j(x)=\cos(\alpha_j r_j^Tx+\beta_j)$, with fixed unit-norm
projections $r_j$ and ten trainable scales and biases. Both maps use
the same Gaussian kernel with $\gamma=1$, kernel-target alignment,
50 finite-difference optimization steps, learning rate 0.3,
minibatch size 16, and a precomputed-kernel SVM.

The datasets are one fixed 80-point two-moons realization (noise 0.15,
generation seed 3), and Wine classes 0 and 1 (130 points, first five
features). For each of ten paired randomized trials, the 60/40
stratified split precedes scaling. Scaling is fit on training inputs
alone; test inputs are transformed and clipped to $[0,\pi]$.
Trials vary split, initialization and minibatch randomness; they are
not independent newly sampled datasets. The models share the training
budget and split protocol, not identical initial feature maps.

\begin{table}[tb]
\caption{Archived matched feature-learning control. Accuracy is mean
$\pm$ sample SD over ten trials, in percent; KTA is the mean before
and after training. These are descriptive small-benchmark results,
not evidence of quantum advantage or a statistical equivalence claim.}
\label{tab:legacy_classical}
\begin{tabular}{llcc}
\toprule
Data & Map & KTA before/after & Accuracy\\
Moons & Classical & .2892/.3974 & $90.63\pm6.75$\\
      & PQK       & .3076/.3335 & $91.25\pm5.47$\\
Wine  & Classical & .2212/.3497 & $91.35\pm3.65$\\
      & PQK       & .3175/.4076 & $93.46\pm2.60$\\
\bottomrule
\end{tabular}
\end{table}

\subsection{Matched embedding placement}
The second control changes only the order of the same gate multiset.
Let D denote a data block and P a parameter block. Chronologically,
the Type II circuit is D-D-P-P-D and the interleaved Type III circuit
is D-P-D-P-D. Both use five qubits, three identical data blocks,
two identical parameter blocks and 20 trainable parameters. Logical
block depth is matched before compilation; equal physical compiled
cost is not assumed. These are fidelity kernels, not the preceding
projected kernels.

The datasets and train-only scaling follow the preceding protocol.
The ten trials pair initial parameters, split and minibatch order.
Training uses 30 KTA steps, learning rate 0.2, batch size 16 and
central finite differences of step $10^{-4}$. Nonzero gradients occur
in both structures (Table~\ref{tab:legacy_order}); interleaving does
not produce a universal accuracy ordering. The results do not assert
that every Type II or Type III circuit is trainable, nor that a
state-QFI rank has been recovered.

\begin{table}[tb]
\caption{Archived embedding-placement control, ten paired trials.
Initial gradient norm and post-training KTA are means. Accuracy is
mean $\pm$ sample SD in percent. Equal means do not establish equal
predictions, statistical equivalence, or equality on other tasks.}
\label{tab:legacy_order}
\begin{tabular}{llccc}
\toprule
Data & Order & $\|\nabla_\theta\mathrm{KTA}\|$ & KTA & Accuracy\\
Moons & II  & .0589 & .3544 & $94.69\pm3.62$\\
      & III & .0561 & .3455 & $94.69\pm3.62$\\
Wine  & II  & .1253 & .3950 & $94.04\pm3.20$\\
      & III & .1086 & .3878 & $94.04\pm3.68$\\
\bottomrule
\end{tabular}
\end{table}

\subsection{Terminal-heavy fidelity candidate reuse}
An earlier fixed candidate set varies the input-independent output
block in $U_\theta(x)=W_\theta E(x)$ across 36 circuits. The fidelity
overlap cancels $W_\theta$, so an embedding-specific key permits reuse
within four conservative groups. Four keys do not prove four minimal
inequivalent kernels: further equivalences between embeddings are
not excluded. This is a static candidate-set audit, not a test of
adaptive search with later data gates.

The saved replay has 1,692 candidate--category records (36 candidates
in 47 categories), each with four support and twenty query inputs.
Evaluating one representative per key avoids 32 of 36 candidate
evaluations, or 88.9\%. The unbatched support--query entry schedule
changes from 135,360 to 15,040 pair tasks. The maximum recorded
kernel discrepancy is $6.00\times10^{-15}$, and all recorded
stabilized AUROCs agree within a group. The algebraic suffix identity
justifies equality; the finite replay only checks this implementation.
Multiplying those task counts by an assumed shot count would be a
workload model, not measured shot savings or equal-precision evidence.

The associated safe reuse procedure is to identify an exactly
certified common suffix, construct a key including the residual
embedding, preprocessing and kernel specification, and cache only
the current candidate's evaluations. Any subsequent data-dependent
extension must be recertified. A state-QFI computation is neither
required for suffix cancellation nor a substitute for this
kernel-equivalence check. Because an ordinary suffix baseline already
finds this reduction, it is not incremental evidence for the new
internal certificate or a contradiction of the later zero-increment
PQK resource surveys.

\subsection{Hardware provenance scope}
The learning and candidate controls in this appendix are classical
calculations or reanalyses of saved numerical records. No hardware
performance or mitigation endpoint is reported: the available
historical hardware summaries do not supply matching backend,
calibration, raw counts, and job identifiers.

%% file: gram_appendix.tex
\section{Coefficient Gram proof and implementation audit}
\label{app:gram}
The three follow-up computations use the previously frozen Kyriienko
IR and source commit \texttt{0f6cb0982fdbe80dcd0ea6cb0acd68c59e362de4}.
Data variables are $(\arccos x_0,\arccos x_1)$ on an open rectangle,
so distinct real Fourier modes satisfy the independence hypothesis.
Heisenberg Pauli propagation carries a Gaussian-rational coefficient
and its $K$ parameter derivatives per data-frequency term. Constant
modes are separated; conjugate-frequency symmetry is checked exactly.
The rational-point coefficient bases have 8, 44, and 72 real
nonconstant columns. These are witness bases, not a claim of minimum
global Fourier support. Modes zero in both value and first derivative
at that point cannot affect its Gram Jacobian.

For $C\in\mathbb R^{m\times A}$, forming all Gram derivatives from
stored coefficient derivatives takes $O(KmA^2)$ arithmetic operations
and at most $O(KA^2)$ output storage; exact coefficient bit lengths and
the preceding Fourier propagation cost are additional. Sparse
Pauli-frequency propagation can grow exponentially. The implemented
gate class is narrower than Corollary~\ref{cor:gramdecision}: fixed
Cliffords, data-only rotations, and single unscaled parameter
rotations in the frozen constructors. It computes a rational-point
rank witness, not a complete symbolic rank solver for all that
corollary's inputs.

The upper-bound audit checks the ordering in Eq.~\eqref{eq:kyriform},
the two exact Clifford-transported endpoint generators, and complete
XYZ orbits on each measured support. In the last Euler block, a
source tangent inducing a left $X$ rotation is
\begin{equation}
 (v_a,v_b,v_c)=
 \left(\frac{\sin c}{\sin b},\cos c,
       -\frac{\sin c\cos b}{\sin b}\right).
\end{equation}
It is independent of the first-block $c_0$ direction. Their products
with the exact witness Gram Jacobian vanish. The global upper proof
uses the output orbits on regular charts, not just these point checks.
Multiplying by $\sin b$ removes the chart denominator and gives the
globally defined analytic field
\begin{equation}
 w=(\sin c,\ \cos c\sin b,\ -\sin c\cos b).
\end{equation}
The identity $D G\,w=0$ extends analytically through chart singularities,
but $w$ itself can vanish, for example at $b=c=0$. Independence of
two null directions is generic, not guaranteed at every parameter
assignment. The moving field is not a constant deletion direction.

Twenty random four-input probes per setting compare reconstructed
Fourier features with the pinned source simulator. The maximum
feature discrepancy is $3.09\times10^{-15}$; distance and analytic
distance-gradient discrepancies are at most $1.29\times10^{-14}$ and
$1.38\times10^{-14}$. Sixty source-coordinate central differences
at step $10^{-6}$ agree within $4.42\times10^{-9}$. These tests check
implementation consistency and do not replace exact minors or the
analytic upper proof. This dedicated implementation was run locally;
the server rerun uses the separate 30-setting Fourier audit to certify
the same three ranks (Table~\ref{tab:reproscope}).

\subsection{An affine-head distinction}
For an unrestricted affine head $w^Tf_\theta(x)+a$, with a freely
adjustable scalar bias and independent $\{1,h_j\}$, the hypothesis
class at fixed $\theta$ is determined by the row space of $C(\theta)$.
Indeed the nonconstant coefficient vectors are exactly
$\{w^TC(\theta):w\in\mathbb R^m\}$. Equality of those sets is
necessary and sufficient for equality of hypothesis classes.
This is not equality of fixed predictors or training trajectories.
Without free bias the offset must be retained; regularization or
constraints on $w$ also change the equivalence contract. Variable
matrix ranks require separate rank strata rather than one fixed
Grassmannian. We do not report a QNN experiment for this observation.

\subsection{Resource accounting is schedule specific}
For the non-cross-gate-cached schedule used in the main text, with
fixed positive $T,N,B,s$, comparison of two arms gives
$\Delta C=2TNB\Delta A$ and $\Delta S=s\Delta C$.
Thus zero circuit-setting reduction is equivalent to zero reduction
in its trainable fused-angle tasks. This is an accounting identity
for that schedule, not a necessity theorem for all compilers,
optimal measurements, wall time, or shots at fixed accuracy. A lower
kernel differential rank alone does not imply a reduced schedule.

%% file: followup_appendix.tex
\section{Extended Fourier audit and scope corrections}
\label{app:followup}
A separate Laurent-state implementation covers all 30 settings in
Table~\ref{tab:extension}. All frozen angles are fixed-frequency:
YZ-CX uses angles that add parameter and data terms, not products; Hubregtsen
and Kyriienko likewise have no trainable data multiplier. The two
underlying data phases depend on distinct input coordinates and
range over an open rectangle. This verifies the finite independent
basis assumptions for the whole extension, not for the separate
24-setting product-angle study.

At exactly specified rational-circle parameter points, nonzero
Gram Jacobian minors modulo two primes give rigorous rational lower
bounds; seven settings also retain rational determinants. The upper
bounds combine exact two-copy invariance fields with completed
symbolic Gram expansions. The 27 previously closed ranks agree,
and the three Kyriienko closures independently agree with the
dedicated computation in Appendix~\ref{app:gram}. The count remains
54 settings, not 84. All 195 active and 57 inactive source-symbol
verdicts agree with the previous structural closure.

Write $\Omega=\sum_j w_j O_j\otimes O_j$. The condition
$[\Omega,P\otimes I+I\otimes P]=0$ preserves the feature inner
product under a common rotation and hence preserves feature
distances. Separate-copy commutation is used for data-dependent
angles. These are sufficient operator certificates, not a complete
test for all distance-preserving transformations. The saved audit
contains 81 fresh-angle symbolic deletion checks and three
symbolic Euler-field checks with perturbed-field negative controls.
This reimplements known structural mechanisms; shared Pauli-arithmetic
helpers and frozen IR mean it is not a fully independent proof system.

Original-circuit symbolic Gram expansion finishes in 16 of 30
settings. Of the earlier budget-limited 32 source entries, all 32
are again certified inactive by two-copy invariance, and nine also
receive a completed symbolic Gram identity. The remaining 23 lack
that second certificate because the 600,000-term budget
(with a 60-second limit) was exceeded.
All 14 unfinished original-circuit expansions record this reason,
not a 60-second timeout. They are
not undecided overall. In particular, a 0.025-second symbolic pass
for the two-qubit full-Pauli Kyriienko case versus the earlier
20-second cutoff is an unmatched diagnostic observation, not a
controlled speedup measurement.

The numerical audit uses 32 random parameter/input-pair probes
per setting. Maximum distance and gradient differences are
$2.85\times10^{-14}$ and $2.58\times10^{-14}$. A further interval-engine
comparison reuses the earlier feasible witnesses. Validation includes
11 unit tests, hashes of 33 manifested artifacts and 30 frozen inputs,
and recomputation of all 60 modular lower certificates. Three theory
diagnostics provide floating-rank and random injectivity consistency
checks, not proofs. The complete computation, including budgeted
symbolic expansions, was rerun on Linux; its certified outputs match
the reference (Table~\ref{tab:reproscope}). No hardware execution is
included.

\input{reproduction_scope}

\subsection{Pointwise metric preservation versus a union of spans}
Let $M=\diag(1,2,3)$ and let $Q$ be the cyclic permutation
$(u_1,u_2,u_3)\mapsto(u_3,u_1,u_2)$. Then
\begin{equation}
 N=Q^TMQ-M=\diag(1,1,-2).
\end{equation}
For the one-column matrix
\begin{equation}
 C(t)=(\sqrt2\cos t,\sqrt2\sin t,1)^T,
\end{equation}
$C(t)^TNC(t)=0$ for all $t$, although the columns as $t$ varies
span $\mathbb R^3$ and $N\ne0$. Thus pointwise preservation does
not imply preservation of the quadratic form on the union's linear
span. Isometry on that span remains sufficient; it is necessary
under stronger conditions, for example when a single column space
already equals it. A small common rescaling of $C$ makes $f_t(x)=xC(t)$,
$|x|<1$, physical mixed-state Bloch features, so this is not merely
a formal large-amplitude counterexample.

\subsection{Classical relaxation without an efficiency claim}
Under Theorem~\ref{thm:gram}, the attainable kernel family lies in
\begin{equation}
 \left\{e^{-\gamma\Delta h^T H\Delta h}:
 H\succeq0,\ \operatorname{rank}H\le\min(m,A)\right\}.
\end{equation}
This follows by setting $H=C(\theta)^TMC(\theta)$. For any common
objective defined on these kernels, its supremum over the circuit
family cannot exceed the supremum over this classical relaxation;
no maximum-attainment assumption is needed. This is representational
containment, not an efficient classical training or simulation result.
The number of Fourier modes and the cost of constructing the metric
may be exponential. Individual explicit sine/cosine modes are not
intrinsically hard to evaluate; polynomial mode count alone would
still not prove efficient extraction or optimization of $H$.

For weight-at-most-$k$ readouts, the original audit used local-rotation
invariance only. Proposition~\ref{prop:gauge}, using the classification
of locality-preserving unitaries in Ref.~\cite{cpr2019locality},
characterizes the connected universal gauge for complete isotropic
weight sectors; it does not classify disconnected or
reachable-family-specific symmetries. The all-state and fixed-family
quantifiers remain distinct.

%% file: reproduction_scope.tex
\paragraph{Cross-environment reproduction.}
The reference computations used Python~3.9.6 and SymPy~1.14.0 on
macOS arm64. The pinned Linux x86\_64 reruns used Python~3.8.10,
NumPy~1.24.4 and SymPy~1.13.3, with Qiskit~1.2.4 for the scaling
study. Jointly, the two full reruns passed all 3,376 strict
comparisons without warnings. These checks compare certificate
contents and provenance; they are not 3,376 independent experiments.
The reruns use the same code, not independent implementations or
formal proof verification.

\begin{table}[t]
\caption{Reproduction scope. Matching certificates establish
cross-environment reproducibility within the listed implementation
scope, not a new census or an independent proof system.}
\label{tab:reproscope}
\begin{tabular}{>{\raggedright\arraybackslash}p{.29\columnwidth}
                >{\raggedright\arraybackslash}p{.63\columnwidth}}
\toprule
Calculation & Reproduction and matched evidence\\
\midrule
30-setting Fourier audit (App.~\ref{app:followup})
& Full Linux rerun, including bounded symbolic expansion: 30 ranks,
7 rational and 60 modular minors; 195 active and 57 inactive entries,
none undecided.\\[4pt]
Kyriienko scaling (App.~\ref{app:kyriscaling})
& Full Linux rerun: all 12 ranks, exact minors, Jacobian payload
hashes and generated IR hashes agree.\\[4pt]
Dedicated three-setting Gram calculation (App.~\ref{app:gram})
& Local execution only. The separate 30-setting audit reproduces the
same three ranks, not this dedicated implementation.\\[4pt]
Earlier extension audit (App.~\ref{app:extension})
& Server replay of saved evidence, not a new discovery run: 30
setting checks, 195 activity witnesses, 27 nonzero minors, structural
verdicts and the nonlinear certificate.\\
\bottomrule
\end{tabular}
\end{table}

%% file: resource_details.tex
\section{Detailed correctness and resource audits}
\label{app:resource_details}
\subsection{Correctness and external-source coverage}
Designed validation covers 360 combinations of case, size, and seed with four
certificate variants, giving 1,440 rows. Maximum kernel and derivative
discrepancies after deletion are $1.33\times10^{-14}$ and
$3.28\times10^{-15}$. Twenty paired synthetic learning runs preserve
objectives and predictions to numerical precision. At eight qubits
and two layers, all 48 parameters of a local full-Pauli-readout control
are certified inactive; its entangling counterpart has 48 \unknown{}
symbols, not 48 proven active ones.

The frozen sQUlearn source-library study \cite{squlearn} executes 34 configurations from ten modules
of a 13-module inventory. Three modules needing additional
configuration remain explicitly excluded. There are 28 trainable and
six parameter-free settings. All 4,590 saved arrays in 102 probes,
including 813 parameter finite-difference probes, were recomputed.
Under the specified isotropic local readout, 16 trainable settings
contain 32 certified inactive symbols. This is conditional coverage
of a selected library and readout, not literature prevalence.

\begin{table}[t]
\caption{Post hoc classification of 32 inactive symbols, summed across
settings. Settings and symbols are not independent discoveries.}
\label{tab:mechanisms}
\begin{tabular}{lr}
\toprule
Mechanism & Symbols\\
Elementary local output tails & 16\\
Transport without overlapping data crossing & 6\\
Affine bias within the same native gate & 6\\
Crosses genuinely later overlapping data gates & 4\\
\bottomrule
\end{tabular}
\end{table}

An elementary wire-local tail baseline already identifies 16 symbols
in six settings. The remaining 16 do not all represent internal data
crossings (Table~\ref{tab:mechanisms}). A local Qiskit~1.2.4 level-three
symbolic-compilation diagnostic preserves the aggregate counts, but
it is not a comparison with all compiler or equivalence techniques.
The inventory and mechanism classification are post hoc.

\subsection{Separating genuine internal work from baseline work}
The follow-up pipeline first compiles, then normalizes the IR, removes
local tails, performs restricted boundary transport, and fuses again
adjacent same-axis rotations. The boundary rule may pass disjoint data
and same-native affine terms, but not a distinct overlapping data gate.
The proposed extra stage applies full supported transport before the
same refusion. This baseline is deliberately stronger than a
commutant test or a strict last-gate scan.

\begin{table}[t]
\caption{Full-study structure after common refusion. The exact
symbol reduction does not imply a lower interpreter gate count.}
\label{tab:strong}
\begin{tabular}{lrr}
\toprule
ChebyshevRx, two layers & Boundary & Full\\
2-qubit retained symbols & 7 & 5\\
2-qubit executed operations & 6 & 6\\
4-qubit retained symbols & 15 & 13\\
4-qubit executed operations & 14 & 14\\
\bottomrule
\end{tabular}
\end{table}

The full study covers 12 configurations, seven arms and three custom
NumPy differentiation backends on WDBC, Ionosphere and Auto MPG.
Five split seeds and five timing repetitions yield 2,700 seven-arm
cold blocks, or 18,900 fits. A two-arm reuse study adds 3,600 fits.
Stored kernel, gradient, parameter-trajectory, objective and prediction
comparisons satisfy the predefined tolerances. Across all stored
comparisons, maximum kernel and gradient discrepancies are
$4.65\times10^{-11}$ and $2.75\times10^{-10}$, respectively.
Selected scalar source-path replay is a separate check, not independent
human review or a second timing replication. The local pilot remains
a design check, not an additional independent sample.
Table~\ref{tab:strong} shows why backend-specific derivative accounting
is indispensable: fusion removes an apparent gate-count saving.

\subsection{From extra symbols to executed work}
\begin{figure}[tbp]
\centering
\includegraphics[width=\textwidth]{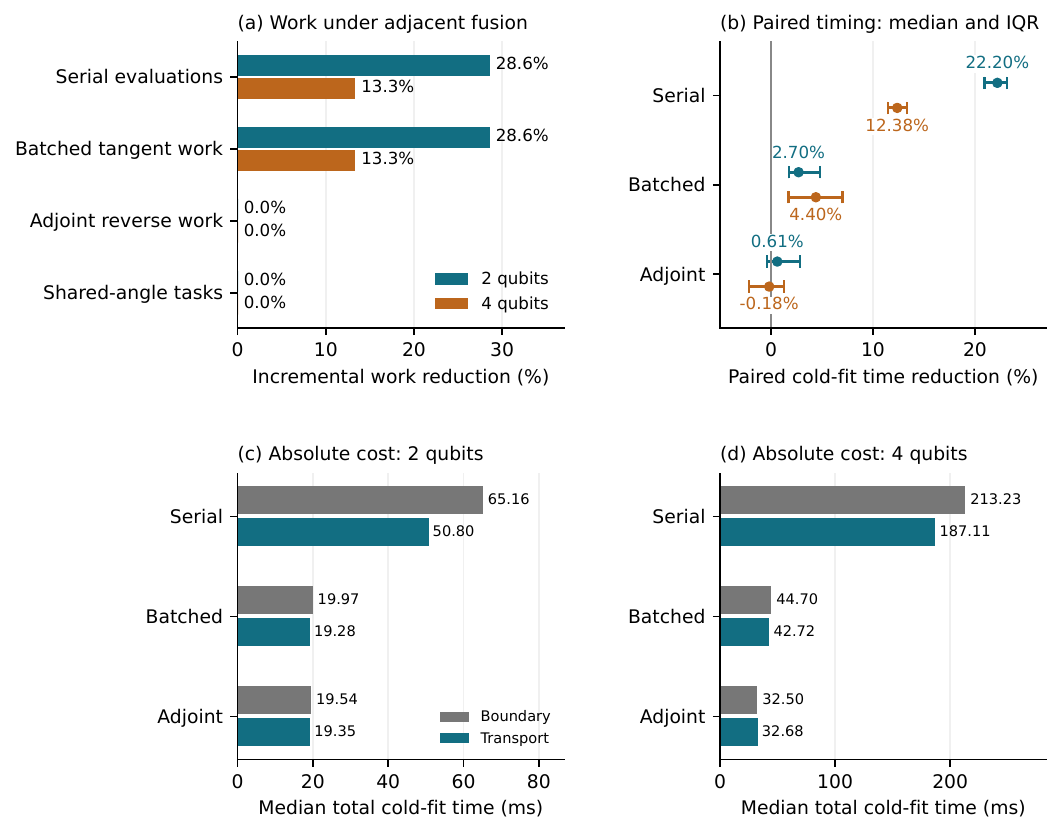}
\caption{Increment over fused boundary simplification for two-layer
ChebyshevRx. Upper left: actual serial evaluator and batched tangent-work
reductions, with unchanged adjoint reverse work and analytically
counted shared-angle tasks. Upper right: paired cold-fit time-reduction
medians and 25th--75th percentiles, from 75 pairs per size and backend
(three datasets, five split seeds, five timing repetitions). Intervals
are IQRs, not confidence intervals. The plot summarizes the distribution;
extreme individual pairs are not shown, but remain in the saved data
and Appendix~\ref{app:full_time}. All times include certification.
The other ten circuit settings have no additional inactive symbol.
Bottom: median absolute cold-fit times for each backend and arm; the
adjoint implementation is substantially faster than transport-serial.
These CPU simulator results are not QPU or fixed-accuracy shot data,
and do not compare against global normalization.}
\label{fig:pilot}
\end{figure}

Figure~\ref{fig:pilot} links the additional certification to backend
work and the repeated full-study timings under the earlier,
adjacent-fusion baseline, together with absolute backend costs.
Serial and batched tangent work decrease in two settings, but
adjoint reverse work does not. Certification-inclusive serial
cold-fit median reductions are 22.20\% and 12.38\%; the fastest
tested backend remains boundary-adjoint rather than transport-serial.
The repetitions concern three problems, not independent new circuit
families. Detailed counters, paired timing distributions and reuse
measurements are retained in Appendix~\ref{app:full_time}; none compares
against the later global-normalization baseline.

\subsection{Gate-angle sharing changes the shift accounting}
Generalized parameter-shift rules exploit Fourier structure and can
reduce derivative-evaluation costs for parameters shared across
multiple gates \cite{wierichs2022general}. Our accounting below is
for a specified gate-angle-sharing schedule, not an optimality claim
over all differentiation rules.
For each fused Pauli rotation, two shifted feature evaluations can
supply its angle derivative. All symbols in that angle reuse the same
derivative through classical chain factors, including input-dependent
coefficients and shared symbols summed over occurrences. One must not
apply a two-shift rule directly to the nonlinear RBF kernel
\cite{schuldgradients}. If $A$ trainable fused angles, $N$ inputs,
$T$ updates and $B$ measurement bases are used, a transparent
non-cross-gate-cached schedule has
\begin{equation}
 C_{\rm train}=TNB(1+2A),\qquad
 S_{\rm train}=s C_{\rm train},
\end{equation}
where $s$ is a fixed shot count per setting. The unshifted features are
reused across all pairwise kernels. Final prediction, mitigation and
precision-driven shot allocation are separate costs. Here $A$ counts
trainable fused-angle tasks in this schedule, not a proven rank of
functionally independent directions. Equal shot counts per setting do
not imply equal estimator accuracy. Fixed-accuracy comparisons require
a specified error target and confidence level, including bias and
variance of the nonlinear kernel estimator; they have not been measured.

For local XYZ readout, $B=3$. In the present two-qubit witness,
$A=4$ for both arms; in the four-qubit witness $A=8$ for both arms.
With $N=64,T=10$, this gives 17,280 and 32,640 circuit settings,
respectively, with zero incremental setting or fixed-shot reduction.
Remaining product terms keep those fused angles trainable.
This is analytic scheduling, not a finite-shot or hardware experiment;
additional cross-gate symmetry optimizations might reduce both arms.
It invalidates inferring QPU savings from the inactive-symbol count
alone. We implement this shared-angle derivative schedule for the
structural survey below and verify its feature Jacobians and objective
gradients against analytic propagation and finite differences. This
implementation validates ideal derivatives, not physical shot costs.

\subsection{A fixed-scope circuit and readout survey}
The follow-up survey retains the previously listed 120 nominal slots:
ten default-constructed source modules, two qubit counts, two layer
counts and three readouts. Eighteen slots duplicate constructors
without a layer argument, leaving 102 distinct slots. The three
readouts are isotropic local XYZ, local Z, and all nonidentity Pauli
words of weight at most two. Modules and readout variants are not
counted as independent circuit families.

To avoid crediting elementary output symmetry as an interior gain,
the baseline additionally handles readout-preserving common-isometry
suffixes. This includes the full two-qubit Pauli readout's invariance
under common output rotations. Both arms use the same compiled circuit,
readout-aware baseline and adjacent fusion; only the extra transport
stage differs. This is a distinct survey baseline, not a retrofit to
the full-study timing data.

The survey finds extra inactive symbols in four slots: the two known
two-layer ChebyshevRx sizes under local XYZ and weight-at-most-two
Pauli readout. Each removes two symbols, but no slot reduces final
fused operations or trainable fused-angle tasks. There are 82 trainable
slots with defined alignment probes, 18 parameter-free slots, and two
trainable local Z slots whose constant centered kernels make alignment
undefined. The latter retain kernel and kernel-derivative checks but
are not counted as successful training runs. All numerical checks
applicable to each slot pass. These negative resource results apply to
the fixed source, constructor and supported-rule scope, not to arbitrary
circuits. They do not prove that a stronger method cannot find savings.
Details and the disclosed alignment-handling amendment appear in
Appendix~\ref{app:survey}.

%% file: additional_scope.tex
\section{Additional optimizer, search, and prediction scope}
\label{app:additional_scope}
The internal transport algorithm is incomplete. For example, the chronological circuit
$R_x(a),R_y(b),R_y(-b),R_x(-a)$ is identically the identity, but the
restricted transport rules with $Z$ readout certify neither symbol.
An ordinary inverse-cancellation compiler can resolve it. Kernel-aware
certification supplements rather than replaces conventional compilation.

Nor does current equivalence license pruning all search descendants.
Preparing $R_y(x)|0\rangle$ followed by $R_z(t)$ leaves the current
$Z$ feature independent of $t$. Appending $R_y(x)$ changes that feature
to $\cos^2x-\sin^2x\cos t$. Candidate extensions require
recertification or a separate preservation theorem. We do not claim
general adaptive-search acceleration.

Trainable circuit angles also do not constitute the entire fitted
predictor. A kernel SVM generally uses
$h(x)=\sum_{i\in\mathrm{SV}}a_i k_\theta(x_i,x)+b$ and must retain
coefficients and reference inputs or their usable feature
representations. Our vision scorer instead uses
$a(x)=1-n_s^{-1}\sum_i k_\theta(s_i,x)$ over normal support examples;
it likewise requires those references. Caching projected features
uses $n_s m$ feature values in addition to the angles, kernel settings
and preprocessing. For four supports and 24 features, that is 96
reference-feature values, not a predictor specified by 48 angles
alone. A classical Gaussian kernel can use the same prototype scorer;
it is not intrinsically tied to a support-vector classifier.
Any memory comparison must separate trainable parameters, stored
references, preprocessing and inference work for matched learners.

Finally, ideal-expectation equality does not guarantee equal noisy
circuits or finite-shot kernel estimators. The source-derived examples
are small and purposefully selected. They do not establish a generic
prevalence rate, quantum advantage, lower hardware shot counts, or
better predictive performance. Kernel-alignment subsampling addresses
cost through a different strategy \cite{sahin2024subsampling}; the
present timings are not a head-to-head superiority result.

The preservation contract permits a compiler to target the specified
kernel rather than every state or feature distinction. The supported
internal rule has one demonstrated circuit family beyond the fused
boundary baseline. Additional circuit sizes are not independent
families. The analysis establishes where work is removed in the tested
implementations and where it is not; it does not establish broad runtime
or hardware savings.

%% file: spectral_context.tex
\section{Relation to spectral and coefficient methods}
\label{app:spectral_context}
Finite Fourier
representations of quantum features are established tools
\cite{schuldFourier}; the factorization $C^TMC$ is also the familiar
pullback of a Mahalanobis metric \cite{weinberger}.
Our use is an exact, readout-specific parameter identifiability
contract and its application to the rank gaps below, not a claim
that either algebraic construction is new.

Fourier coefficient correlations already characterize ansatz-dependent
constraints \cite{fingerprints}. Spectral tensor-network work represents
quantum kernels by Fourier coefficient tensors and uses an empirical
Fourier Gram matrix for data-dependent alignment \cite{spectralTN}.
That data matrix is not the parameter-dependent $C(\theta)^TMC(\theta)$
here, and tensor bond rank is not the differential rank in
Eq.~\eqref{eq:gramrank}. Fourier-geometric gate-placement criteria
also exploit adjoint frequency planes \cite{fourierDesign}.
These overlaps motivate careful separation of representation,
coefficient reachability, and exact parameter identifiability.
Campbell, Del Debbio, and Wallden's circuit harmonic matrix also
eliminates data dependence from coefficient-space analysis
\cite{CHM}. It represents joint harmonics of inputs and parameters and
constructs pointwise coefficient Jacobians and QNTKs, as well as
parameter-averaged statistics. Our distinction is therefore not
``pointwise versus averaged'': $G=C(\theta)^TMC(\theta)$ describes
fixed-metric feature distances, whereas their harmonic QNTK is a
Gram matrix of coefficient derivatives. We use this distinction to
certify distance-equivalence fibers and their differential dimension.
In the four-qubit Hubregtsen example of Sec.~\ref{sec:gram}, the
coefficient Jacobian has rank four, whereas $D_\theta\operatorname{vech}G$
has rank two.

%% file: transport_implementation.tex
\section{Implementation of the restricted transport rule}
\label{app:transport_impl}
This appendix records the implementation of
Theorem~\ref{thm:transport} and the scope of its guarantee.

The implementation supports fixed Hadamards and CNOTs for Clifford
transport. Unsupported noncommuting rotations block the rule. It scans
parameter-only occurrences backwards, records each transport and
readout check, deletes a certified occurrence, and restarts. With $L$
gates, $m$ observables and length-$n$ Pauli words, a straightforward
implementation has conservative complexity $O(L^3n+L^2mn)$, without
statevector evaluation. Equal metric weights are checked exactly,
not by a numerical symmetry tolerance. The operational steps are:
\begin{enumerate}
\item Translate source gates to the supported IR, preserving every
symbol, coefficient, input expression, and shared occurrence.
\item Apply the common compiler, boundary rules, and same-axis fusion
used by the baseline; retain provenance for data-crossing classification.
\item Transport an eligible parameter-only generator through supported
gates and test exact closure and metric preservation.
\item Delete a successful occurrence and restart; block unsupported
transport rather than inferring activity or inactivity numerically.
\item Mark a symbol inactive only when all its occurrences are absent,
then build backend work from the final fused program.
\end{enumerate}
Source-to-IR correctness is a precondition of the guarantee, not a
consequence of finite-sample tests. Additive angle splitting is valid
for rotations about the same generator; splitting a noncommuting
sum is not permitted. Product angles such as
$\theta x$ are not common-output candidates. Additive same-axis
angles may be exactly decomposed, but that standard identity is not
an independent theoretical novelty.

Pointwise equality implies equal retained kernel derivatives.
A deterministic optimizer preserves its retained-parameter trajectory
if updates depend only on these kernels and derivatives with identical
retained state, schedules and tie rules. Regularizers, stopping rules,
optimizer couplings, or random-number consumption involving deleted
coordinates require separate treatment. This is not a bitwise
floating-point guarantee. Source-adapter numerical validation also
does not replace a formal source-to-IR proof.

%% file: kyriienko_scaling_appendix.tex
\section{Kyriienko-type scaling certificates}
\label{app:kyriscaling}
This appendix records the computation behind
Table~\ref{tab:kyriscaling} and the status of the all-size question.
The $n=6,8$ settings extend the default constructor of the census
(two inputs, one encoding and one variational layer) to larger sizes;
they are not part of the 54 settings. Only even $n$ were computed.
The reference calculation and its pinned-environment Linux rerun
produce identical certificates (Table~\ref{tab:reproscope}).

\paragraph{Source route and validation.}
Circuits come from the census's vendored sQUlearn commit
(Appendix~\ref{app:gram}),
\texttt{KyriienkoEncodingCircuit(num\_qubits=$n$)} with default
options, Qiskit~1.2.4 transpilation (optimization level 3, seed 71,
approximation degree 1, one process), and the census IR adapter.
As a validation gate, the same route regenerates the six frozen
$n=2,4$ IR files byte for byte (equal SHA-256 hashes) and reproduces
their census ranks: $4,0,4$ at $n=2$ and $10,10,8$ at $n=4$ for local
XYZ, weight $\le2$, and local Z. For $n=6,8$ the IR agrees gate by
gate with a circuit written directly from the source formula: encoding
gate $m$ is $R_y[2(\lfloor m/2\rfloor+1)\arccos x_{m\bmod2}]$ on qubit
$m\bmod n$. For even $n$, qubit $q$ therefore encodes $x_{q\bmod2}$
with total frequency $\kappa_q=4\lfloor q/2\rfloor+n+4$, namely
$\kappa=(10,10,14,14,18,18)$ for $n=6$ and
$(12,12,16,16,20,20,24,24)$ for $n=8$. Exact checks confirm
Eq.~\eqref{eq:kyriform}, fixed frequencies, unit readout weights, one
unit-coefficient occurrence per parameter, and commutation of $E$ with
$Z_0$ and $X_{n-1}$.

\paragraph{Exact lower bounds.}
Parameters are $\theta_k=4\arctan u_k$ with $u_k=(k+2)/(k+5)$, so
$e^{i\theta_k/2}=(1+iu_k)^2/(1+u_k^2)$ is Gaussian rational. Inputs
are $x_b=\cos(2\arctan s_b)$ for ten pairs $(s_0,s_1)$:
$(\frac12,\frac53)$, $(\frac45,\frac34)$, $(\frac12,3)$,
$(\frac43,\frac13)$, $(3,\frac54)$, $(\frac52,4)$, $(4,\frac53)$,
$(\frac35,\frac13)$, $(\frac13,\frac54)$, $(\frac45,\frac52)$.
Since $s_b>0$, $\arccos x_b=2\arctan s_b$ exactly, and even
$\kappa_q$ make every data half-angle an integer multiple. The 45
unordered pairs of these inputs give rows
$\partial_{\theta_k}d^2(x_i,x_j)$, computed exactly from
Gaussian-integer statevectors and tangents. Scaling each row by a
positive rational to integers preserves nonvanishing of minors.
A minor is selected by elimination modulo $2^{61}-1$. Its determinant
is computed exactly by fraction-free Bareiss elimination, agrees with
an independent SymPy \texttt{DomainMatrix} determinant, and is nonzero
modulo $2^{61}-1$, $2^{62}-57$, and $2^{63}-25$
(Table~\ref{tab:kyriminors}). A replay from the stored integer
Jacobians reproduces all twelve determinants, including $n=2,4$.
Each minor, with the 45 input pairs fixed, is a real-analytic
function of $\theta$ on the connected parameter domain. Being nonzero
at one point, it vanishes only on a proper analytic subset, so it is
nonzero on an open dense set of full measure; this gives the generic
lower bound.

\begin{table}[b]
\caption{New exact minors ($n=6,8$). $m$ counts measured Pauli words.
Omitted columns are angles $c_q$ of Eq.~\eqref{eq:kyriform}.
``Digits'' refers to the row-scaled integer determinant. The last
column gives rounded decimal approximations of the exact, unscaled
rational minors.}
\label{tab:kyriminors}
\begin{tabular}{rlrrlrr}
\toprule
$n$ & Readout & $m$ & Order & Omitted & Digits & Minor (approx.)\\
6 & XYZ & 18 & 16 & $c_0,c_5$ & 5773 & $-0.99$\\
6 & $\le2$ & 153 & 16 & $c_0,c_5$ & 7093 & $1.6\times10^{4}$\\
6 & Z & 6 & 12 & all $c_q$ & 2823 & $1.7\times10^{-6}$\\
8 & XYZ & 24 & 22 & $c_0,c_7$ & 10464 & $-0.094$\\
8 & $\le2$ & 276 & 22 & $c_0,c_7$ & 15632 & $-1.5\times10^{7}$\\
8 & Z & 8 & 16 & all $c_q$ & 5346 & $-7.4\times10^{-3}$\\
\bottomrule
\end{tabular}
\end{table}

\paragraph{Upper-bound and consistency checks.}
On the generated IR, $c_0$ is deleted exactly: its transported
generator $P=Z_0$ satisfies $[\Omega,P\otimes I+I\otimes P]=0$ in exact
Pauli arithmetic (Appendix~\ref{app:followup}). The analytic Euler
field $w$ of Appendix~\ref{app:gram} satisfies
$\sum_kw_k\partial_kU=-\frac i2\sin b\,XU$ symbolically for the last
block, and a field with one doubled component is rejected. For local
Z, the $n$ angles $c_q$ are deleted with Z-string output generators,
as in Proposition~\ref{prop:kyribounds}(ii). At the certificate point
these upper-bound directions have exact rank $2$ (XYZ, and weight
$\le2$ with $k<n$) or $n$ (Z), and every exact null vector annihilates
all 45 rows. This
one-point consistency check is not an additional proof.

\paragraph{Floating cross-checks.}
At the certificate point, exact features agree with a floating IR
engine within $9.3\times10^{-15}$, with an IR-free dense statevector
within $2.2\times10^{-15}$, and with Qiskit's source statevector
within $1.9\times10^{-15}$; exact and floating gradients agree within
$2.2\times10^{-15}$. At four random parameter points with
$\max(3K,60)$ random pairs, SVD ranks at relative tolerance $10^{-9}$
equal the certified values in all twelve settings. For $r>0$,
$\sigma_{r+1}/\sigma_1\le4.4\times10^{-16}$ and
$\sigma_r/\sigma_1\ge7.1\times10^{-6}$; in the rank-zero setting,
$\sigma_1$ is below $1.3\times10^{-15}$ times the largest sampled
$d^2$. These are implementation checks, not certificates.

\paragraph{Cost and environment.}
The run used one process and one thread on macOS arm64 with
Python~3.9.6, Qiskit~1.2.4, NumPy~1.24.4, SymPy~1.14.0, and
SciPy~1.10.1. The frozen census requirements pin SymPy~1.13.3; this
reference run used SymPy~1.14.0, and the replay below used the pinned
version.
The twelve settings took $32.4$~s of wall time
with $225$~MB peak resident memory. Each $n=6$ setting took at most
$1.2$~s. The $n=8$ settings took $16.5$~s (local XYZ, including
$12.5$~s of exact states shared with the other readouts), $10.3$~s
(weight $\le2$), and $1.2$~s (local Z). Bareiss and SymPy determinants
took at most $2.2$ and $3.6$~s. The arithmetic stack (Python integers,
NumPy object arrays, SymPy) is not formally verified; agreement of two
determinant implementations, three residues, and the floating engines
is an implementation check. The $n=6,8$ IR also depends on Qiskit
transpilation; the source-formula comparison and the Qiskit-free
statevector check it independently.
The Linux reproduction is summarized in Table~\ref{tab:reproscope}.
A Python~3.8 compatibility shim handled multi-argument
\texttt{math.gcd} calls without modifying the calculation source;
IR agreement is for the same Qiskit~1.2.4 transpiler.

\paragraph{Numerical scan.}
We estimate $r$ for unit-weight local XYZ by statevector
simulation, applying the parameter-shift rule to the feature
expectations and the chain rule to $d^2$ over
random input pairs at random parameter points. Every size
$n=2,\ldots,12$ gives $r=3n-2$ (Table~\ref{tab:kyriscan}). For
$n=9$--$12$, the smallest retained singular value lies between $0.268$
and $0.343$, and the next one is below $7\times10^{-15}$. Complete
weight-at-most-$k$ readout gives $r=3n-2$ at
$(n,k)=(3,2),(4,2),(5,2),(4,3),(5,3),(5,4)$. These ranks are numerical,
not certified.

For $(n,k)=(5,3)$, the readout contains $375$ Pauli observables and
the circuit has $15$ trainable angles. Five trials (seeds
$3,42,123,777,2024$) each sample one parameter point uniformly from
$[-2\pi,2\pi]^{15}$ and $120$ input pairs independently from
$[-0.95,0.95]^2$. Each $120\times15$ distance Jacobian has numerical
rank $13$ at thresholds $\max(\tau\sigma_1,10^{-9})$, for
$\tau=10^{-6},10^{-7},10^{-8},10^{-9},10^{-10}$. Across the trials,
$\sigma_{13}$ ranges from $0.239$ to $7.274$, whereas
$\sigma_{14}<5.6\times10^{-14}$. A dense-Pauli calculation agrees
with the bit-mask implementation on every sampled Jacobian to within
$2.2\times10^{-14}$ in maximum absolute error. On three input pairs
per seed, the pinned sQUlearn circuit evaluated with Qiskit agrees
on the derivatives to within $1.1\times10^{-14}$; a centered
finite-difference check with step $10^{-5}$ agrees to within
$5.0\times10^{-9}$. The script, frozen protocol, sampled arrays,
and singular values are included in
\texttt{experiments/kyriienko\_weight3/}; the repository command
\texttt{python reproduce.py kyriienko-weight3} repeats the calculation.
These checks complete the six-pair numerical scan, but do not supply
an exact lower-bound certificate or remove the block-cover obstruction
at $n=5$.

\begin{table}[b]
\caption{Local XYZ ranks for the Kyriienko-type family. ``Num.'' is a
floating-point rank and ``Exact'' a certified value. (DS) is the rank
condition of Eq.~\eqref{eq:kyrids}, including the constant mode,
tested by floating-point sampled rank (not a certificate): Y holds;
N fails, through dependent cover modes at $n=5,7$ and a
four-dimensional intersection at $n=12$; NT denotes not tested.}
\label{tab:kyriscan}
\setlength{\tabcolsep}{3pt}
\begin{tabular}{lrrrrrrrrrrr}
\toprule
$n$ & 2&3&4&5&6&7&8&9&10&11&12\\
$3n-2$ & 4&7&10&13&16&19&22&25&28&31&34\\
Num. & 4&7&10&13&16&19&22&25&28&31&34\\
Exact & 4&NT&10&NT&16&NT&22&NT&NT&NT&NT\\
(DS) & Y&NT&Y&N&Y&N&Y&NT&Y&NT&N\\
\bottomrule
\end{tabular}
\end{table}

\paragraph{Block-cover route and its obstruction.}
A lower bound reduces to local blocks through
Theorem~\ref{thm:gram}. Write $P_i=E^\dagger O_iE$. The encoded
product state expands in data modes $\Sigma_a\in\{I,X,Z\}^{\otimes n}$;
the mode function of $\Sigma_a$ is a product of sines and cosines of
the total encoding angles on $\mathrm{supp}\,\Sigma_a$.
Because the Euler layer is local, feature $i$ couples only to modes
with $\mathrm{supp}\,\Sigma_a=\mathrm{supp}\,P_i$. Hence the mode Gram
matrix is block diagonal by support, and each block depends only on
the Euler angles on that support; this step is proved. For $n\ge3$, exact
rational-point minors give generic block ranks 6 for an interior
even-layer pair, 5 for each endpoint pair (null axis $Z_0$ or
$X_{n-1}$), and 2 for the endpoint singleton of odd $n$.
At $n=2$, both endpoints occupy the same pair. Its pullback types
are $\{XX,YX,ZY,ZZ\}$ and its Gram block has rank four: a
rational-point Jacobian has rank four, while the two endpoint
gauge directions bound it above by four. Thus this pair is counted
once, not as two rank-five endpoint blocks.
The even-layer pairs, plus the odd-$n$ singleton, cover the chain,
and their ranks sum to $3n-2$ (checked for $2\le n\le12$).
Let $H_{\rm cov}$ list
the $N_{\rm cov}=2n$ mode functions on these cover supports and
$H_{\rm oth}$ all other mode functions that enter the features. Writing
$\operatorname{rank}_D$ for the linear rank of functions on the actual
input domain $D$, the route requires
\begin{equation}
 \operatorname{rank}_D[1,H_{\rm cov},H_{\rm oth}]
 =N_{\rm cov}+\operatorname{rank}_D[1,H_{\rm oth}].
 \tag{DS}\label{eq:kyrids}
\end{equation}
Equivalently, the cover-mode functions are linearly independent and
their span meets $\mathrm{span}\{1,H_{\rm oth}\}$ only at zero. This
suffices. Choose $B\subseteq H_{\rm oth}$ such that $\{1\}\cup B$ is a
basis of $\mathrm{span}\{1,H_{\rm oth}\}$. Then $h=(H_{\rm cov},B)$
satisfies the independence hypothesis of Theorem~\ref{thm:gram}, and
rewriting the remaining functions of $H_{\rm oth}$ changes only the
coefficients of $B$ and the offset. The cover-mode coefficients, and
hence the cover blocks, therefore remain coordinates of $G$ in an
independent basis. The cover blocks depend on disjoint Euler-angle
sets, so their generic ranks add, giving $r\ge3n-2$. Sampled
floating-point rank tests (tolerance $10^{-9}$, constant column
included) indicate (DS) for even $n=2,\ldots,10$. They are not exact
certificates, so this route adds no certified size beyond
Table~\ref{tab:kyriscaling}. (DS) fails in two ways. At $n=5,7$, both
inputs enter every qubit and the cover modes are already dependent
(sampled ranks 8 of 10 and 10 of 14). At $n=12$, the cover modes are
independent, but a higher-order Chebyshev resonance makes their span
meet $\mathrm{span}\{1,H_{\rm oth}\}$ in four dimensions. Both
failures violate Eq.~\eqref{eq:kyrids}, and these tests are again
sampled floating-point ranks, not certificates. The direct numerical
rank remains $3n-2$ at these sizes,
so the obstruction concerns this proof route rather than the
conjectured value. An all-size argument would need uniform control of
these resonances or a mode-free proof that no reachable-state isometry
exists beyond the two gauge axes.

%% file: recovery_appendix.tex
\section{Metric recovery: proofs, exact evidence, and scope}
\label{app:recovery}
\subsection{Quantifiers in the metric recovery statement}
Proposition~\ref{prop:metricrecover} concerns a fixed parameter point
with a nonconstant feature map, and quantifies over \emph{all} fixed
positive definite metrics. To justify its linear-algebra step, from
$ab^T+ba^T=0$ and $a\ne0$, multiply by $a$ to obtain
$a(a^Tb)+b\norm{a}^2=0$. Inner product with $a$ gives $a^Tb=0$,
and hence $b=0$. A direction that changes features only by a common
translation is invisible to every metric. At a constant feature map,
however, every squared-distance first derivative vanishes even if
the directional feature derivative is nonconstant.

For the generic-metric observation in Sec.~\ref{sec:recovery}, fix
$\theta$ and consider the linear map
$v\mapsto C^TMH_v+H_v^TMC$ in finite coordinates. Each matrix entry
is linear in the independent entries of $M$. A nonzero maximal-rank
minor is thus a nonzero polynomial, whose zero set has measure zero
in the chosen open metric family. Apply this argument in full
symmetric coordinates for dense metrics or in diagonal coordinates
for diagonal metrics. A finite union of exceptional sets still has
measure zero. Positive rescaling of $M$ does not change rank, so
trace normalization is harmless. This is not a uniform guarantee at
every parameter point, a conditioning bound, or an efficient
maximum-rank algorithm. In particular, exchanging the quantifiers
``for every direction there exists a metric'' and ``there exists one
metric for every direction'' is invalid.

\subsection{Exact fixed-readout ceiling and correlation recovery}
Writing $(a,b,c,d)=(\sin\theta_0,\sin\theta_1,
\sin\theta_2,\sin\theta_3)$, the local Z coefficient matrix, up to
an overall sign, is
\begin{equation}
 C=\begin{pmatrix}a&0\\0&b\\c&0\\0&d\end{pmatrix}.
\end{equation}
For diagonal $M$, $G=C^TMC$ has two diagonal entries and its
Jacobian in $(a,b)$ has minor $4ab w_0w_1\ne0$ generically.
For the dense metric in Eq.~\eqref{eq:recoverymetric}, evaluate at
\begin{align}
 (\sin\theta_0,\ldots,\sin\theta_3)&=(3/5,5/13,8/17,7/25),\\
 (\cos\theta_0,\ldots,\cos\theta_3)&=(4/5,12/13,15/17,24/25).
\end{align}
The Jacobian of $(G_{00},G_{01},G_{11})$ in
$(\theta_0,\theta_1,\theta_2)$ has exact determinant
\begin{equation}
 -\frac{835727837184}{978271035584375}\ne0.
\end{equation}
Together with $A(A+1)/2=3$, this proves maximal fixed-dense-metric
rank three. The centered features themselves have generic rank four.

Before the diagonal controlled-rotation tail, the state is a product,
and that tail commutes with all Z strings. Thus $f_{qr}(x)=f_q(x)f_r(x)$
for the added ZZ features. All these features remain independent of
the four tail symbols, giving rank at most four. A saved diagonal
metric and nonzero $4\times4$ Fourier Gram minors certify the matching
lower bound. There is also a direct analytic lower bound for any fixed
positive diagonal metric on these ten features. Put
$u=\cos x_0$ and $v=\cos x_1$. The five nonconstant data functions
$u,v,u^2,v^2,uv$, together with 1, are independent on the input
domain. Four diagonal entries of their coefficient Gram matrix are
\begin{equation}
 \begin{aligned}
 \beta_u&=w_a a^2+w_c c^2,&\beta_{u^2}&=w_{ac}a^2c^2,\\
 \beta_v&=w_b b^2+w_d d^2,&\beta_{v^2}&=w_{bd}b^2d^2.
 \end{aligned}
\end{equation}
Here $w_{ac}$ and $w_{bd}$ are the positive weights of $Z_0Z_2$
and $Z_1Z_3$. The $(a,c)$ Jacobian block for the first pair has
determinant $4acw_{ac}(w_a a^2-w_c c^2)$, and the $(b,d)$ block
has the analogous generically nonzero determinant. Multiplying
by the four cosine derivatives converts to the angle Jacobian and
remains generically nonzero. Thus the rank is four. This argument
also shows that these two same-input-group correlations suffice
under diagonal weighting; the experiment retained all six ZZ
correlations and does not test minimal feature selection.
The two stored modular certificates independently reduce rational coefficients
at a valid rational-phase parameter point; nonzero modular minors
provide lower bounds, not upper bounds from modular zeros. Their
saved integer matrices were checked again together with the other
40 minor certificates. The upper bound follows from the exact absence
of the tail symbols, not from sampling.

\begin{table}[h]
\centering
\caption{Exact metric-redesign certificates on ten reused source
circuits. The original isotropic ranks come from the earlier
certificates; each redesigned local XYZ metric uses evenly spaced
weights from 1 to 4, normalized to unit trace. The new rank equals
the source count $K$, with a nonzero exact minor and universal upper
bound $K$. This is not an expanded prevalence census.}
\label{tab:metricexactall}
\begin{tabular}{lrrrr}
\toprule
Source family & Qubits & Layers & Isotropic rank & Redesigned rank $=K$\\
\midrule
Hubregtsen & 2 & 1 & 0 & 2\\
Hubregtsen & 2 & 2 & 2 & 4\\
Hubregtsen & 4 & 1 & 8 & 8\\
Hubregtsen & 4 & 2 & 16 & 16\\
Kyriienko & 2 & 1 & 4 & 6\\
Kyriienko & 4 & 1 & 10 & 12\\
YZ-CX & 2 & 1 & 3 & 4\\
YZ-CX & 2 & 2 & 7 & 8\\
YZ-CX & 4 & 1 & 6 & 8\\
YZ-CX & 4 & 2 & 14 & 16\\
\bottomrule
\end{tabular}
\end{table}

The initial metric survey used 30 readout settings on these ten
circuits, with three design and five held-out parameter points and
96 input pairs per point. A fixed diagonal metric increased sampled
rank in 16 settings; the sampled dense-metric arms increased it in
17. Five validation points and three relative rank tolerances
($10^{-8},10^{-9},10^{-10}$, with absolute floor $10^{-11}$)
agreed. These counts remain numerical diagnostics. Exact certificates
were added for all ten local XYZ designs in
Table~\ref{tab:metricexactall} and ten selected one-basis designs.
The source and design files record which verification was added after
the numerical pilot; we do not present it as independent preregistration.

The one-basis search used Pauli words of weight at most two and
joint compatible parities, not one new circuit per observable.
Its selected designs reached full source rank in all ten cases,
but the best of the three uniform X, Y, or Z bases also reached
full sampled rank in all ten. Uniform X and uniform Y each did so
in eight cases. This strong control prevents attributing the result
to a generally superior combinatorial readout-search algorithm.

\section{Finite-shot design protocol and complete budget comparison}
\label{app:shotprotocol}
\subsection{Fixed design, common targets, and estimators}
The follow-up inventory contains three source circuits and four
readout cases, not four independent families: two-qubit, one-layer
YZ-CX/local XYZ and Kyriienko/local XYZ, and four-qubit, one-layer
Hubregtsen/local Z and Z+ZZ. Three fixed seed panels each use three
design points and five held-out points, with 16 disjoint input pairs
at each point, drawn uniformly from $(-0.95,0.95)^2$. The original
raw or arccos encoding is preserved. No labels, classification
datasets, or predictive metrics enter this experiment.

For each point, center the parameter derivatives of the reference
features by subtracting one input's derivatives. Their sampled row
space defines the target parameter subspace. Both Hubregtsen cases
use the \emph{local Z reference space}, of dimension four, to avoid
changing the target when adding ZZ features. The component invisible
to the initial isotropic kernel is recorded separately. A canonical
orthonormal basis is obtained from the subspace projector by
projected-coordinate Gram--Schmidt, with a positive largest component.
The basis, inputs, parameters, metrics and selection indices are
stored before the server run. No test-time SVD sign or favorable
direction selection is required.

The candidate bank has the identity, evenly spaced diagonal weights
from 1 to 4, sixteen random diagonals, and sixteen random dense
positive definite matrices. All have unit trace and condition number
at most four. Both selectors maximize the worst design-point target
rank, then the mean logarithm of the smallest retained singular value
of either $JQ$ or $W$ at 1,024 shots. The first random draw is an
untuned baseline, and every candidate's held-out diagnostics are
retained. The metric is fixed thereafter across points and budgets.
Thus noise-aware selection is an ideal-model finite-bank heuristic;
it is not trained on shot outcomes or task labels.

Let $\widehat a$ and $\widehat b$ be independent half-shot estimates
of the same feature difference $\delta$, each with covariance $S$.
The estimator $\widehat d^2=\widehat a^T M\widehat b$ obeys
\begin{align}
 \mathbb E\widehat d^2&=\delta^TM\delta,\label{eq:splitmean}\\
 \operatorname{Var}(\widehat d^2)
 &=2\delta^TMSM\delta+\Tr(MSMS).\label{eq:splitvariance}
\end{align}
To see this, write $\widehat a=\delta+e$ and
$\widehat b=\delta+e'$, with independent zero-mean errors. The two
linear terms have total variance $2\delta^TMSM\delta$; the bilinear
term has variance $\Tr(MSMS)$ and zero covariance with the linear
terms. Joint outcome covariances, including ZZ correlations, enter
$S$. Input pairs are disjoint so their estimates are independent.
Neither clipping negative distance estimates nor exponentiating
them is part of this unbiased-distance analysis.

For finite displacements $\theta\pm0.05v$ along \emph{each} frozen
target basis direction, the two state estimates use independent
draws. Their exact distance variances add. Every method receives
its own ideal matched linear contrast, proportional to the exact
pairwise signal divided by its variance and normalized to unit
Euclidean norm. This contrast is fixed before Monte Carlo sampling.
It treats methods symmetrically but uses oracle model information:
the resulting SNR is not calibrated detection power or a practical
shot-only identification algorithm. It is also distinct from the
local-limit singular-value statistic in Table~\ref{tab:shotdesign}.

\subsection{Measurement accounting and uncertainty}
Each observable is assigned to its first compatible product basis.
Within-basis covariances are preserved by joint multinomial sampling.
For requested budget $B$ and $L$ bases, each half/basis gets
$\lfloor B/(2L)\rfloor$ shots, so the actual budget is
$2L\lfloor B/(2L)\rfloor$ per input and parameter state. Local XYZ
uses three bases and hence actual budgets $252,1020,4092,16380$;
local Z and Z+ZZ use one basis and the requested budgets exactly.
For one direction and one replicate, the 16 pairs and two parameter
states require 64 times that per-input budget. Multiplication by
256 replicates and by the number of target directions gives the
simulation workload. Metrics reuse the same outcomes, and matched
Hubregtsen Z/Z+ZZ cases reuse the same ZZZZ draws. Observable count
is not substituted for basis count or total shots.

There are 60 panel/readout/probe records, not 60 independent
scientific problems. We report each panel separately and do not
interpret repeated probes, different budgets, or shared metric arms
as independent discoveries. The pooled variance-ratio percentiles
summarize calibration across correlated checks, not confidence
intervals. The largest absolute mean-bias/Monte-Carlo-SE diagnostic
is 4.07 among the many checked cells; it is not a single preselected
hypothesis test. Analytic unbiasedness is given by
Eq.~\eqref{eq:splitmean}, while empirical means, standard deviations,
and all samples are saved for inspection.

\begin{table}[htbp]
\centering
\caption{Complete requested-budget comparison. Triples are the
three seed-panel medians of five paired retained-sensitivity ratios
(noise-aware/baseline). D: fixed diagonal; R: rank-based selector;
U: first untuned random dense metric. Ratios are not pooled across
panels. A dash for local Z versus D avoids comparing different
ranks (three versus two); both remain below the four-dimensional
reference space. This table reports the local ideal-model diagnostic,
not the empirical matched-contrast SNR.}
\label{tab:allshotbudgets}
\small
\begin{tabular}{llccc}
\toprule
Case & Budget & Versus D & Versus R & Versus U\\
\midrule
YZ-CX & 256 & 2.037,1.689,2.346 & 1.000,1.328,1.612 & 1.000,1.312,2.054\\
 & 1024 & 1.837,1.665,2.336 & 1.000,1.304,1.608 & 1.000,1.279,2.110\\
 & 4096 & 1.767,1.653,2.329 & 1.000,1.303,1.608 & 1.000,1.274,2.135\\
 & 16384 & 1.748,1.651,2.327 & 1.000,1.304,1.608 & 1.000,1.273,2.142\\[3pt]
Kyriienko & 256 & 1.729,2.028,3.289 & 1.000,1.224,1.104 & 1.014,1.592,2.008\\
 & 1024 & 1.746,2.026,3.231 & 1.000,1.233,1.102 & 1.014,1.605,2.007\\
 & 4096 & 1.750,2.025,3.216 & 1.000,1.235,1.101 & 1.014,1.608,2.007\\
 & 16384 & 1.752,2.025,3.213 & 1.000,1.236,1.101 & 1.014,1.609,2.008\\[3pt]
Hubregtsen Z & 256 & -- & 1.000,1.000,1.000 & 0.800,0.818,1.009\\
 & 1024 & -- & 1.000,1.000,1.000 & 0.848,0.820,1.073\\
 & 4096 & -- & 1.000,1.000,1.000 & 0.869,0.822,1.119\\
 & 16384 & -- & 1.000,1.000,1.000 & 0.876,0.825,1.139\\[3pt]
Hubregtsen Z+ZZ & 256 & 1.320,1.669,1.069 & 1.000,1.000,1.000 & 0.775,1.000,0.698\\
 & 1024 & 1.264,1.485,0.970 & 1.000,1.000,1.000 & 0.774,1.000,0.671\\
 & 4096 & 1.219,1.316,0.913 & 1.000,1.000,1.000 & 0.770,1.000,0.658\\
 & 16384 & 1.216,1.221,0.891 & 1.000,1.000,1.000 & 0.765,1.000,0.654\\
\bottomrule
\end{tabular}
\end{table}

\subsection{Reproduction and what remains unproved}
The Linux full run used Python~3.8.10, NumPy~1.24.4 and
SymPy~1.13.3. A full macOS replay reproduced all 60 records and
1,658,880 saved Monte Carlo contrast values within relative tolerance
$10^{-9}$ and absolute tolerance $10^{-11}$; the maximum absolute
sample discrepancy was $1.04\times10^{-13}$. All 21 recorded package
file hashes matched, and sampled ranks were stable at the three
specified tolerances. Ten unit tests cover covariance, unbiased
variance, frozen splits, the common target, and shared Z/ZZ counts.
This is same-code cross-platform reproducibility, not independent
theoretical verification or device validation.

The preceding recovery survey also had a full Linux rerun. Its
exact certificates and rank diagnostics agreed. In its initial
finite-shot pilot, an unfrozen SVD direction reversed sign for one
YZ-CX case across platforms; this changed the allocation of random
draws to the two states, not the direction axis. Replaying the saved
server direction reproduced the statistics. The present follow-up
freezes the full direction basis and metric choices. Its symmetric
oracle contrasts are computed from exact model probabilities before
drawing shot samples, rather than fitted to the sampled outcomes.
An RNG seed alone was not treated as sufficient for reproducibility.

These ideal-model experiments neither compare supervised metric
learning nor establish predictive generalization, optimal measurement
selection, calibrated identification at fixed error, or QPU shot
savings. In particular, better retained sensitivity can coexist with
an unidentified direction, and full rank can coexist with a small
weakest sensitivity. Those distinctions, rather than a general
acceleration claim, connect the design study to the exact
identifiability analysis.

%% file: product_control.tex
\section{Classical product-feature estimator control}
\label{app:productcontrol}
This control was added after the finite-shot study to separate a
nonlinear representation change from the way that representation
is estimated. It uses the same frozen Hubregtsen circuit and ideal
Z-outcome distribution, not a new circuit family or hardware dataset.
Let $Z_q\in\{-1,1\}$ have mean $\mu_q$ at one input and parameter
state. The diagonal tail preserves the product-state computational
basis probabilities, so the four coordinates $Z_q$ are independent.

\subsection{Unbiasedness and conditional variance}
For $N$ shots in one independent half, define
$\bar Z_q=N^{-1}\sum_sZ_q^{(s)}$. Direct ZZ estimation uses
$N^{-1}\sum_sZ_q^{(s)}Z_r^{(s)}$, whereas the classical lift uses
$\bar Z_q\bar Z_r$. Independence across qubits makes both unbiased
for $\mu_q\mu_r$. Their variances are respectively
\begin{align}
 V_{\rm direct}&=(1-\mu_q^2\mu_r^2)/N,\\
 V_{\rm product}&=\frac{\mu_q^2(1-\mu_r^2)+\mu_r^2(1-\mu_q^2)}{N}
 +\frac{(1-\mu_q^2)(1-\mu_r^2)}{N^2}.
\end{align}
Their difference is
$(1-\mu_q^2)(1-\mu_r^2)(1/N-1/N^2)\ge0$.
This scalar observation alone is not sufficient for a dense metric,
because covariances between different features also matter.

The full comparison follows by conditioning. Given each qubit's
sample count in a half, its outcome sequence is uniformly permuted;
different qubit sequences remain conditionally independent. Therefore
the conditional expectation of each direct parity average equals the
product of its marginal means. Condition separately on the counts
of every input, parameter state, and independent half. The two halves
remain independent, and bilinearity gives
\begin{equation}
 \mathbb E[\widehat d^2_{\rm direct}\mid\text{marginal counts}]
 =\widehat d^2_{\rm product}.
\end{equation}
The same holds for any fixed linear contrast of distance differences.
The law of total variance proves nonincreasing variance, for any
fixed metric, without changing the target signal. This is the usual
conditional-expectation variance reduction, not a quantum advantage.
It assumes factorization of the Z-outcome distribution. For instance,
a distribution supported equally on $++$ and $--$ has zero local
means and unit ZZ mean, so a product-of-means substitute is incorrect.

\subsection{Exact covariance and frozen comparison}
For a feature indexed by a nonempty qubit set $I$, put
$\widehat f_I=\prod_{q\in I}\bar Z_q$. Independence gives
\begin{equation}
 \mathbb E[\widehat f_I\widehat f_J]
 =\prod_{q\in I\cap J}\left(\mu_q^2+\frac{1-\mu_q^2}{N}\right)
  \prod_{q\in I\triangle J}\mu_q.
\end{equation}
Subtracting $\prod_{q\in I}\mu_q\prod_{q\in J}\mu_q$ gives the
complete feature covariance. Summing the independent two-input
covariances gives $S$ in Eq.~\eqref{eq:splitvariance}. That variance
formula only needs independent, unbiased halves; it does not require
the feature estimator itself to be linear in outcomes.

We retain all three frozen panels, five held-out points, four target
directions, six metric arms and four budgets, with 256 replicates.
The 1,440 comparison records share data and metrics and are not
independent discoveries. No new metric or favorable contrast is
selected: the direct-parity oracle contrast is fixed for both arms.
At each state, both estimates use exactly the same bitstring counts
and total shot budget. Table~\ref{tab:productcontrol} reports the
frozen noise-aware arm; the data retain all six arms.

\begin{table}[h]
\centering
\caption{Product/direct exact contrast-variance ratios, summarized
separately by panel. Each cell is the median of 20 ratios (five
held-out points and four fixed directions); smaller is better.
The metric, direct-parity contrast and underlying shots are identical
between estimators. These are model-based variances, not confidence
intervals, hardware-shot reductions or predictive gains.}
\label{tab:productcontrol}
\begin{tabular}{rrrr}
\toprule
Shots per input/state & Panel 0 & Panel 1 & Panel 2\\
\midrule
256 & 0.864 & 0.789 & 0.838\\
1,024 & 0.884 & 0.819 & 0.858\\
4,096 & 0.895 & 0.829 & 0.869\\
16,384 & 0.899 & 0.838 & 0.872\\
\bottomrule
\end{tabular}
\end{table}

The product-estimator empirical/exact variance ratios have pooled
median $0.990$ and 5th--95th percentiles $[0.869,1.120]$ across
the correlated comparison records. At 4,096 shots, panel-median
empirical mean-squared-error ratios are $0.890,0.827,0.907$.
These are descriptive Monte Carlo checks; unbiasedness follows from
the factorization argument, not a significance test. Dense-state
probabilities agree with product probabilities within
$1.7\times10^{-15}$, and exact-feature products agree within
$1.8\times10^{-15}$. Direct-parity samples match the saved Linux
study within $5.6\times10^{-14}$. The new arm was run locally with
Python 3.9.6 and NumPy 1.24.4; server reproduction remains pending.
Four unit tests include exact enumeration of all 256 two-shot,
four-qubit outcome sequences and their conditional expectations.
The separate control bundle retains its pre-execution scope,
frozen-plan hash, code, per-record variances and paired samples.

%% file: one_product_appendix.tex
\section{Exact certificate for one product and boundaries of the design class}
\label{app:oneproduct}
Theorem~\ref{thm:oneproduct} is proved by algebra in the main text,
not by a numerical rank estimate. The accompanying supplement
contains \texttt{verification\_math/verify\_one\_product.py} and
\texttt{verification\_math/test\_verify\_one\_product.py}.
Seventeen tests in exact arithmetic check the metric eigenvalues,
coefficient Gram, amplitude determinant, angle chain rule, witness,
normalization to unit trace, independence on the actual domain,
the ceiling for the original features, and counterexamples to
stronger claims. Direct evaluation of
the four input pairs from $(0,0)$ to $(\pi/6,0)$, $(\pi/4,0)$,
$(0,\pi/6)$ and $(\pi/6,\pi/6)$ also gives an invertible exact
map from the four Gram entries to squared distances. All these inputs
lie in the frozen circuit's domain $(-1,1)^2$.
The script reports the exact values and tests nonvanishing; the
symbolic computation is not a formal verification by a proof assistant.

\paragraph{Basis convention.}
Our witness $-9/128$ is for $h=(u,v,u^2)$ with
$u=\cos x_0$, $v=\cos x_1$. Although $u^2$ has a constant Fourier
component, $1,u,v,u^2$ are independent, as the theorem requires.
In the pure Fourier basis $(\cos x_0,\cos x_1,\cos 2x_0)$,
the last feature is $ac/2+(ac/2)\cos 2x_0$. Separating its offset
divides the last Gram entry by four and gives the minor $-9/512$
(or $-9/1229312$ at unit trace). Both conventions certify rank four.

\paragraph{A diagonal metric and one pairwise product.}
With a positive diagonal metric on the original means and one
additional $f_i f_j$, the latter is a single data monomial
$p(\theta)g(x)$, where $g$ is $u^2$, $uv$ or $v^2$.
All distances then depend on just three coefficient combinations,
\[
 w_a a^2+w_c c^2,\qquad
 w_b b^2+w_d d^2,\qquad w_p p(\theta)^2.
\]
Consequently their differential rank is at most three. This claim
allows repeated indices (squares), but concerns one product, not
one arbitrary scalar function.

\paragraph{Why a broader impossibility claim for diagonal metrics is false.}
The single appended scalar
$g=f_0f_2+f_1f_3=ac u^2+bd v^2$ is already sufficient with
any fixed positive diagonal metric. In basis $(u,v,u^2,v^2)$,
four Gram entries are
\[
 w_a a^2+w_c c^2,\quad w_g a^2c^2,\quad
 w_b b^2+w_d d^2,\quad w_g b^2d^2.
\]
Their Jacobian in order $(a,b,c,d)$ has determinant
\[
 -16abcd\,w_g^2(w_a a^2-w_c c^2)(w_b b^2-w_d d^2),
\]
which is not identically zero. This provides an explicit boundary
on what the diagonal construction with two products proves. It is
not a new result about measurement information, and this estimator
of a sum of products is not one of the four experiment arms in
Table~\ref{tab:minimal_product_matched}.

\paragraph{Local recovery is not global recovery.}
At the two positive amplitude points exhibited after
Theorem~\ref{thm:oneproduct}, the unnormalized Gram entries are
$(3/4,1/8,3/8,1/16)$ and the amplitude determinants are $-1/8$
and $1/16$, respectively. Thus even two regular points can have
identical kernels on all input pairs. At $\theta=0$ the derivative
vanishes entirely. Exact generic rank four must therefore not be
read as rank four everywhere or globally unique parameter estimation.

%% file: census_traceability.tex
\section{Census evidence and reproduction routes}
\label{sec:census-traceability}
The census contains 24 product angle settings and 30 disjoint settings
with fixed frequencies. It does not apply the fixed finite Fourier
basis theorem to all 54 cases.
Table~\ref{tab:census-routes} separates their assumptions and evidence.
The supplement's \texttt{census\_54\_mapping.csv} and
\texttt{census\_54\_mapping.json} give one row per setting, its exact
rank, frozen input and certificate paths and SHA-256 hashes, the
JSON fields for the upper proof and lower witness, and the commands
for replay and fresh computation. Paths are relative to the public
repository at commit
\texttt{d53cdfe9ee254c0d687ebec28f13584218f5ff04}.
The index is a new supplement, not a file claimed to exist in that commit.

\begin{table}[htbp]
\caption{Two certificate routes in the same census of 54 settings.
The ranks are those in Tables~\ref{tab:dimension} and
\ref{tab:extension}; repeated verification does not add settings.
Empty minors for rank zero are not witnesses of positive rank.}
\label{tab:census-routes}
\begin{tabular}{>{\raggedright\arraybackslash}p{.22\linewidth}
                >{\raggedright\arraybackslash}p{.34\linewidth}
                >{\raggedright\arraybackslash}p{.35\linewidth}}
\toprule
Settings and semantic contract & Upper bound & Lower bound and replay scope\\
\midrule
24 ChebyshevRx and ParamZ settings; shared products of trainable
parameters and data, not a fixed finite Fourier basis
& Exact formal coefficient identities and constant null directions
(App.~\ref{app:dimension})
& 22 nonzero minors on feasible input pairs, with interval enclosures;
two exactly constant kernels. Public \texttt{core-audit} recomputes
the identities and replays the witnesses at 65 digits.\\[4pt]
30 YZ-CX, Hubregtsen, and Kyriienko settings; fixed frequencies and
independent data modes on an open domain
& Exact invariance fields for two copies and completed symbolic
Gram span bounds (App.~\ref{app:followup}); the index records which
bound closes each setting
& Two modular records per setting: 54 nonempty minors and six empty
minors for rank zero. Seven rational records include one empty minor.
Public \texttt{core-audit} only aggregates these stored Gram records;
\texttt{core-full} recomputes them.\\
\bottomrule
\end{tabular}
\end{table}

A nonzero reduced minor is an exact lower certificate because the
parameter point lies on rational circles and the rational coefficients
have denominators invertible at the chosen prime: reduction is a ring
homomorphism on this localized coefficient ring. It is neither a
numerical rank estimate nor a proof of an upper bound.
The index checker verifies the stored denominator conditions and
agreement of tables and hashes. Its optional modular replay
regenerates all 60 modular records with the released Laurent engine
and checks the saved minors, but does not regenerate the symbolic
upper proofs. For fresh generation of both census routes, the
supported public command is
\texttt{python reproduce.py core-full -{}-workers 2}.
These are software certificates for the ideal model, not QPU
measurements or verification by an independent formal proof system.